\documentclass[a4paper,twocolumn,traditabstract]{aa}
\usepackage{graphicx}
\usepackage[breaklinks, colorlinks, citecolor=blue]{hyperref}
\usepackage{natbib}
\usepackage{stfloats}
\usepackage{subfigure}
\usepackage{amsmath}
\usepackage{epsfig}
\usepackage{psfrag}
\usepackage{multirow}
\usepackage{color}
\usepackage{txfonts}
\newcommand{\ie}{{i.e.}}

\usepackage{url}
 
\newcommand{\healpix}{{\tt HEALPix}}

\def\pdeg{\ifmmode $\setbox0=\hbox{$^{\circ}$}\rlap{\hskip.11\wd0 .}$^{\circ}
          \else \setbox0=\hbox{$^{\circ}$}\rlap{\hskip.11\wd0 .}$^{\circ}$\fi}

\def\setsymbol#1#2{\expandafter\def\csname #1\endcsname{#2}}
\def\getsymbol#1{\csname #1\endcsname}

\def\Planck{\textit{Planck}}

\def\all2013resultspapers{\nocite{planck2013-p01, planck2013-p02, planck2013-p02a, planck2013-p02d, planck2013-p02b, planck2013-p03, planck2013-p03c, planck2013-p03f, planck2013-p03d, planck2013-p03e, planck2013-p01a, planck2013-p06, planck2013-p03a, planck2013-pip88, planck2013-p08, planck2013-p11, planck2013-p12, planck2013-p13, planck2013-p14, planck2013-p15, planck2013-p05b, planck2013-p17, planck2013-p09, planck2013-p09a, planck2013-p20, planck2013-p19, planck2013-pipaberration, planck2013-p05, planck2013-p05a, planck2013-pip56, planck2013-p06b}}

\newbox\tablebox    \newdimen\tablewidth
\def\leaderfil{\leaders\hbox to 5pt{\hss.\hss}\hfil}
\def\tablenote#1 #2\par{\begingroup \parindent=0.8em
    \abovedisplayshortskip=0pt\belowdisplayshortskip=0pt
    \noindent
    $$\hss\vbox{\hsize\tablewidth \hangindent=\parindent \hangafter=1 \noindent
    \hbox to \parindent{$^#1$\hss}\strut#2\strut\par}\hss$$
    \endgroup}

\def\L2{\ifmmode L_2\else $L_2$\fi}

\def\DeltaT{\ifmmode \Delta T\else $\Delta T$\fi}
\def\deltat{\ifmmode \Delta t\else $\Delta t$\fi}
\def\fknee{\ifmmode f_{\rm knee}\else $f_{\rm knee}$\fi}
\def\Fmax{\ifmmode F_{\rm max}\else $F_{\rm max}$\fi}
\def\solar{\ifmmode{\rm M}_{\mathord\odot}\else${\rm M}_{\mathord\odot}$\fi}
\def\Msolar{\ifmmode{\rm M}_{\mathord\odot}\else${\rm M}_{\mathord\odot}$\fi}
\def\Lsolar{\ifmmode{\rm L}_{\mathord\odot}\else${\rm L}_{\mathord\odot}$\fi}

\def\inv{\ifmmode^{-1}\else$^{-1}$\fi}
\def\mo{\ifmmode^{-1}\else$^{-1}$\fi}
\def\sup#1{\ifmmode ^{\rm #1}\else $^{\rm #1}$\fi}
\def\expo#1{\ifmmode \times 10^{#1}\else $\times 10^{#1}$\fi}
\def\,{\thinspace}
\def\lsim{\mathrel{\raise .4ex\hbox{\rlap{$<$}\lower 1.2ex\hbox{$\sim$}}}}
\def\gsim{\mathrel{\raise .4ex\hbox{\rlap{$>$}\lower 1.2ex\hbox{$\sim$}}}}

\def\simprop{\mathrel{\raise .4ex\hbox{\rlap{$\propto$}\lower 1.2ex\hbox{$\sim$}}}}
\def\deg{\ifmmode^\circ\else$^\circ$\fi}
\def\pdeg{\ifmmode $\setbox0=\hbox{$^{\circ}$}\rlap{\hskip.11\wd0 .}$^{\circ}
          \else \setbox0=\hbox{$^{\circ}$}\rlap{\hskip.11\wd0 .}$^{\circ}$\fi}
\def\arcs{\ifmmode {^{\scriptstyle\prime\prime}}
          \else $^{\scriptstyle\prime\prime}$\fi}
\def\arcm{\ifmmode {^{\scriptstyle\prime}}
          \else $^{\scriptstyle\prime}$\fi}
\newdimen\sa  \newdimen\sb
\def\parcs{\sa=.07em \sb=.03em
     \ifmmode \hbox{\rlap{.}}^{\scriptstyle\prime\kern -\sb\prime}\hbox{\kern -\sa}
     \else \rlap{.}$^{\scriptstyle\prime\kern -\sb\prime}$\kern -\sa\fi}
\def\parcm{\sa=.08em \sb=.03em
     \ifmmode \hbox{\rlap{.}\kern\sa}^{\scriptstyle\prime}\hbox{\kern-\sb}
     \else \rlap{.}\kern\sa$^{\scriptstyle\prime}$\kern-\sb\fi}
\def\ra[#1 #2 #3.#4]{#1\sup{h}#2\sup{m}#3\sup{s}\llap.#4}
\def\dec[#1 #2 #3.#4]{#1\deg#2\arcm#3\arcs\llap.#4}
\def\deco[#1 #2 #3]{#1\deg#2\arcm#3\arcs}
\def\rra[#1 #2]{#1\sup{h}#2\sup{m}}

\def\dots{\relax\ifmmode \ldots\else $\ldots$\fi}
\def\WHzsr{\ifmmode $W\,Hz\mo\,sr\mo$\else W\,Hz\mo\,sr\mo\fi}
\def\mHz{\ifmmode $\,mHz$\else \,mHz\fi}
\def\GHz{\ifmmode $\,GHz$\else \,GHz\fi}
\def\mKs{\ifmmode $\,mK\,s$^{1/2}\else \,mK\,s$^{1/2}$\fi}
\def\muKs{\ifmmode \,\mu$K\,s$^{1/2}\else \,$\mu$K\,s$^{1/2}$\fi}
\def\muKRJs{\ifmmode \,\mu$K$_{\rm RJ}$\,s$^{1/2}\else \,$\mu$K$_{\rm RJ}$\,s$^{1/2}$\fi}
\def\muKHz{\ifmmode \,\mu$K\,Hz$^{-1/2}\else \,$\mu$K\,Hz$^{-1/2}$\fi}
\def\MJysr{\ifmmode \,$MJy\,sr\mo$\else \,MJy\,sr\mo\fi}
\def\MJysrmK{\ifmmode \,$MJy\,sr\mo$\,mK$_{\rm CMB}\mo\else \,MJy\,sr\mo\,mK$_{\rm CMB}\mo$\fi}
\def\microns{\ifmmode \,\mu$m$\else \,$\mu$m\fi}

\def\muK{\ifmmode \,\mu$K$\else \,$\mu$\hbox{K}\fi}
\def\microK{\ifmmode \,\mu$K$\else \,$\mu$\hbox{K}\fi}
\def\muW{\ifmmode \,\mu$W$\else \,$\mu$\hbox{W}\fi}
\def\kms{\ifmmode $\,km\,s$^{-1}\else \,km\,s$^{-1}$\fi}
\def\kmsMpc{\ifmmode $\,\kms\,Mpc\mo$\else \,\kms\,Mpc\mo\fi}
\setsymbol{LFI:center:frequency:70GHz:units}{70.3\,GHz}
\setsymbol{LFI:center:frequency:44GHz:units}{44.1\,GHz}
\setsymbol{LFI:center:frequency:30GHz:units}{28.5\,GHz}

\setsymbol{LFI:center:frequency:70GHz}{70.3}
\setsymbol{LFI:center:frequency:44GHz}{44.1}
\setsymbol{LFI:center:frequency:30GHz}{28.5}

\setsymbol{LFI:center:frequency:LFI18:Rad:M:units}{71.7\GHz}
\setsymbol{LFI:center:frequency:LFI19:Rad:M:units}{67.5\GHz}
\setsymbol{LFI:center:frequency:LFI20:Rad:M:units}{69.2\GHz}
\setsymbol{LFI:center:frequency:LFI21:Rad:M:units}{70.4\GHz}
\setsymbol{LFI:center:frequency:LFI22:Rad:M:units}{71.5\GHz}
\setsymbol{LFI:center:frequency:LFI23:Rad:M:units}{70.8\GHz}
\setsymbol{LFI:center:frequency:LFI24:Rad:M:units}{44.4\GHz}
\setsymbol{LFI:center:frequency:LFI25:Rad:M:units}{44.0\GHz}
\setsymbol{LFI:center:frequency:LFI26:Rad:M:units}{43.9\GHz}
\setsymbol{LFI:center:frequency:LFI27:Rad:M:units}{28.3\GHz}
\setsymbol{LFI:center:frequency:LFI28:Rad:M:units}{28.8\GHz}
\setsymbol{LFI:center:frequency:LFI18:Rad:S:units}{70.1\GHz}
\setsymbol{LFI:center:frequency:LFI19:Rad:S:units}{69.6\GHz}
\setsymbol{LFI:center:frequency:LFI20:Rad:S:units}{69.5\GHz}
\setsymbol{LFI:center:frequency:LFI21:Rad:S:units}{69.5\GHz}
\setsymbol{LFI:center:frequency:LFI22:Rad:S:units}{72.8\GHz}
\setsymbol{LFI:center:frequency:LFI23:Rad:S:units}{71.3\GHz}
\setsymbol{LFI:center:frequency:LFI24:Rad:S:units}{44.1\GHz}
\setsymbol{LFI:center:frequency:LFI25:Rad:S:units}{44.1\GHz}
\setsymbol{LFI:center:frequency:LFI26:Rad:S:units}{44.1\GHz}
\setsymbol{LFI:center:frequency:LFI27:Rad:S:units}{28.5\GHz}
\setsymbol{LFI:center:frequency:LFI28:Rad:S:units}{28.2\GHz}

\setsymbol{LFI:center:frequency:LFI18:Rad:M}{71.7}
\setsymbol{LFI:center:frequency:LFI19:Rad:M}{67.5}
\setsymbol{LFI:center:frequency:LFI20:Rad:M}{69.2}
\setsymbol{LFI:center:frequency:LFI21:Rad:M}{70.4}
\setsymbol{LFI:center:frequency:LFI22:Rad:M}{71.5}
\setsymbol{LFI:center:frequency:LFI23:Rad:M}{70.8}
\setsymbol{LFI:center:frequency:LFI24:Rad:M}{44.4}
\setsymbol{LFI:center:frequency:LFI25:Rad:M}{44.0}
\setsymbol{LFI:center:frequency:LFI26:Rad:M}{43.9}
\setsymbol{LFI:center:frequency:LFI27:Rad:M}{28.3}
\setsymbol{LFI:center:frequency:LFI28:Rad:M}{28.8}
\setsymbol{LFI:center:frequency:LFI18:Rad:S}{70.1}
\setsymbol{LFI:center:frequency:LFI19:Rad:S}{69.6}
\setsymbol{LFI:center:frequency:LFI20:Rad:S}{69.5}
\setsymbol{LFI:center:frequency:LFI21:Rad:S}{69.5}
\setsymbol{LFI:center:frequency:LFI22:Rad:S}{72.8}
\setsymbol{LFI:center:frequency:LFI23:Rad:S}{71.3}
\setsymbol{LFI:center:frequency:LFI24:Rad:S}{44.1}
\setsymbol{LFI:center:frequency:LFI25:Rad:S}{44.1}
\setsymbol{LFI:center:frequency:LFI26:Rad:S}{44.1}
\setsymbol{LFI:center:frequency:LFI27:Rad:S}{28.5}
\setsymbol{LFI:center:frequency:LFI28:Rad:S}{28.2}

\setsymbol{LFI:white:noise:sensitivity:70GHz:units}{134.7\muKs}
\setsymbol{LFI:white:noise:sensitivity:44GHz:units}{164.7\muKs}
\setsymbol{LFI:white:noise:sensitivity:30GHz:units}{143.4\muKs}

\setsymbol{LFI:white:noise:sensitivity:70GHz}{134.7}
\setsymbol{LFI:white:noise:sensitivity:44GHz}{164.7}
\setsymbol{LFI:white:noise:sensitivity:30GHz}{143.4}

\setsymbol{LFI:white:noise:sensitivity:LFI18:Rad:M:units}{512.0\muKs}
\setsymbol{LFI:white:noise:sensitivity:LFI19:Rad:M:units}{581.4\muKs}
\setsymbol{LFI:white:noise:sensitivity:LFI20:Rad:M:units}{590.8\muKs}
\setsymbol{LFI:white:noise:sensitivity:LFI21:Rad:M:units}{455.2\muKs}
\setsymbol{LFI:white:noise:sensitivity:LFI22:Rad:M:units}{492.0\muKs}
\setsymbol{LFI:white:noise:sensitivity:LFI23:Rad:M:units}{507.7\muKs}
\setsymbol{LFI:white:noise:sensitivity:LFI24:Rad:M:units}{462.2\muKs}
\setsymbol{LFI:white:noise:sensitivity:LFI25:Rad:M:units}{413.6\muKs}
\setsymbol{LFI:white:noise:sensitivity:LFI26:Rad:M:units}{478.6\muKs}
\setsymbol{LFI:white:noise:sensitivity:LFI27:Rad:M:units}{277.7\muKs}
\setsymbol{LFI:white:noise:sensitivity:LFI28:Rad:M:units}{312.3\muKs}
\setsymbol{LFI:white:noise:sensitivity:LFI18:Rad:S:units}{465.7\muKs}
\setsymbol{LFI:white:noise:sensitivity:LFI19:Rad:S:units}{555.6\muKs}
\setsymbol{LFI:white:noise:sensitivity:LFI20:Rad:S:units}{623.2\muKs}
\setsymbol{LFI:white:noise:sensitivity:LFI21:Rad:S:units}{564.1\muKs}
\setsymbol{LFI:white:noise:sensitivity:LFI22:Rad:S:units}{534.4\muKs}
\setsymbol{LFI:white:noise:sensitivity:LFI23:Rad:S:units}{542.4\muKs}
\setsymbol{LFI:white:noise:sensitivity:LFI24:Rad:S:units}{399.2\muKs}
\setsymbol{LFI:white:noise:sensitivity:LFI25:Rad:S:units}{392.6\muKs}
\setsymbol{LFI:white:noise:sensitivity:LFI26:Rad:S:units}{418.6\muKs}
\setsymbol{LFI:white:noise:sensitivity:LFI27:Rad:S:units}{302.9\muKs}
\setsymbol{LFI:white:noise:sensitivity:LFI28:Rad:S:units}{285.3\muKs}

\setsymbol{LFI:white:noise:sensitivity:LFI18:Rad:M}{512.0}
\setsymbol{LFI:white:noise:sensitivity:LFI19:Rad:M}{581.4}
\setsymbol{LFI:white:noise:sensitivity:LFI20:Rad:M}{590.8}
\setsymbol{LFI:white:noise:sensitivity:LFI21:Rad:M}{455.2}
\setsymbol{LFI:white:noise:sensitivity:LFI22:Rad:M}{492.0}
\setsymbol{LFI:white:noise:sensitivity:LFI23:Rad:M}{507.7}
\setsymbol{LFI:white:noise:sensitivity:LFI24:Rad:M}{462.2}
\setsymbol{LFI:white:noise:sensitivity:LFI25:Rad:M}{413.6}
\setsymbol{LFI:white:noise:sensitivity:LFI26:Rad:M}{478.6}
\setsymbol{LFI:white:noise:sensitivity:LFI27:Rad:M}{277.7}
\setsymbol{LFI:white:noise:sensitivity:LFI28:Rad:M}{312.3}
\setsymbol{LFI:white:noise:sensitivity:LFI18:Rad:S}{465.7}
\setsymbol{LFI:white:noise:sensitivity:LFI19:Rad:S}{555.6}
\setsymbol{LFI:white:noise:sensitivity:LFI20:Rad:S}{623.2}
\setsymbol{LFI:white:noise:sensitivity:LFI21:Rad:S}{564.1}
\setsymbol{LFI:white:noise:sensitivity:LFI22:Rad:S}{534.4}
\setsymbol{LFI:white:noise:sensitivity:LFI23:Rad:S}{542.4}
\setsymbol{LFI:white:noise:sensitivity:LFI24:Rad:S}{399.2}
\setsymbol{LFI:white:noise:sensitivity:LFI25:Rad:S}{392.6}
\setsymbol{LFI:white:noise:sensitivity:LFI26:Rad:S}{418.6}
\setsymbol{LFI:white:noise:sensitivity:LFI27:Rad:S}{302.9}
\setsymbol{LFI:white:noise:sensitivity:LFI28:Rad:S}{285.3}

\setsymbol{LFI:knee:frequency:70GHz:units}{29.5\mHz}
\setsymbol{LFI:knee:frequency:44GHz:units}{56.2\mHz}
\setsymbol{LFI:knee:frequency:30GHz:units}{113.7\mHz}

\setsymbol{LFI:knee:frequency:70GHz}{29.5}
\setsymbol{LFI:knee:frequency:44GHz}{56.2}
\setsymbol{LFI:knee:frequency:30GHz}{113.7}

\setsymbol{LFI:knee:frequency:LFI18:Rad:M:units}{16.3\mHz}
\setsymbol{LFI:knee:frequency:LFI19:Rad:M:units}{15.1\mHz}
\setsymbol{LFI:knee:frequency:LFI20:Rad:M:units}{18.7\mHz}
\setsymbol{LFI:knee:frequency:LFI21:Rad:M:units}{37.2\mHz}
\setsymbol{LFI:knee:frequency:LFI22:Rad:M:units}{12.7\mHz}
\setsymbol{LFI:knee:frequency:LFI23:Rad:M:units}{34.6\mHz}
\setsymbol{LFI:knee:frequency:LFI24:Rad:M:units}{46.2\mHz}
\setsymbol{LFI:knee:frequency:LFI25:Rad:M:units}{24.9\mHz}
\setsymbol{LFI:knee:frequency:LFI26:Rad:M:units}{67.6\mHz}
\setsymbol{LFI:knee:frequency:LFI27:Rad:M:units}{187.4\mHz}
\setsymbol{LFI:knee:frequency:LFI28:Rad:M:units}{122.2\mHz}
\setsymbol{LFI:knee:frequency:LFI18:Rad:S:units}{17.7\mHz}
\setsymbol{LFI:knee:frequency:LFI19:Rad:S:units}{22.0\mHz}
\setsymbol{LFI:knee:frequency:LFI20:Rad:S:units}{8.7\mHz}
\setsymbol{LFI:knee:frequency:LFI21:Rad:S:units}{25.9\mHz}
\setsymbol{LFI:knee:frequency:LFI22:Rad:S:units}{15.8\mHz}
\setsymbol{LFI:knee:frequency:LFI23:Rad:S:units}{129.8\mHz}
\setsymbol{LFI:knee:frequency:LFI24:Rad:S:units}{100.9\mHz}
\setsymbol{LFI:knee:frequency:LFI25:Rad:S:units}{38.9\mHz}
\setsymbol{LFI:knee:frequency:LFI26:Rad:S:units}{58.9\mHz}
\setsymbol{LFI:knee:frequency:LFI27:Rad:S:units}{104.4\mHz}
\setsymbol{LFI:knee:frequency:LFI28:Rad:S:units}{40.7\mHz}

\setsymbol{LFI:knee:frequency:LFI18:Rad:M}{16.3}
\setsymbol{LFI:knee:frequency:LFI19:Rad:M}{15.1}
\setsymbol{LFI:knee:frequency:LFI20:Rad:M}{18.7}
\setsymbol{LFI:knee:frequency:LFI21:Rad:M}{37.2}
\setsymbol{LFI:knee:frequency:LFI22:Rad:M}{12.7}
\setsymbol{LFI:knee:frequency:LFI23:Rad:M}{34.6}
\setsymbol{LFI:knee:frequency:LFI24:Rad:M}{46.2}
\setsymbol{LFI:knee:frequency:LFI25:Rad:M}{24.9}
\setsymbol{LFI:knee:frequency:LFI26:Rad:M}{67.6}
\setsymbol{LFI:knee:frequency:LFI27:Rad:M}{187.4}
\setsymbol{LFI:knee:frequency:LFI28:Rad:M}{122.2}
\setsymbol{LFI:knee:frequency:LFI18:Rad:S}{17.7}
\setsymbol{LFI:knee:frequency:LFI19:Rad:S}{22.0}
\setsymbol{LFI:knee:frequency:LFI20:Rad:S}{8.7}
\setsymbol{LFI:knee:frequency:LFI21:Rad:S}{25.9}
\setsymbol{LFI:knee:frequency:LFI22:Rad:S}{15.8}
\setsymbol{LFI:knee:frequency:LFI23:Rad:S}{129.8}
\setsymbol{LFI:knee:frequency:LFI24:Rad:S}{100.9}
\setsymbol{LFI:knee:frequency:LFI25:Rad:S}{38.9}
\setsymbol{LFI:knee:frequency:LFI26:Rad:S}{58.9}
\setsymbol{LFI:knee:frequency:LFI27:Rad:S}{104.4}
\setsymbol{LFI:knee:frequency:LFI28:Rad:S}{40.7}

\setsymbol{LFI:slope:70GHz:units}{$-1.03$\mHz}
\setsymbol{LFI:slope:44GHz:units}{$-0.89$\mHz}
\setsymbol{LFI:slope:30GHz:units}{$-0.87$\mHz}

\setsymbol{LFI:slope:70GHz}{$-1.03$}
\setsymbol{LFI:slope:44GHz}{$-0.89$}
\setsymbol{LFI:slope:30GHz}{$-0.87$}

\setsymbol{LFI:slope:LFI18:Rad:M:units}{$-1.04$\mHz}
\setsymbol{LFI:slope:LFI19:Rad:M:units}{$-1.09$\mHz}
\setsymbol{LFI:slope:LFI20:Rad:M:units}{$-0.69$\mHz}
\setsymbol{LFI:slope:LFI21:Rad:M:units}{$-1.56$\mHz}
\setsymbol{LFI:slope:LFI22:Rad:M:units}{$-1.01$\mHz}
\setsymbol{LFI:slope:LFI23:Rad:M:units}{$-0.96$\mHz}
\setsymbol{LFI:slope:LFI24:Rad:M:units}{$-0.83$\mHz}
\setsymbol{LFI:slope:LFI25:Rad:M:units}{$-0.91$\mHz}
\setsymbol{LFI:slope:LFI26:Rad:M:units}{$-0.95$\mHz}
\setsymbol{LFI:slope:LFI27:Rad:M:units}{$-0.87$\mHz}
\setsymbol{LFI:slope:LFI28:Rad:M:units}{$-0.88$\mHz}
\setsymbol{LFI:slope:LFI18:Rad:S:units}{$-1.15$\mHz}
\setsymbol{LFI:slope:LFI19:Rad:S:units}{$-1.00$\mHz}
\setsymbol{LFI:slope:LFI20:Rad:S:units}{$-0.95$\mHz}
\setsymbol{LFI:slope:LFI21:Rad:S:units}{$-0.92$\mHz}
\setsymbol{LFI:slope:LFI22:Rad:S:units}{$-1.01$\mHz}
\setsymbol{LFI:slope:LFI23:Rad:S:units}{$-0.95$\mHz}
\setsymbol{LFI:slope:LFI24:Rad:S:units}{$-0.73$\mHz}
\setsymbol{LFI:slope:LFI25:Rad:S:units}{$-1.16$\mHz}
\setsymbol{LFI:slope:LFI26:Rad:S:units}{$-0.79$\mHz}
\setsymbol{LFI:slope:LFI27:Rad:S:units}{$-0.82$\mHz}
\setsymbol{LFI:slope:LFI28:Rad:S:units}{$-0.91$\mHz}

\setsymbol{LFI:slope:LFI18:Rad:M}{$-1.04$}
\setsymbol{LFI:slope:LFI19:Rad:M}{$-1.09$}
\setsymbol{LFI:slope:LFI20:Rad:M}{$-0.69$}
\setsymbol{LFI:slope:LFI21:Rad:M}{$-1.56$}
\setsymbol{LFI:slope:LFI22:Rad:M}{$-1.01$}
\setsymbol{LFI:slope:LFI23:Rad:M}{$-0.96$}
\setsymbol{LFI:slope:LFI24:Rad:M}{$-0.83$}
\setsymbol{LFI:slope:LFI25:Rad:M}{$-0.91$}
\setsymbol{LFI:slope:LFI26:Rad:M}{$-0.95$}
\setsymbol{LFI:slope:LFI27:Rad:M}{$-0.87$}
\setsymbol{LFI:slope:LFI28:Rad:M}{$-0.88$}
\setsymbol{LFI:slope:LFI18:Rad:S}{$-1.15$}
\setsymbol{LFI:slope:LFI19:Rad:S}{$-1.00$}
\setsymbol{LFI:slope:LFI20:Rad:S}{$-0.95$}
\setsymbol{LFI:slope:LFI21:Rad:S}{$-0.92$}
\setsymbol{LFI:slope:LFI22:Rad:S}{$-1.01$}
\setsymbol{LFI:slope:LFI23:Rad:S}{$-0.95$}
\setsymbol{LFI:slope:LFI24:Rad:S}{$-0.73$}
\setsymbol{LFI:slope:LFI25:Rad:S}{$-1.16$}
\setsymbol{LFI:slope:LFI26:Rad:S}{$-0.79$}
\setsymbol{LFI:slope:LFI27:Rad:S}{$-0.82$}
\setsymbol{LFI:slope:LFI28:Rad:S}{$-0.91$}

\setsymbol{LFI:FWHM:70GHz:units}{13\parcm01}
\setsymbol{LFI:FWHM:44GHz:units}{27\parcm92}
\setsymbol{LFI:FWHM:30GHz:units}{32\parcm65}

\setsymbol{LFI:FWHM:70GHz}{13.01}
\setsymbol{LFI:FWHM:44GHz}{27.92}
\setsymbol{LFI:FWHM:30GHz}{32.65}

\setsymbol{LFI:FWHM:LFI18:units}{13\parcm39}
\setsymbol{LFI:FWHM:LFI19:units}{13\parcm01}
\setsymbol{LFI:FWHM:LFI20:units}{12\parcm75}
\setsymbol{LFI:FWHM:LFI21:units}{12\parcm74}
\setsymbol{LFI:FWHM:LFI22:units}{12\parcm87}
\setsymbol{LFI:FWHM:LFI23:units}{13\parcm27}
\setsymbol{LFI:FWHM:LFI24:units}{22\parcm98}
\setsymbol{LFI:FWHM:LFI25:units}{30\parcm46}
\setsymbol{LFI:FWHM:LFI26:units}{30\parcm31}
\setsymbol{LFI:FWHM:LFI27:units}{32\parcm65}
\setsymbol{LFI:FWHM:LFI28:units}{32\parcm66}

\setsymbol{LFI:FWHM:LFI18}{13.39}
\setsymbol{LFI:FWHM:LFI19}{13.01}
\setsymbol{LFI:FWHM:LFI20}{12.75}
\setsymbol{LFI:FWHM:LFI21}{12.74}
\setsymbol{LFI:FWHM:LFI22}{12.87}
\setsymbol{LFI:FWHM:LFI23}{13.27}
\setsymbol{LFI:FWHM:LFI24}{22.98}
\setsymbol{LFI:FWHM:LFI25}{30.46}
\setsymbol{LFI:FWHM:LFI26}{30.31}
\setsymbol{LFI:FWHM:LFI27}{32.65}
\setsymbol{LFI:FWHM:LFI28}{32.66}

\setsymbol{LFI:FWHM:uncertainty:LFI18:units}{0.170\arcm}
\setsymbol{LFI:FWHM:uncertainty:LFI19:units}{0.174\arcm}
\setsymbol{LFI:FWHM:uncertainty:LFI20:units}{0.170\arcm}
\setsymbol{LFI:FWHM:uncertainty:LFI21:units}{0.156\arcm}
\setsymbol{LFI:FWHM:uncertainty:LFI22:units}{0.164\arcm}
\setsymbol{LFI:FWHM:uncertainty:LFI23:units}{0.171\arcm}
\setsymbol{LFI:FWHM:uncertainty:LFI24:units}{0.652\arcm}
\setsymbol{LFI:FWHM:uncertainty:LFI25:units}{1.075\arcm}
\setsymbol{LFI:FWHM:uncertainty:LFI26:units}{1.131\arcm}
\setsymbol{LFI:FWHM:uncertainty:LFI27:units}{1.266\arcm}
\setsymbol{LFI:FWHM:uncertainty:LFI28:units}{1.287\arcm}

\setsymbol{LFI:FWHM:uncertainty:LFI18}{0.170}
\setsymbol{LFI:FWHM:uncertainty:LFI19}{0.174}
\setsymbol{LFI:FWHM:uncertainty:LFI20}{0.170}
\setsymbol{LFI:FWHM:uncertainty:LFI21}{0.156}
\setsymbol{LFI:FWHM:uncertainty:LFI22}{0.164}
\setsymbol{LFI:FWHM:uncertainty:LFI23}{0.171}
\setsymbol{LFI:FWHM:uncertainty:LFI24}{0.652}
\setsymbol{LFI:FWHM:uncertainty:LFI25}{1.075}
\setsymbol{LFI:FWHM:uncertainty:LFI26}{1.131}
\setsymbol{LFI:FWHM:uncertainty:LFI27}{1.266}
\setsymbol{LFI:FWHM:uncertainty:LFI28}{1.287}

\setsymbol{HFI:center:frequency:100GHz:units}{100\,GHz}
\setsymbol{HFI:center:frequency:143GHz:units}{143\,GHz}
\setsymbol{HFI:center:frequency:217GHz:units}{217\,GHz}
\setsymbol{HFI:center:frequency:353GHz:units}{353\,GHz}
\setsymbol{HFI:center:frequency:545GHz:units}{545\,GHz}
\setsymbol{HFI:center:frequency:857GHz:units}{857\,GHz}

\setsymbol{HFI:center:frequency:100GHz}{100}
\setsymbol{HFI:center:frequency:143GHz}{143}
\setsymbol{HFI:center:frequency:217GHz}{217}
\setsymbol{HFI:center:frequency:353GHz}{353}
\setsymbol{HFI:center:frequency:545GHz}{545}
\setsymbol{HFI:center:frequency:857GHz}{857}

\setsymbol{HFI:Ndetectors:100GHz}{8}
\setsymbol{HFI:Ndetectors:143GHz}{11}
\setsymbol{HFI:Ndetectors:217GHz}{12}
\setsymbol{HFI:Ndetectors:353GHz}{12}
\setsymbol{HFI:Ndetectors:545GHz}{3}
\setsymbol{HFI:Ndetectors:857GHz}{4}

\setsymbol{HFI:FWHM:Maps:100GHz:units}{9\parcm88}
\setsymbol{HFI:FWHM:Maps:143GHz:units}{7\parcm18}
\setsymbol{HFI:FWHM:Maps:217GHz:units}{4\parcm87}
\setsymbol{HFI:FWHM:Maps:353GHz:units}{4\parcm65}
\setsymbol{HFI:FWHM:Maps:545GHz:units}{4\parcm72}
\setsymbol{HFI:FWHM:Maps:857GHz:units}{4\parcm39}
\setsymbol{HFI:FWHM:Maps:100GHz}{9.88}
\setsymbol{HFI:FWHM:Maps:143GHz}{7.18}
\setsymbol{HFI:FWHM:Maps:217GHz}{4.87}
\setsymbol{HFI:FWHM:Maps:353GHz}{4.65}
\setsymbol{HFI:FWHM:Maps:545GHz}{4.72}
\setsymbol{HFI:FWHM:Maps:857GHz}{4.39}

\setsymbol{HFI:beam:ellipticity:Maps:100GHz}{1.15}
\setsymbol{HFI:beam:ellipticity:Maps:143GHz}{1.01}
\setsymbol{HFI:beam:ellipticity:Maps:217GHz}{1.06}
\setsymbol{HFI:beam:ellipticity:Maps:353GHz}{1.05}
\setsymbol{HFI:beam:ellipticity:Maps:545GHz}{1.14}
\setsymbol{HFI:beam:ellipticity:Maps:857GHz}{1.19}

\setsymbol{HFI:FWHM:Mars:100GHz:units}{9\parcm37}
\setsymbol{HFI:FWHM:Mars:143GHz:units}{7\parcm04}
\setsymbol{HFI:FWHM:Mars:217GHz:units}{4\parcm68}
\setsymbol{HFI:FWHM:Mars:353GHz:units}{4\parcm43}
\setsymbol{HFI:FWHM:Mars:545GHz:units}{3\parcm80}
\setsymbol{HFI:FWHM:Mars:857GHz:units}{3\parcm67}

\setsymbol{HFI:FWHM:Mars:100GHz}{9.37}
\setsymbol{HFI:FWHM:Mars:143GHz}{7.04}
\setsymbol{HFI:FWHM:Mars:217GHz}{4.68}
\setsymbol{HFI:FWHM:Mars:353GHz}{4.43}
\setsymbol{HFI:FWHM:Mars:545GHz}{3.80}
\setsymbol{HFI:FWHM:Mars:857GHz}{3.67}

\setsymbol{HFI:beam:ellipticity:Mars:100GHz}{1.18}
\setsymbol{HFI:beam:ellipticity:Mars:143GHz}{1.03}
\setsymbol{HFI:beam:ellipticity:Mars:217GHz}{1.14}
\setsymbol{HFI:beam:ellipticity:Mars:353GHz}{1.09}
\setsymbol{HFI:beam:ellipticity:Mars:545GHz}{1.25}
\setsymbol{HFI:beam:ellipticity:Mars:857GHz}{1.03}

\setsymbol{HFI:CMB:relative:calibration:100GHz}{$\lsim 1\%$}
\setsymbol{HFI:CMB:relative:calibration:143GHz}{$\lsim 1\%$}
\setsymbol{HFI:CMB:relative:calibration:217GHz}{$\lsim 1\%$}
\setsymbol{HFI:CMB:relative:calibration:353GHz}{$\lsim 1\%$}
\setsymbol{HFI:CMB:relative:calibration:545GHz}{}
\setsymbol{HFI:CMB:relative:calibration:857GHz}{}

\setsymbol{HFI:CMB:absolute:calibration:100GHz}{$\lsim 2\%$}
\setsymbol{HFI:CMB:absolute:calibration:143GHz}{$\lsim 2\%$}
\setsymbol{HFI:CMB:absolute:calibration:217GHz}{$\lsim 2\%$}
\setsymbol{HFI:CMB:absolute:calibration:353GHz}{$\lsim 2\%$}
\setsymbol{HFI:CMB:absolute:calibration:545GHz}{}
\setsymbol{HFI:CMB:absolute:calibration:857GHz}{}

\setsymbol{HFI:FIRAS:gain:calibration:accuracy:statistical:100GHz}{}
\setsymbol{HFI:FIRAS:gain:calibration:accuracy:statistical:143GHz}{}
\setsymbol{HFI:FIRAS:gain:calibration:accuracy:statistical:217GHz}{}
\setsymbol{HFI:FIRAS:gain:calibration:accuracy:statistical:353GHz}{2.5\%}
\setsymbol{HFI:FIRAS:gain:calibration:accuracy:statistical:545GHz}{1\%}
\setsymbol{HFI:FIRAS:gain:calibration:accuracy:statistical:857GHz}{0.5\%}

\setsymbol{HFI:FIRAS:gain:calibration:accuracy:systematic:100GHz}{}
\setsymbol{HFI:FIRAS:gain:calibration:accuracy:systematic:143GHz}{}
\setsymbol{HFI:FIRAS:gain:calibration:accuracy:systematic:217GHz}{}
\setsymbol{HFI:FIRAS:gain:calibration:accuracy:systematic:353GHz}{}
\setsymbol{HFI:FIRAS:gain:calibration:accuracy:systematic:545GHz}{7\%}
\setsymbol{HFI:FIRAS:gain:calibration:accuracy:systematic:857GHz}{7\%}

\setsymbol{HFI:FIRAS:zero:point:accuracy:100GHz:units}{0.8\MJysr}
\setsymbol{HFI:FIRAS:zero:point:accuracy:143GHz:units}{}
\setsymbol{HFI:FIRAS:zero:point:accuracy:217GHz:units}{}
\setsymbol{HFI:FIRAS:zero:point:accuracy:353GHz:units}{1.4\MJysr}
\setsymbol{HFI:FIRAS:zero:point:accuracy:545GHz:units}{2.2\MJysr}
\setsymbol{HFI:FIRAS:zero:point:accuracy:857GHz:units}{1.7\MJysr}

\setsymbol{HFI:FIRAS:zero:point:accuracy:100GHz}{0.8}
\setsymbol{HFI:FIRAS:zero:point:accuracy:143GHz}{}
\setsymbol{HFI:FIRAS:zero:point:accuracy:217GHz}{}
\setsymbol{HFI:FIRAS:zero:point:accuracy:353GHz}{1.4}
\setsymbol{HFI:FIRAS:zero:point:accuracy:545GHz}{2.2}
\setsymbol{HFI:FIRAS:zero:point:accuracy:857GHz}{1.7}

\setsymbol{HFI:unit:conversion:100GHz:units}{0.2415\MJysrmK}
\setsymbol{HFI:unit:conversion:143GHz:units}{0.3694\MJysrmK}
\setsymbol{HFI:unit:conversion:217GHz:units}{0.4811\MJysrmK}
\setsymbol{HFI:unit:conversion:353GHz:units}{0.2883\MJysrmK}
\setsymbol{HFI:unit:conversion:545GHz:units}{0.05826\MJysrmK}
\setsymbol{HFI:unit:conversion:857GHz:units}{0.002238\MJysrmK}

\setsymbol{HFI:unit:conversion:100GHz}{0.2415}
\setsymbol{HFI:unit:conversion:143GHz}{0.3694}
\setsymbol{HFI:unit:conversion:217GHz}{0.4811}
\setsymbol{HFI:unit:conversion:353GHz}{0.2883}
\setsymbol{HFI:unit:conversion:545GHz}{0.05826}
\setsymbol{HFI:unit:conversion:857GHz}{0.002238}

\setsymbol{HFI:colour:correction:alpha=-2:V1.01:100GHz}{0.9893}
\setsymbol{HFI:colour:correction:alpha=-2:V1.01:143GHz}{0.9759}
\setsymbol{HFI:colour:correction:alpha=-2:V1.01:217GHz}{1.0007}
\setsymbol{HFI:colour:correction:alpha=-2:V1.01:353GHz}{1.0028}
\setsymbol{HFI:colour:correction:alpha=-2:V1.01:545GHz}{1.0019}
\setsymbol{HFI:colour:correction:alpha=-2:V1.01:857GHz}{0.9889}

\setsymbol{HFI:colour:correction:alpha=0:V1.01:100GHz}{1.0008}
\setsymbol{HFI:colour:correction:alpha=0:V1.01:143GHz}{1.0148}
\setsymbol{HFI:colour:correction:alpha=0:V1.01:217GHz}{0.9909}
\setsymbol{HFI:colour:correction:alpha=0:V1.01:353GHz}{0.9888}
\setsymbol{HFI:colour:correction:alpha=0:V1.01:545GHz}{0.9878}
\setsymbol{HFI:colour:correction:alpha=0:V1.01:857GHz}{1.0014}

\providecommand{\sorthelp}[1]{}

\begin{document}

\title{Polarization measurements analysis}
\subtitle{II. Best estimators of polarization fraction and angle}
\author{L. Montier\inst{1,2}, S. Plaszczynski\inst{3}, F. Levrier\inst{4},  M. Tristram\inst{3}, D. Alina\inst{1,2}, I. Ristorcelli\inst{1,2}, J.-P. Bernard\inst{1,2}, V. Guillet\inst{5}}
\institute{ 
\inst{1} Universit\'{e} de Toulouse, UPS-OMP, IRAP, F-31028 Toulouse cedex 4, France\\
\inst{2} CNRS, IRAP, 9 Av. colonel Roche, BP 44346, F-31028 Toulouse cedex 4, France\\
\inst{3} Laboratoire de l'Acc\'{e}l\'{e}rateur Lin\'{e}aire, Universit\'{e} Paris-Sud 11, CNRS/IN2P3, Orsay, France\\
\inst{4} LERMA/LRA - ENS Paris et Observatoire de Paris, 24 rue Lhormond, 75231 Paris Cedex 05, France\\
\inst{5} Institut d'Astrophysique Spatiale, CNRS (UMR8617) Universit\'{e} Paris-Sud 11, B\^{a}timent 121, Orsay, France
}

\authorrunning{Montier et al.}
\titlerunning{Polarization measurements analysis II}

\abstract{ 
With the forthcoming release of high precision polarization measurements, such as from the {\it Planck\/} satellite, 
it becomes critical to evaluate the performance of estimators for the polarization fraction and angle.
These two physical quantities suffer from a well-known bias in the presence of measurement noise, as has been described in part I of this series. 
In this paper, part II of the series, we explore the extent to which various estimators may correct the bias. Traditional frequentist estimators of the polarization
fraction are compared with two recent estimators: one inspired by a Bayesian analysis and a second following an asymptotic method. 
We investigate the sensitivity of these estimators to the asymmetry of the covariance matrix which may vary over large datasets.
We present for the first time a comparison among polarization angle estimators, and evaluate the statistical bias on the angle 
that appears when the covariance matrix exhibits effective ellipticity. 
We also address the question of the accuracy of  
the polarization fraction and angle uncertainty estimators. The methods linked to the credible intervals and to the variance estimates are tested against 
the robust confidence interval method. From this pool of polarization fraction and angle estimators, we build recipes 
adapted to  different use-cases: we provide the best
estimators to build a mask, to compute large maps of the polarization fraction and angle, and to deal with low 
signal-to-noise data. More generally, we show that the traditional estimators suffer from discontinuous  distributions at low signal-to-noise ratio, 
while the asymptotic and Bayesian methods do not. Attention is given to the shape of the output distribution of the estimators, 
and is compared with a Gaussian distribution. In this regard, the new asymptotic method presents the best performance, while the Bayesian
output distribution is shown to be strongly asymmetric with a sharp cut at low signal-to-noise ratio. 
Finally, we present an optimization of the estimator derived from the Bayesian analysis using adapted priors.  
}

\maketitle

\section{Introduction}
\label{sec:introduction}

\begin{figure*}[tp!]
\begin{equation}
f(I,p,\psi\, | \,I_0,p_0,\psi_0, \tens{\Sigma}) =    \frac{2|p|\,I^2} {\sqrt{(2\pi)^3} \sigma^3} \, 
\mathrm{exp} \left \lgroup - \frac{1}{2} 
\left[ \begin{array}{c} 
I -I_0  \\
p \, I\,  \cos2\psi-p_0\,I_0\cos2\psi_0 \\
p \, I\, \sin2\psi-p_0\,I_0\sin2\psi_0 \\\end{array}
\right] ^{T}
\tens{\Sigma}^{-1}
\left[ \begin{array}{c} 
I-I_0 \\
p\,I\,\cos2\psi-p_0\,I_0\,\cos2\psi_0\\
p\,I\,\sin2\psi-p_0\,I_0\,\sin2\psi_0\\\end{array}
\right]
\right \rgroup \, ,
\tag{1}
\label{eq:f_ipphi}
\end{equation}
\end{figure*}

\begin{figure*}[tp!]
\begin{equation}
f_{2D}(p,\psi\, | \, I_0\,p_0,\psi_0, \tens{\Sigma}_p) =    \frac{p} {\pi \sigma_p^2} \, 
\mathrm{exp} \left \lgroup - \frac{1}{2} 
\left( p^2 
\left[ \begin{array}{c} 
 \cos2\psi \\
\sin2\psi\\\end{array}
\right] ^{T}
\tens{\Sigma}_{p}^{-1}
\left[ \begin{array}{c} 
\cos2\psi\\
\sin2\psi\\\end{array}
\right]
 - 2 pp_0 
 \left[ \begin{array}{c} 
 \cos2\psi \\
\sin2\psi \\\end{array}
\right] ^{T}
\tens{\tens{\Sigma}}_{p}^{-1}
\left[ \begin{array}{c} 
\cos2\psi_0\\
\sin2\psi_0\\\end{array}
\right]
 +p_0^2
 \left[ \begin{array}{c} 
\cos2\psi_0 \\
\sin2\psi_0 \\\end{array}
\right] ^{T}
\tens{\Sigma}_{p}^{-1}
\left[ \begin{array}{c} 
\cos2\psi_0\\
\sin2\psi_0\\\end{array}
\right]
\right)
\right \rgroup \, ,
\label{eq:f_2d_polar}
\tag{2}
\end{equation}
\end{figure*}

The complexity of polarization measurement analysis has been described by \citet{Serkowski1958} when discussing the presence 
of a systematic bias in optical measurements of linear polarization from stars, and then
by \citet{Wardle1974} addressing the same issue in the field of radio astronomy. 
The bias of polarization measurements happens when one is interested in
the polarization intensity $P\equiv   \sqrt{(Q^2 + U^2)}$ 
or the polarization fraction $p\equiv P/I$ and the polarization angle $\psi=1/2\,\mathrm{atan}(U/Q)$ 
(where $I$, $Q$ and $U$ are the Stokes parameters),
quantities which become systematically biased in the presence of noise. 
Working with the Stokes parameters $Q$ and $U$ as far as possible avoids this kind of  bias.  Once a physical modeling of $p$ and $\psi$ is available, and can be translated into $Q$ and $U$, a likelihood analysis can be performed directly on the Stokes parameters. For the other cases, where no modeling is available,
\citet{Simmons1985} proposed the first compilation and comparison of
methods to deal with the problem of getting unbiased polarization estimates of the polarization fraction and angle, with 
their associated uncertainties. 
Then \citet{Naghizadeh1993} extended the work of \citet{Simmons1985} to the
characterisation of the polarization angle uncertainties, and \citet{Vaillancourt2006} proposed  a method to build confidence limits on polarization fraction measurements.  More recently, \citet{Quinn2012} suggested using a Bayesian approach to get polarization estimates with low bias.
In all these studies the authors made strong assumptions:
no noise on the intensity $I$ and no correlation between
the $Q$ and $U$ components, which were also assumed to have equal noise properties.
\citet[hereafter PMA I, ][]{Montier2014}
have quantified the impact of the 
asymmetry and the correlation between the $Q$ and $U$ noise components 
on the bias of the polarization fraction and angle measurements. They have shown that 
the asymmetry of the noise properties can not be systematically neglected as is usually done, and that the 
uncertainty of the intensity may significantly affect the polarization measurements in the low signal-to-noise (SNR) regime.

In the context of the new generation of polarization data, such as  {\it Planck\/}\footnote{\Planck~(\url{http://www.esa.int/Planck}) is a
project of the European Space Agency (ESA) with instruments
  provided by two scientific consortia funded by ESA member states (in
  particular the lead countries France and Italy), with contributions
  from NASA (USA) and telescope reflectors provided by a collaboration
  between ESA and a scientific consortium led and funded by Denmark.}
   \citep{planck2011-1.1}, 
Blast-Pol \citep[The Balloon-borne Large Aperture Submillimeter Telescope for Polarimetry, ][]{Fissel2010}, 
PILOT \citep{Bernard2007} or ALMA  \citep{Perez2013}, 
which benefit from a much better control of the noise properties, 
it is essential to take into account the full covariance matrix
when deriving the polarization measurement estimates.
In recent works no correction for the bias of the polarization
 fraction were applied \citep[e.g.][]{Dotson2010}, or
only high SNR data were used for analysis ($>$3) 
 to avoid these issues \citep[e.g.][]{Vaillancourt2012}.
Two issues are immediately apparent. First, this choice of the SNR threshold may not be relevant for all measurements,
and the asymmetry between the orthogonal Stokes noise components 
 could affect the threshold choice. Secondly, the question remains of how to 
 deal with low signal-to-noise data.  
 Using simply the measurements of the polarization parameters (we will call them the ``na\"ive'' ones)
as estimators of the true values leads to
very poor performance, as they lack any information on the noise power. 
Instead, we would like to perform some transformation on the
polarization parameters, in order to remove bias and improve the variance.

This work is the second of a series on the  'Analysis of polarization measurements'. 
Its aim is to describe how to recover from a measurement ($p$, $\psi$) the true polarization 
fraction $p_0$ and polarization angle $\psi_0$ with their associated uncertainties, 
taking into account the full covariance matrix $\tens{\Sigma}$. 
We will compare the performance of the various estimators available, 
and study the impact of the correlation and ellipticity of the covariance matrix on these estimates. 
We stress that we adopt a frequentist approach to investigate the properties of these estimators, even 
when dealing with the method inspired by the Bayesian analysis. This means that the estimators are defined  as
 single value estimates, instead of considering the probability density function (pdf) as the proper estimate, as it is usually done in Bayesian methods. 
The performance of these estimators will be evaluated using three main criteria:
the minimum bias, the smallest  risk function, and the shape of the distribution of the output estimates. The choice of the most appropriate estimator may vary with the application at hand,
and a compromise among them may be chosen to achieve good overall performance.   
Throughout this work we will make the following two assumptions: i) circular polarization is assumed to be negligible, and ii) 
the noise on Stokes parameters is assumed to be Gaussian. We also define four regimes of the covariance matrix 
to quantify its asymmetry, in terms of effective ellipticity ($\varepsilon_\mathrm{eff}$) as described in PMA I: 
the {\it extreme} (1$<$$\varepsilon_\mathrm{eff}$$<$2), 
the {\it low} (1$<$$\varepsilon_\mathrm{eff}$$<$1.1),  
the {\it tiny}  (1$<$$\varepsilon_\mathrm{eff}$$<$1.01)
and the canonical ($\varepsilon_\mathrm{eff}$=1)  regimes.

The paper is organized as follows: we first review in Sect.~\ref{sec:estimators}
the expression and the limitations of the polarization estimators, which are extended to take into account the 
full covariance matrix. We discuss in Sect.~\ref{sec:uncertainties} the meaning of the polarization uncertainties
and we present the different uncertainty estimators.
We then compare the performance of the estimators of the polarization fraction in Sect.~\ref{sec:comparison_p_estimators}, 
and of the polarization angle in Sect.~\ref{sec:comparison_estimators_psi}. In Sect.~\ref{sec:3Dcase}, we discuss some aspects of the problem when the total intensity $I$ is not perfectly known. We conclude with general recipes in Sect.~\ref{sec:conclusion}.

\section{Polarization estimators}
\label{sec:estimators}


Early work on polarization estimators was based on the \citet{Rice1945} distribution which provides
the probability to find a measurement $p$, for a given true value $p_0$ and the noise estimate $\sigma_p$ of the $Q$ and $U$
Stokes parameters. The noise values of the Stokes parameters were assumed to be equal ($\sigma_p$=$\sigma_{\rm Q}$/$I_0$=$\sigma_{\rm U}$/$I_0$), and the total intensity was assumed to be perfectly known, $I=I_0$.
 As we would like to include the full covariance matrix, 
 we use the generalized expression of the pdf from PMA I, which provides the probability
 to get the measurements ($I$,$p$, $\psi$), given the true values ($I_0$, $p_0$,$\psi_0$) and the covariance matrix $\tens{\Sigma}$.  
Following the notations of PMA I, the expression of the pdf in 3D, including the intensity terms, denoted
$f(I, p,\psi|I_0,p_0,\psi_0,\tens{\Sigma})$, is given by Eq.~\ref{eq:f_ipphi}, and the pdf in 2D, 
$f_{2D}(p,\psi| I_0, p_0,\psi_0,\tens{\Sigma}_p)$, by Eq.~\ref{eq:f_2d_polar} when the intensity $I_0$ is assumed to be perfectly known.
\setcounter{equation}{2}
We also note the introduction of the covariance matrix reduced in 2D, 
\begin{equation}
\tens{\Sigma}_p= \frac{1}{I_0^2}  \left(\begin{array}{cc}
 \sigma_{\rm Q}^2 & \sigma_{\rm QU} \\
 \sigma_{\rm QU} & \sigma_{\rm U}^2 \\
\end{array}\right) 
\quad = \quad
 \frac{ \sigma_{p,G}^2 } {\sqrt{1-\rho^2 }} \left \lgroup \begin{array}{cc}
\varepsilon & \rho  \\
\rho & 1/\varepsilon \\
\end{array}\right \rgroup \, ,
\end{equation}
where $\varepsilon=\sigma_{\rm Q} / \sigma_{\rm U}$ is the ellipticity and $\rho=\sigma_{\rm QU}/\sigma_{\rm Q}\sigma_{\rm U}$ is the correlation between the $Q$ and $U$ noise components, 
leading to an effective ellipticity given by: 
\begin{equation}
\varepsilon_\mathrm{eff} = \sqrt{\frac{1 + \varepsilon^2 + \sqrt{(\varepsilon^2-1)^2 + 4\rho^2\varepsilon^2}}{1 + \varepsilon^2 - \sqrt{(\varepsilon^2-1)^2 + 4\rho^2\varepsilon^2}}} \, .
\end{equation}
With these notations we have  $\mathrm{Det}(\tens{\Sigma}_p)=\sigma_{p,\mathrm{G}}^4$ and 
\begin{equation}
\sigma_{p,\mathrm{G}}^2 = \frac{\sigma_{\rm Q}^2}{I_0^2}\, \frac{\sqrt{1-\rho^2}}{\varepsilon}\, ,
\end{equation}
which represents the 
equivalent radius of a circular Gaussian distribution with the same integrated area as the elliptical one. 
 We also define $\sigma_p$=$\sigma_{\rm Q}$/$I_0$=$\sigma_{\rm U}$/$I_0$ when $\varepsilon_\mathrm{eff}$=1.
Finally the pdfs of $p$ and $\psi$, $f_p$ and $f_{\psi}$, 
are obtained by marginalization of $f_{2D}$ over $\psi$ and $p$, respectively. The expressions for the 
1D pdfs $f_p$ and $f_{\psi}$ depend on the full set of initial parameters ($I_0$, $p_0$, $\psi_0$) in the general case, unlike the case
under the canonical simplifications (see appendix~C of PMA I for fully developed analytical expressions).

We describe below the various estimators of the polarization fraction and angle listed in Table~\ref{tab:list_esti}. 
We stress that most of the expressions derived in this work have been obtained when restricting the 
analysis in the 2D case, assuming furthermore  that the true intensity $I_0$ is 
perfectly known, except for the Bayesian estimator where we present a 3D development (see Sect.~\ref{sec:3Dcase}).

\begin{table}
\caption{List of the acronyms of the estimators used in this work. The parameters to which each estimator applies, independently (/) or 
simultaneously (\&), are given in the last column.}
\center 
\begin{tabular}{llc}
\hline
\hline
Acronym & Description & Parameters \\
\hline
ML & Maximum Likelihood & $p$ / $\psi$ \\
MP & Most Probable  in 1D & $p$ / $\psi$ \\
MP2 & Most Probable in 2D & $p$ \& $\psi$ \\
AS & Asymptotic & $p$ \\
MAS & Modified Asymptotic & $p$ \\
MAP & Maximum A Posteriori & $p$ / $\psi$ \\
MAP2 & Maximum A Posteriori in 2D & $p$ \& $\psi$ \\
MB & Mean posterior Bayesian  & $I$ \& $p$ \& $\psi$ \\
\hline
\end{tabular}
\label{tab:list_esti}
\end{table}

\subsection{Maximum Likelihood estimators}
\label{sec:ML}

The Maximum Likelihood (ML) estimators are defined as the values of $p_0$  and $\psi_0$ which maximize the
pdf calculated at the polarization measurements $p$ and $\psi$. 
When computed using the
2D pdf $f_{2D}$ to fit $p_0$ and $\psi_0$ simultaneously, this estimator gives back the measurements, 
whatever the bias and the covariance matrix are, and is inefficient at correcting the bias of the data. 

After marginalization of the pdf $f_{2D}$ over $\psi$,
the 1D ML estimator of $p_0$, $\hat{p}_{\text{ML}}$,  is now defined by
 \begin{equation} 
0 = \frac{\partial f_{p}}{ \partial p_0} \Big (p\, | \, p_0,\psi_0,  \tens{\Sigma}_p \Big) _{\big | {p_0=\hat{p}_{\text{ML}}}}  \, .
\end{equation}
Note that the expression of $f_p$ is independent of the measurement $\psi$, but still theoretically 
depends on the true value $\psi_0$ which is unknown. In the canonical case ($\varepsilon_\mathrm{eff}$=1) $\psi_0$ 
disappears from the expression, but it must be considered as a nuisance parameter in the general case. 
One way to proceed in such a case is to compute the mean of the solutions $\hat{p}_{\text{ML}}$ for $\psi_0$ varying in the range $-\pi/2$ to $\pi/2$.
As already stressed by \citet{Simmons1985}, this estimator yields a zero estimate below a certain threshold of the measurement $p$,
which implies a strong discontinuity in the resulting distribution of this $p_0$ estimator. 
Nevertheless, contrary to the 2D ML 
estimators, the $p$ ML estimator does not give back the initial measurements, 
and is often used to build polarization estimates. 

Similarly, the 1D ML estimator of $\psi_0$, $\hat{\psi}_{\text{ML}}$,  is given after marginalization of $f_{2D}$ over $p$ by
 \begin{equation} 
0 = \frac{\partial f_{\psi}}{ \partial \psi_0} \Big (\psi \, | \, p_0, \psi_0, \tens{\Sigma}_p \Big) _{\big | {\psi_0=\hat{\psi}_{\text{ML}}}}  \, .
\label{eq:ml1_psi}
\end{equation}
As mentioned for the ML estimator $\hat{p}_{\text{ML}}$, 
the unknown parameter $p_0$ in the above expression has to be considered as a nuisance parameter when solving Eq.~\ref{eq:ml1_psi}.
We stress that because the canonical simplifications have always been assumed in the literature, 
bias on the $\psi$ measurements has not been previously considered
and the $\hat{\psi}_{\text{ML}}$ estimator has not yet been used and qualified to correct this kind of bias.
This analysis is done in Sect.~\ref{sec:comparison_estimators_psi}.

\subsection{Most Probable estimators}

The Most Probable estimators  of $p_0$ and $\psi_0$ are the values for which the 
pdf $f_{2D}$ reaches its maximum at the 
measurements values ($p$,$\psi$). Notice that the Most Probable estimators ensure that
the measurement values  ($p$,$\psi$) are the most probable values of the pdf computed for this choice of $p_0$ and $\psi_0$, \ie 
they take the maximum probability among all possible measurements with this set of $p_0$ and $\psi_0$.
As a comparison the ML estimators ensure that the measurement values  ($p$,$\psi$) 
take the maximum probability for this choice of $p_0$ and $\psi_0$, compared to the probability of the same
measurement values ($p$,$\psi$) for all other possible sets of $p_0$ and $\psi_0$.

The 2D Most Probable estimators (MP2), $\hat{p}_{\text{MP2}}$  and $\hat{\psi}_{\text{MP2}}$,  
are defined as the values of $p_0$ and $\psi_0$ simultaneously satisfying the two following relations:
\begin{equation} 
\label{eq:mp_1}
0=\frac{\partial f_{2D}}{ \partial p} \Big (p,\psi \, | \, p_0, \psi_0, \tens{\Sigma}_p \Big) _{\Bigg |{\begin{array}{l}p_0=\hat{p}_{\text{MP2}} \\ \psi_0=\hat{\psi}_{\text{MP2}} \end{array}}}  
\end{equation}
and
\begin{equation} 
\label{eq:mp_2}
0=\frac{\partial f_{2D}}{\partial \psi} \Big  (p,\psi \, | \, p_0, \psi_0, \tens{\Sigma}_p \Big )_{\Bigg |{\begin{array}{l}p_0=\hat{p}_{\text{MP2}}\\ \psi_0=\hat{\psi}_{\text{MP2}} \end{array}}} \, .
\end{equation}
These relations can be solved, using the fully developed expression of $f_{2D}$ including the terms of the inverse matrix $\tens{\Sigma}_p^{-1}$, 
as provided in Appendix~\ref{sec:most_probable_detail}. When canonical simplifications are assumed, 
this yields
\begin{eqnarray}
\hat{\psi}_{\text{MP2}} & = & \psi \, , \nonumber \\
\hat{p}_{\text{MP2}} & = & \Bigg \{  \begin{array}{ll}
(p - \sigma_p^2 / p)  & \, \, \, \mathrm{for} \, \, p > \sigma_p \\
0 &  \, \, \, \mathrm{for} \, \,  p \le \sigma_p
\end{array} \, ,
\end{eqnarray}
as found in \citet{Quinn2012}. We observe that the MP2 estimate of the polarization fraction is systematically
 lower than the measurements, so that this estimator
tends to over-correct $p$, as it will be shown in Sect.~\ref{sec:comparison_p_estimators}.

After marginalization over $p$ or $\psi$, the 1D Most Probable (MP) estimators, $\hat{p}_{\text{MP}}$  and $\hat{\psi}_{\text{MP}}$,  are defined independently by:
 \begin{equation} 
0 = \frac{\partial f_p}{ \partial p} \Big (p\, | \, p_0, \psi_0, \tens{\Sigma}_p \Big) _{\big | {p_0=\hat{p}_{\text{MP}}}} 
\end{equation}
and
 \begin{equation} 
0 = \frac{\partial f_{\psi}}{ \partial \psi} \Big (\psi \, | p_0 \, \psi_0, \tens{\Sigma}_p \Big) _{\big | {\psi_0=\hat{\psi}_{\text{MP}}}}\, .
\end{equation}
The 1D and 2D estimators are not expected to provide the same estimates. Under the canonical assumptions,
the MP estimator of $p$ is commonly known as the Wardle and Kronberg's \citep{Wardle1974} estimator.
 
As mentioned earlier, the MP estimator yields a zero estimate below
a certain threshold of $p$ \citep{Simmons1985}, 
which implies a strong discontinuity in the resulting distribution of these estimators
for low SNR measurements.

\subsection{Asymptotic estimator}
\label{sec:as}

The Asymptotic estimator (AS) of the polarization fraction $p$ is usually defined 
in the canonical case by
\begin{equation}
  \label{eq:AS}
  \hat p_{\text{AS}}= \Bigg \{ \begin{array}{lcl} 
  \sqrt{p^2-\sigma_p^2} & p > \sigma_p \\
\quad 0 & p \le \sigma_p 
  \end{array} \, .
\end{equation}
The output distribution of the AS estimator appears as the asymptotic limit of the \citet{Rice1945} distribution when 
$p/\sigma_p$ tends to $\infty$, just as the ML and MP estimators, and given by
\begin{equation}
  \label{eq:limit}
 \mathrm{pdf} \left( \dfrac{p}{\sigma_p} \right) \rightarrow {\cal N}\left( \sqrt{\left(\dfrac{p_0}{\sigma_p}\right)^2+1},1 \right) \, ,
\end{equation}
where  ${\cal N}(\mu,\sigma)$ denotes the Gaussian distribution of mean
$\mu$ and variance $\sigma^2$. As with the previously presented estimators, this one suffers from 
a strong discontinuity at $\hat{p}_{\text{AS}}$=0.

In the general case, when the canonical simplification is not assumed, 
it has been shown by \citet[][hereafter P14]{Plaszczynski2014} that the 
expression of the asymptotic estimator
can be extended to a general expression by changing the term $\sigma_p^2$ in Eq.~\ref{eq:AS} into a 'noise-bias' parameter $b^2$  defined by
\begin{equation}
  \label{eq:sigp_equiv}
  b^2 =  \frac{\sigma_{\rm U}^{\prime2}\cos^2(2\psi_0-\theta)+ \sigma_{\rm Q}^{\prime2}\sin^2(2\psi_0-\theta) }{I_0^2} \, ,
\end{equation}
where $\theta$ represents the position angle of the iso-probability
bi-variate distribution, and $\sigma_{\rm U}^{\prime2},\sigma_{\rm Q}^{\prime2}$
the  rotated variances
\begin{eqnarray}
  \label{eq:rot}
    \theta &=& \frac{1}{2}\mathrm{atan} \left( \frac{2\rho \sigma_{\rm Q} \sigma_{\rm U}}{\sigma_{\rm Q}^2-\sigma_{\rm U}^2}\right)\, , \\
    \label{eq:sigmaqprime}
   \sigma_{\rm Q}^{\prime2}&=&\sigma_{\rm Q}^2\cos^2\theta+\sigma_{\rm U}^2\sin^2\theta+\rho \sigma_{\rm Q}  \sigma_{\rm U} \sin2\theta \, , \\
    \label{eq:sigmauprime}
  \sigma_{\rm U}^{\prime2}&=&\sigma_{\rm Q}^2\sin^2\theta+\sigma_{\rm U}^2\cos^2\theta-\rho  \sigma_{\rm Q}  \sigma_{\rm U} \sin2\theta \, .
\end{eqnarray}
and $\psi_0$ is the true polarization angle, which can be approximated asymptotically by the na\"ive measurement $\psi$ 
or, even better, by the estimate $\hat{\psi}_{\text{ML}}$ of Sect.~\ref{sec:ML}.
It has been shown that this equivalent 'noise-bias' $b^2$ ensures the minimal bias of $\hat{p}_{\text{AS}}$.

\subsection{Discontinuous estimators}
\label{sec:discontinuous_estimators}

\begin{figure}
  \centering
  \psfrag{xtitle}{$\hat{p}/p_0$}
  \includegraphics[width=9cm]{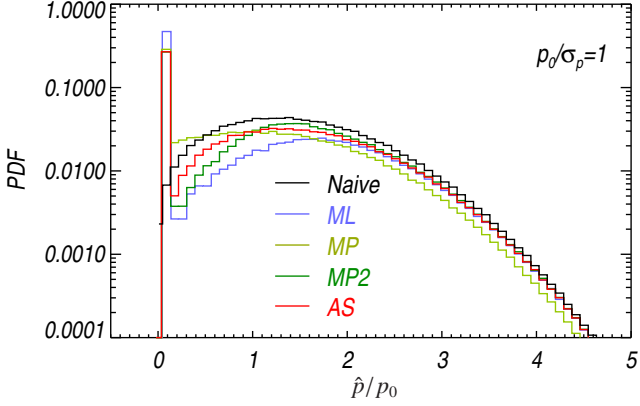}
  \caption{Distributions of $\hat{p}$ estimates obtained with the standard estimators: na\"ive (black), ML (blue), MP (light green), MP2 (green) and AS (red). 
  We assume the covariance matrix to be canonical, and a SNR of $p_0/\sigma_p$=1.}
\label{fig:histo_all}
\end{figure}

\begin{figure}
  \centering
  \includegraphics[width=9cm]{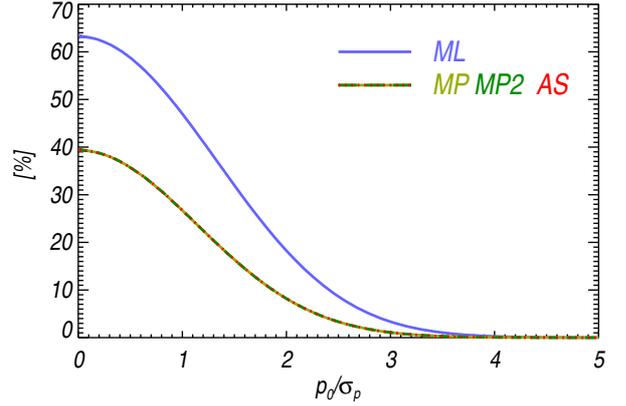}
  \caption{Statistical fraction of null estimates of $\hat{p}$ provided by the ML, MP, MP2 and AS estimators
  applied on Monte-Carlo measurements, as a function of the 
  SNR, in the canonical case.}
\label{fig:fracnull}
\end{figure}

The estimators of $\hat{p}$ introduced above (ML, MP and AS)
 exhibit  a common feature: below
some cutoff value the estimator yields exactly zero. 
This means that the estimator distribution is discontinuous and is
a mixture of a discrete one (at $\hat{p}$=0) and a continuous one (for $\hat{p}>0$), This type of distribution is illustrated 
in Fig.~\ref{fig:histo_all} for a SNR $p_0/\sigma_{p}$=1 and a canonical covariance matrix.
The distribution of the na\"ive measurements is built using a Monte-Carlo simulation, starting from true polarization parameters
$p_0$ and $\psi_0$. The other three distributions of $\hat{p}$ are obtained after applying the ML, MP and AS estimators.
A non negligible fraction of the measurements provide null estimates of $\hat{p}$.
As shown in Fig.~\ref{fig:fracnull}, this fraction of null estimates reaches 40\% at low SNR with the MP and AS estimators, 
and more than 50\% with the ML estimator for SNR$<$1. It converges to 0\% for SNR $>$4.

If taken into account as reliable estimate of $\hat{p}$, null estimates will somewhat artificially lower the statistical bias of the $\hat{p}$ 
estimates compared to the true value $p_0$, as detailed in Sect.~\ref{sec:comparison_p_estimators}.
A null value of these estimators should be understood as an indicator of the low SNR of this measurement, 
which has in fact to be included into any further analysis as an upper limit value.
In practice, the user seldom has various realizations at hand. Using these estimators then leads to a result with upper limits mixed with non-zero estimates in the analysis.
Such complications may be especially hard to handle when studying polarized maps of the interstellar medium.
On the other hand, it would be disastrous to omit those estimates in any statistical analysis, 
since weakly-polarized points would be systematically rejected.
To avoid such complications, we explore below other estimators which avoid this issue and 
lead to continuous distributions. This is especially important in the range of SNR between 2 and 3, where the discontinuous estimators still
yield up to 20\% of null estimates.

\begin{figure*}[bp!]
\begin{equation}
B(I_0,p_0,\psi_0\, | \,I,p,\psi, \tens{\Sigma}) \quad \propto  \quad \sqrt{\frac{Det(\tens{\Sigma}^{-1})}{2(\pi)^3}}\, 
\mathrm{exp} \left \lgroup - \frac{1}{2} 
\left[ \begin{array}{c} 
I -I_0  \\
p \, I\,  \cos(2\psi)-p_0\,I_0\cos(2\psi_0) \\
p \, I\, \sin(2\psi)-p_0\,I_0\sin(2\psi_0) \\\end{array}
\right] ^{T}
\tens{\Sigma}^{-1}
\left[ \begin{array}{c} 
I-I_0 \\
p\,I\,\cos(2\psi)-p_0\,I_0\,\cos(2\psi_0)\\
p\,I\,\sin(2\psi)-p_0\,I_0\,\sin(2\psi_0)\\\end{array}
\right]
\right \rgroup \, ,
\tag{23}
\label{eq:B_ipphi}
\end{equation}
\end{figure*}

\begin{figure*}[bp!]
\begin{equation}
B_{2D}(p_0,\psi_0\, | \,p,\psi, \tens{\Sigma}_p)\quad  \propto  \quad \frac{1}{\pi \sigma_{p,G}^2}  \, 
\mathrm{exp} \left \lgroup - \frac{1}{2} 
\left[ \begin{array}{c} 
p \,  \cos(2\psi)-p_0 \cos(2\psi_0) \\
p \, \sin(2\psi)-p_0 \sin(2\psi_0) \\\end{array}
\right] ^{T}
\tens{\Sigma}_{p}^{-1}
\left[ \begin{array}{c} 
p\,\cos(2\psi)-p_0\,\cos(2\psi_0)\\
p\,\sin(2\psi)-p_0\,\sin(2\psi_0)\\\end{array}
\right]
\right \rgroup \, ,
\label{eq:B_2d_polar}
\tag{25}
\end{equation}
\end{figure*}

\subsection{Modified Asymptotic estimator}
\label{sec:mas_estimator}

A novel powerful asymptotic estimator has been introduced by \citet{Plaszczynski2014} to correct the 
discontinuous distribution of the standard estimators while still keeping the
asymptotic properties. It has been derived from the first order development of the Asymptotic estimator, 
which has been modified to ensure positivity, smoothness and asymptotical convergence at high SNR. 
The Modified Asymptotic (MAS) estimator  is defined as follows:
\begin{equation}
  \label{eq:mas1}
  \hat{p}_{\text{MAS}}=p- b^2 \cdot \frac{1-e^{-p^2 / b^2}}{2p} \, ,
\end{equation}
where the 'noise-bias' $b^2$ is given by Eq.~\ref{eq:sigp_equiv} and computed using a polarization angle assessed from each sample using the asymptotic
estimator $\psi$.

P14 also provides a sample  estimate of the variance of
the estimator that is shown to represent asymptotically the absolute risk function (defined in Sec.~\ref{sec:variance_risk}) of
the estimator:
\begin{equation}
    \sigma^2_{\hat{p}, MAS}  =\frac{\sigma_{\rm Q}^{\prime2}\cos^2(2\psi-\theta)+ \sigma_{\rm U}^{\prime2}\sin^2(2\psi-\theta)}{I_0^2} \, .
\end{equation}

This estimator focuses on getting a ``good'' distribution, that
transforms smoothly from a Rayleigh-like to a Gaussian one, the latter
being reached  in the canonical case for an SNR of about 2.

\subsection{Bayesian estimators}
\label{sec:bayesian_estimators}

The pdfs introduced in Sect.~\ref{sec:estimators} provide the probability
to observe a set of polarization measurements ($I$, $p$, $\psi$) 
given the true polarization parameters ($I_0$, $p_0$, $\psi_0$) and the covariance matrix $\tens{\Sigma}$.
Because we are interested in the opposite, i.e. getting an estimate of the true polarization parameters given  a 
measurement and the knowledge of the noise properties, we use Bayes Theorem to build the
posterior distribution. The posterior pdf $B$
is given in the 3D case by
\begin{eqnarray}
\label{eq:bayesian_3d}
& & B( I_0, p_0, \psi_0\, | \,I,p,\psi, \tens{\Sigma}) =  \\
& & \frac{f(I,p,\psi\, | \,I_0, p_0, \psi_0, \tens{\Sigma}) \cdot \kappa (I_0, p_0, \psi_0)}{ \int_{0}^{+\infty} \int_{0}^{1} \int_{-\pi/2}^{\pi/2} f(I,p,\psi\, | \,I'_0, p'_0, \psi'_0, \tens{\Sigma})\,  \kappa (I'_0, p'_0, \psi'_0) \, dI'_0dp'_0d\psi'_0}\, , \nonumber
\end{eqnarray}
where $\kappa(I_0,p_0,\psi_0)$ is the prior distribution, 
which represents the a priori knowledge of the true polarization parameters and has to be positive everywhere 
and normalized to 1 over the definition ranges of $I_0$, $p_0$ and $\psi_0$.
 When no a priori knowledge is provided, we have to properly define a 'flat' prior, or non-informative prior, which
encodes the ignorance of the prior. A class of non-informative priors is given by the Jeffreys' prior \citep{Jeffrey1939} where 
the ignorance is defined under symmetry transformations that leave the prior invariant.
As discussed by \citet{Quinn2012} for the two dimensional case, this kind of prior can be built 
as a uniform prior in cartesian space ($Q_0$,$U_0$) or in polar space ($p_0$, $\psi_0$), 
both expressing the ignorance of location.
We will prefer the latter, uniform in polar space, which ensures uniform sampling even for small values of $p_0$. 
While $p_0$ and $\psi_0$ are only defined on a finite range ($[0,1]$ and $[-\pi/2,\pi/2]$, respectively), 
the intensity $I_0$ may be infinite in theory, which leads to an issue when defining the ignorance prior. 
In practice, an approximation of the ignorance prior for $I_0$ will be chosen as a top-hat centered on the measurement $I$ and chosen to be
 sufficiently wide to cover the wings of the distribution until it becomes negligible. Such uniform priors lead to the 
 expression of $B$ given in Eq.~\ref{eq:B_ipphi}, where the normalization factor has been omitted.
 \setcounter{equation}{23}
We emphasize that the definition of the ignorance prior introduced above becomes data-dependent, which is not strictly following
the Bayesian approach. Furthermore, the question of the definition range of the prior and the introduction
 of non-flat priors will be discussed in Sect.~\ref{sec:priors}, in the context of 
 comparing the performance of the estimators inspired by the Bayesian approach.

Similarly, the posterior pdf in 2D (i.e., when the total intensity is perfectly know, $I=I_0$) is defined by
 \begin{equation}
 \hspace{-0.3cm}
B_{2D}( p_0, \psi_0\, | \,p,\psi, \tens{\Sigma}_p)=\frac{f_{2D}(p,\psi\, | \,p_0, \psi_0, \tens{\Sigma}_p) \cdot \kappa (p_0, \psi_0)}{ \int\limits_{0}^{1} \int\limits_{-\pi/2}^{+\pi/2} f_{2D}(p,\psi | p'_0, \psi'_0, \tens{\Sigma}_p)\,  \kappa (p'_0, \psi'_0) \, dp'_0d\psi'_0} \, .
\label{eq:b_2d}
\end{equation}
\setcounter{equation}{25}
The analytical expressions of the posterior pdf $B_{2D}$ with a flat prior
is given in Eq.~\ref{eq:B_2d_polar}, where the normalization factors have been omitted and the intensity has been assumed perfectly known
($I=I_0$). Illustrations of this posterior pdf are presented in Appendix~\ref{sec:pdf_posterior}. 
We also introduce $B_p$ and $B_{\psi}$ the Bayesian posterior pdfs of $p$ and $\psi$ in 1D, respectively, and 
defined as the marginalization of the  $B_{2D}$ over $\psi$ and $p$, respectively.

We use the Bayesian posterior pdf in 2D $B_{2D}$  to build two
frequentist estimators:  the MAP and the MB.

The MAP2 and MAP estimators in 2D and 1D, respectively, are simply defined as the ($p_0$, $\psi_0$) values corresponding 
to the maximum of the posterior pdf, $B_{2D}$, and $B_p$ and $B_{\psi}$, respectively. 
We recall that these estimators
match exactly the ML estimators of Sect.~\ref{sec:estimators} in one and two dimensions, respectively, 
when a flat prior is assumed. Hence the MAP2 estimators yield back the polarization measurements, 
whereas the MAP estimators provide a simple way to compute the ML estimates.

The Mean Bayesian Posterior (MB) estimators are defined as the first order moments of the posterior pdf:
\begin{equation}
\hat{p}_{\text{MB}} \equiv   \int_{-\pi/2}^{+\pi/2} \int_{0}^{1} p_0 B_{2D}(p_0,\psi_0\, | \, p, \psi,  \tens{\Sigma}_p ) dp_0d\psi_0 
\end{equation}
and 
\begin{equation}
\hat{\psi}_{\text{MB}} \equiv  \int_{\psi-\pi/2}^{\psi+\pi/2} \int_{0}^{1} \psi_0 B_{2D}(p_0,\psi_0\, | \, p, \psi,  \tens{\Sigma}_p ) dp_0d\psi_0 \, .
\end{equation}
Notice that in the definition of $\hat{\psi}_{\text{MB}} $ the integral over $\psi_0$ is performed over a range centered on the measurement $\psi$. 
This has to be done to take into account the circularity of the posterior pdf over the $\psi_0$ dimension. 
Note  that $B_{2D}(p_0,\psi_0\, | \, p, \psi,  \tens{\Sigma}_p ) =  B_{2D}(p_0,\psi_0+\pi \, | \, p, \psi,  \tens{\Sigma}_p )$.

  We stress that the frequentist estimators inspired by a Bayesian approach, $\hat{p}_{\text{MB}}$ and $\hat{\psi}_{\text{MB}}$, introduced above in the 2D case can be 
 easily extended to the 3D case by integrating the pdf $B( I_0, p_0, \psi_0\, | \,I,p,\psi, \tens{\Sigma})$ 
 of Eq.~\ref{eq:bayesian_3d} over the $I$, $p$ and $\psi$ dimensions. 
 This is extremely powerful when the uncertainty of the intensity $I$ has to be taken into account in the estimate of the polarization parameters, 
 which is highly recommended in some circumstances, such as a low SNR on $I$ ($<$5) 
 or the presence of an unpolarized component on the line of sight (see Sect.~\ref{sec:3Dcase} and PMA I for more details).

\section{Uncertainties}
\label{sec:uncertainties}

\subsection{Variance and risk function}
\label{sec:variance_risk}

It is important not to confuse the variance (noted $\mathsf{V}$) of an estimator with its absolute
risk function (noted $\mathsf{R}$).  For any distribution of the random $p$ variable the definitions
are : 
\begin{eqnarray}
  \label{eq:var_risk}
  \mathsf{V}&\equiv&E\left[\left(X-E[X]\right)^2\right] \, , \\
 \mathsf{R}&\equiv& E\left[\left(X- X_0\right)^2\right] \, ,
\end{eqnarray} 
where $E[X]$ is the expectation of the random variable $X$ and $X_0$ is the true value. Introducing the  absolute  bias in $E[X]=X_0+\mathsf{B}$ and
expanding both relations, the link between the variance and the absolute risk function  is
simply:
\begin{equation}
  \label{eq:var_risk_link}
 \mathsf{V}=\mathsf{R}-\mathsf{B}^2 \, .
\end{equation}

Therefore, for a constant absolute risk function, the variance decreases with the absolute bias
and both are equal when the estimator is unbiased. The variance does not require knowing 
the true value of the random variable, which makes it useful to provide an uncertainty estimate, but
it has to be used extremely carefully in the presence of bias. In such cases, 
the variance will always underestimate the uncertainty.

Furthermore, it is known that the variance is not appropriate for providing uncertainties with non-Gaussian distributions, 
which is the case for the polarization fraction and angle. In such circumstances, the confidence intervals (see Sect.~\ref{sec:confidence_intervals}) are the preferred method for obtaining robust uncertainties.
The variance, however, is often used as a proxy of the uncertainty in the high regime of the SNR.
We will detail in Sect.~\ref{sec:uncertainty_comparison} and \ref{sec:uncertainty_psi_comparison} in which conditions this can still be applied.

\subsection{Posterior uncertainties}
\label{sec:credible_intervals}

One of the main benefits of the Bayesian approach is to provide simple estimates of the uncertainties associated with the 
polarization estimates. One option is to build credible intervals around the MAP estimates as first proposed by \citet{Vaillancourt2006}, 
and the other option is to use the variance of the pdf.

Given a polarization measurement ($p$, $\psi$) and the posterior pdf $B_{2D}(p_0,\psi_0 | p, \psi, \tens{\Sigma}_p)$, 
the lower and upper limits of the $\lambda$\% credible intervals are defined as
 the lower and upper limits of  $p_0$ and $\psi_0$ ranging the iso-probability region $\Omega(\lambda, p, \psi)$ over which
 the integral of $B$ equals $\lambda$\%, so that
 \begin{equation}
\iint_{\Omega(\lambda,p,\psi)} B_{2D} (p_0, \psi_0 \, | \, p, \psi, \tens{\Sigma}_p) \, dp_0d\psi_0= \frac{\lambda}{100} \, .
\end{equation}
These intervals, $[p^{\rm low}_{\text{MAP2}},p^{\rm up}_{\text{MAP2}}]$ and 
$[\psi^{\rm low}_{\text{MAP2}},\psi^{\rm up}_{\text{MAP2}}]$, estimated from the 2D expression of $B_{2D}$ are defined around the
MAP2 estimates $\hat{p}_{\text{MAP2}}$ and $\hat{\psi}_{\text{MAP2}}$, which are
equal to the measurements ($p$, $\psi$). 
 
A similar definition can be given in the one-dimensional case, which leads to different results.
The lower and upper limits, $p^{\rm low}_{\text{MAP}}$ and  $p^{\rm up}_{\text{MAP}}$, around $\hat{p}_{\text{MAP}}$ are defined as follows
\begin{equation}
\int_{p^{\rm low}_{\text{MAP}}}^{p^{\rm up}_{\text{MAP}}} B_p (p_0 \, | \, p, \tens{\Sigma}_p) \, dp_0=  \frac{\lambda}{100} \, ,
\end{equation}
 with the constraint that the posterior probability function is identical for $p^{\rm low}_{\text{MAP}}$ and  $p^{\rm up}_{\text{MAP}}$. Similarly, the lower and upper limits, 
$\psi^{\rm low}_{\text{MAP}}$ and  $\psi^{\rm up}_{\text{MAP}}$, around $\hat{\psi}_{\text{MAP}}$ are given by
\begin{equation}
\int_{\psi^{\rm low}_{\text{MAP}}}^{\psi^{\rm up}_{\text{MAP}}} B_{\psi} (\psi_0 \, | \, \psi, \tens{\Sigma}_p)\,  d\psi_0= \frac{\lambda}{100} \, .
\end{equation}
We recall  that this integral has to be computed around the measurement value $\hat{\psi}_{\text{MAP}}$  to take into account the circularity of the posterior pdf with the polarization angle. 
Notice that the credible intervals built in 1D or 2D are not supposed to be identical, as 
($\hat{p}_{\text{MAP2}}$, $\hat{\psi}_{\text{MAP2}}$) and 
($\hat{p}_{\text{MAP}}$, $\hat{\psi}_{\text{MAP}}$) are not equal in the general case.

The second definition of the uncertainty comes from the second moment of the 1D posterior probability
density functions $B_p$ and $B_\psi$, as follows:
\begin{equation}
\quad \quad \quad \sigma_{p,\mathrm{MB}}^2 \equiv   \int_{0}^{1} (p_0- \hat{p}_{\text{MB}})^2 B_p(p_0\, | \, p,  \tens{\Sigma}_p ) \, dp_0\, ,
\end{equation}
and 
\begin{equation}
\label{eq:sigma_psi_mb}
\quad \quad  \quad \sigma_{\psi,\mathrm{MB}}^2  \equiv  \int_{\psi-\pi/2}^{\psi+\pi/2} (\psi_0-\hat{\psi}_{\text{MB}} ) ^2 B_{\psi}(\psi_0\, | \, \psi,  \tens{\Sigma}_p )\, d\psi_0 \, .
\end{equation}
The operation of subtraction between the two polarization angles must be done with care, restricting the 
the maximum distance to $\pi/2$. At very low SNR, i.e. an almost flat uniform pdf,
the uncertainty reaches the upper limit
$\sigma_{\psi,\mathrm{MB}} \le \pi/\sqrt{12} \, \mathrm{rad} = 51.^{\circ}96$. 
We stress that these 1-$\sigma$ estimates may not be associated with the usual 68\% confidence 
 intervals of the normal distribution,  because of the asymmetrical shape of the posterior distribution and because of the circularity of the angular variable.

\subsection{Confidence intervals}
\label{sec:confidence_intervals}

So far we have considered point estimation of the true $p_0$ value
which is somewhat tricky in the low SNR regime because of the 
non-Gaussian nature of the estimator distribution. 

A different approach, that takes into account the entire shape of the
distribution is to build confidence regions (or intervals), which
allows at some given significance level, to obtain bounds on the true value
given some estimator value.

\citet{Simmons1985} have built the so-called Neyman ``confidence belt'' for
the na\"ive estimator in the canonical case. PMA I proposed the
construction of two-dimensional ($p_0,\psi_0$) intervals, for the
general covariance matrix case.
The classical construction suffers from a standard issue: at very low SNR the confidence
interval lies entirely in the unphysical $p<0$ region, and both
previous studies provide over-conservative regions. 

 P14 has implemented the Feldman-Cousins prescription
\citep{Feldman1998} which is based on using a likelihood ratio
criterium in the Neyman construction. This allows building
intervals that always lie in the physical region without ever being conservative.
They provided these intervals for the MAS estimator
including analytical approximations to the upper and lower limits for
68, 95 and 99.5\% significance levels.

\begin{figure}[t!p]
\vspace{1.2cm}
  \centering
  \begin{tabular}{c}
  \includegraphics[width=.5\textwidth]{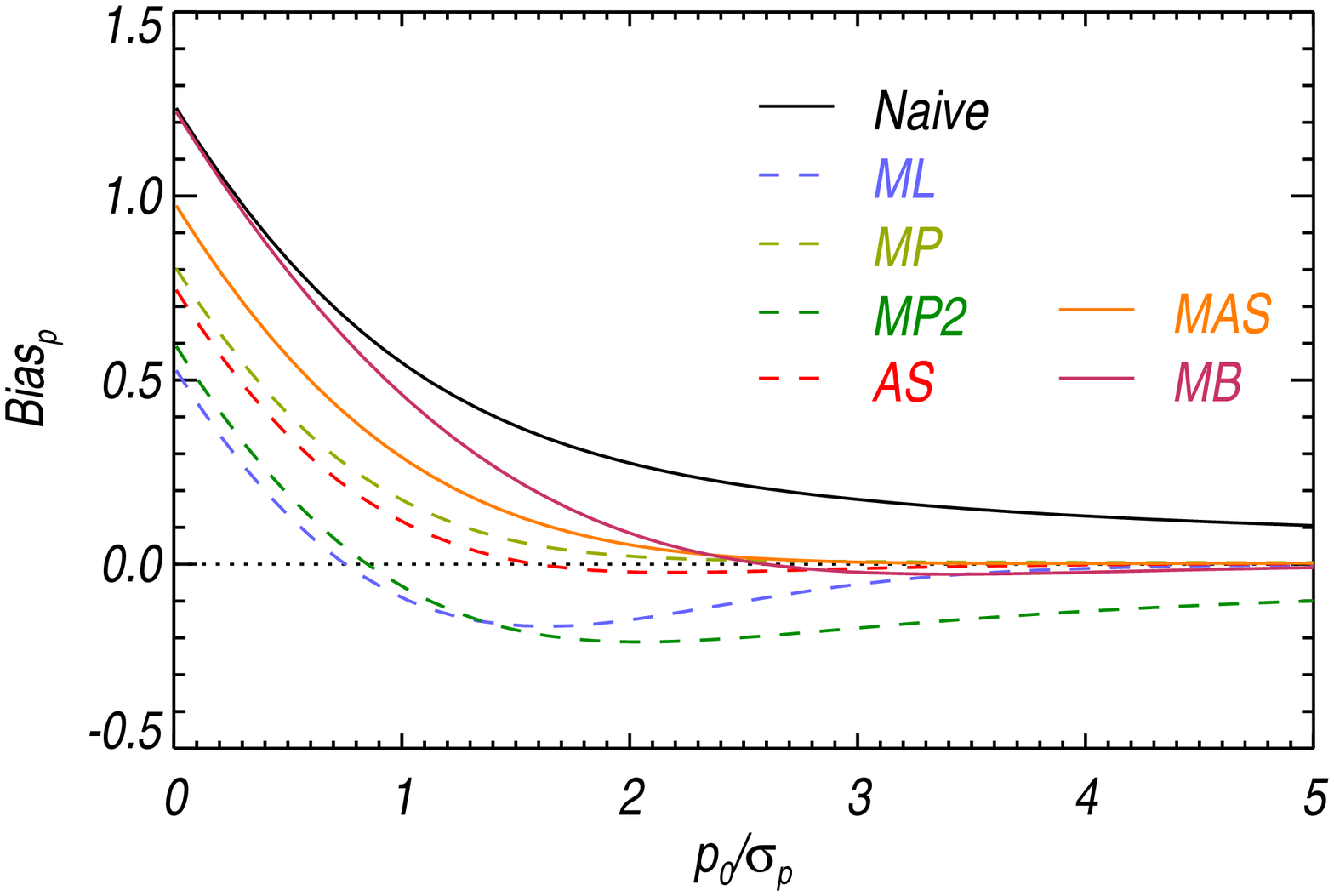}\\
  \includegraphics[width=.5\textwidth]{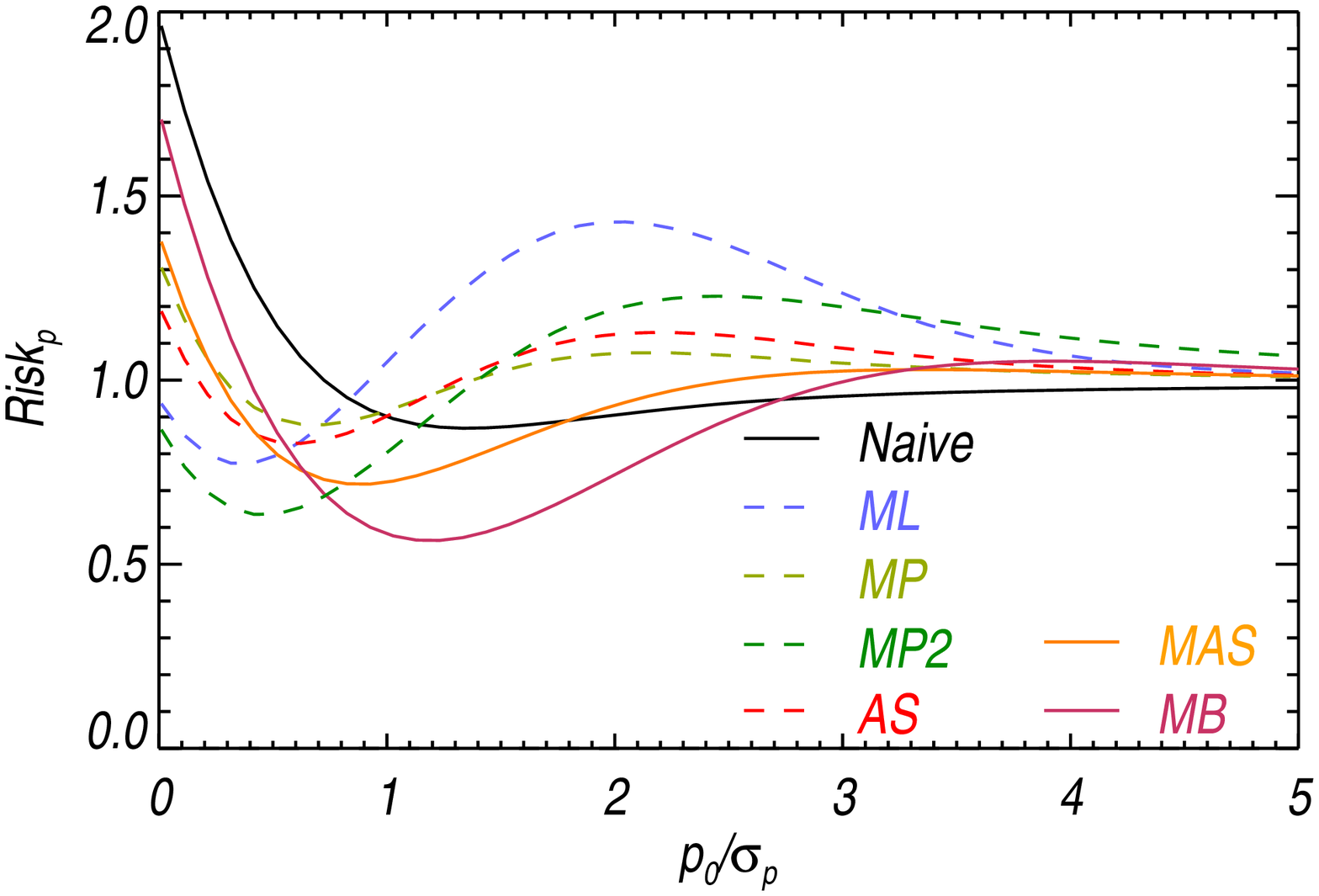} \\
  \includegraphics[width=.5\textwidth]{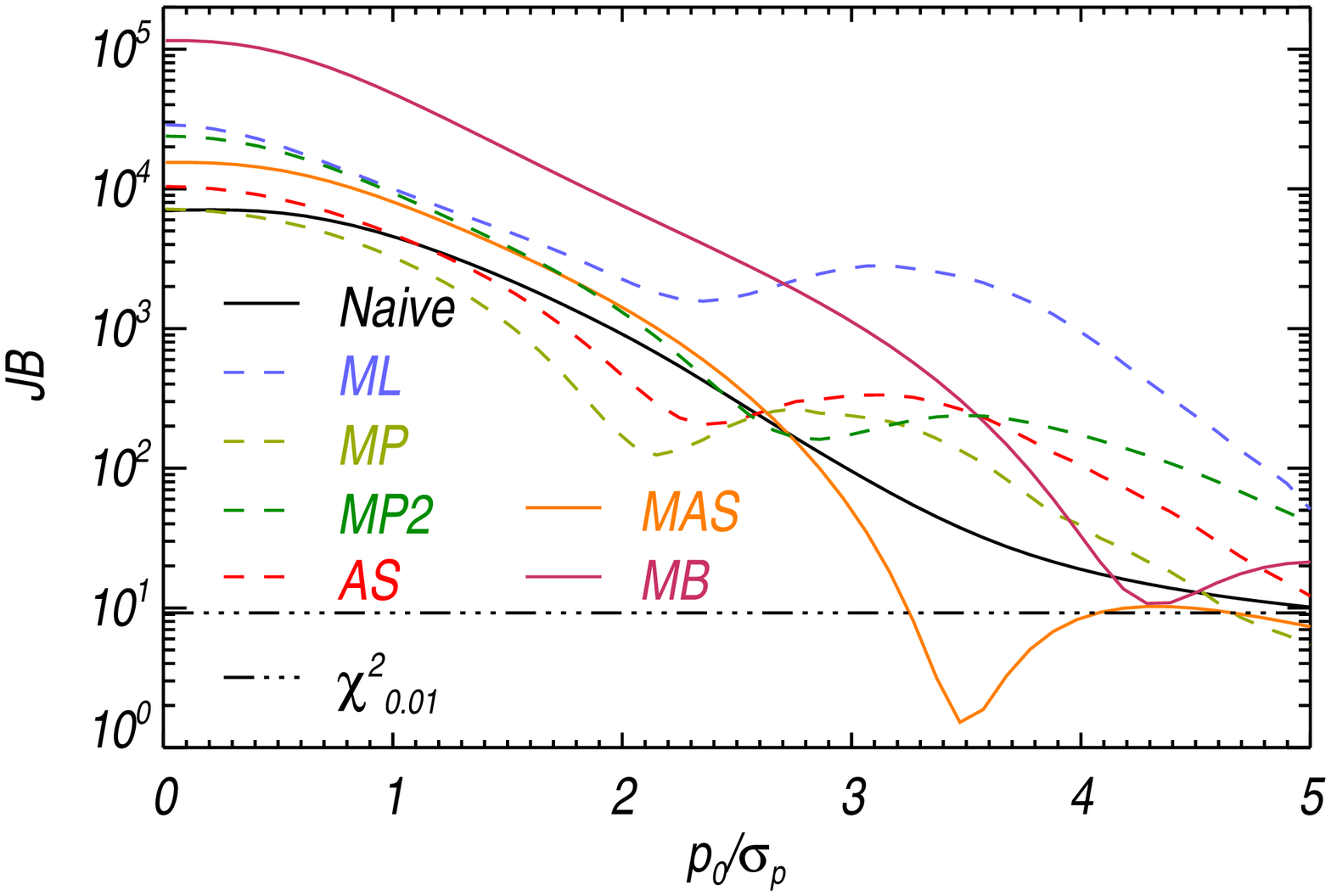} 
\end{tabular}
\caption{Comparison of the average relative bias (top), risk function  (middle) and Jarque-Bera test (bottom)
of the pure measurements (na\"ive, black), ML (dashed blue), 
MP (dashed light green),  MP2 (dashed green), AS (dashed red),
  MAS (orange) and MB (pink) $\hat{p}$ estimators in the canonical case, 
  as a function of the the SNR $p_0/\sigma_p$.
  The dashed lines stand for the discontinuous estimators presenting a peak of their output distribution at $\hat{p}$=0.}
\label{fig:comparison_p_estimator}
\end{figure}

\begin{figure}[t!p]
\vspace{1.2cm}
  \centering
  \begin{tabular}{c}
    \psfrag{xtitle}{$\hat{p}/p_0$}
  \includegraphics[width=.5\textwidth]{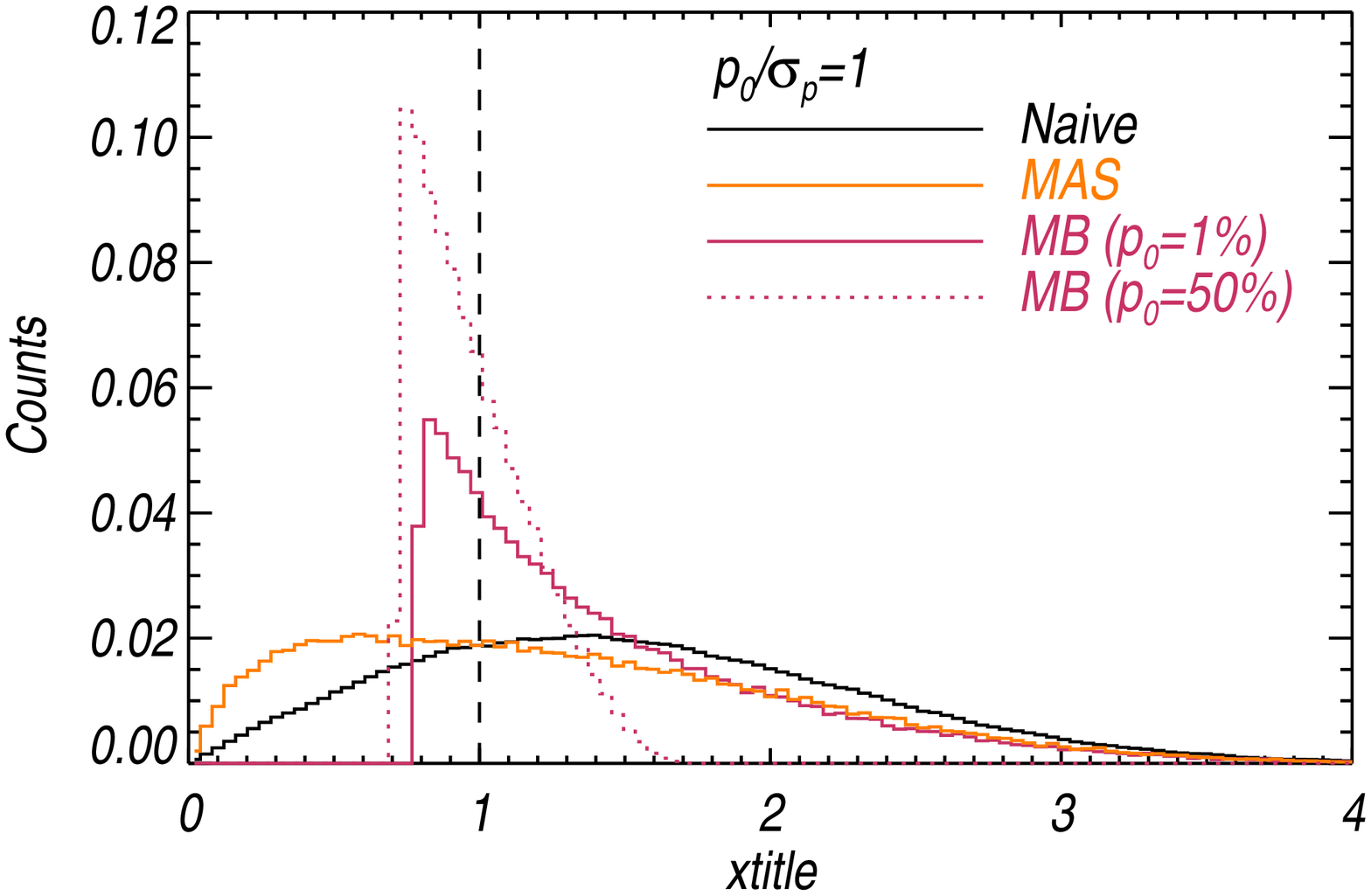}  \\
    \psfrag{xtitle}{$\hat{p}/p_0$}
  \includegraphics[width=.5\textwidth]{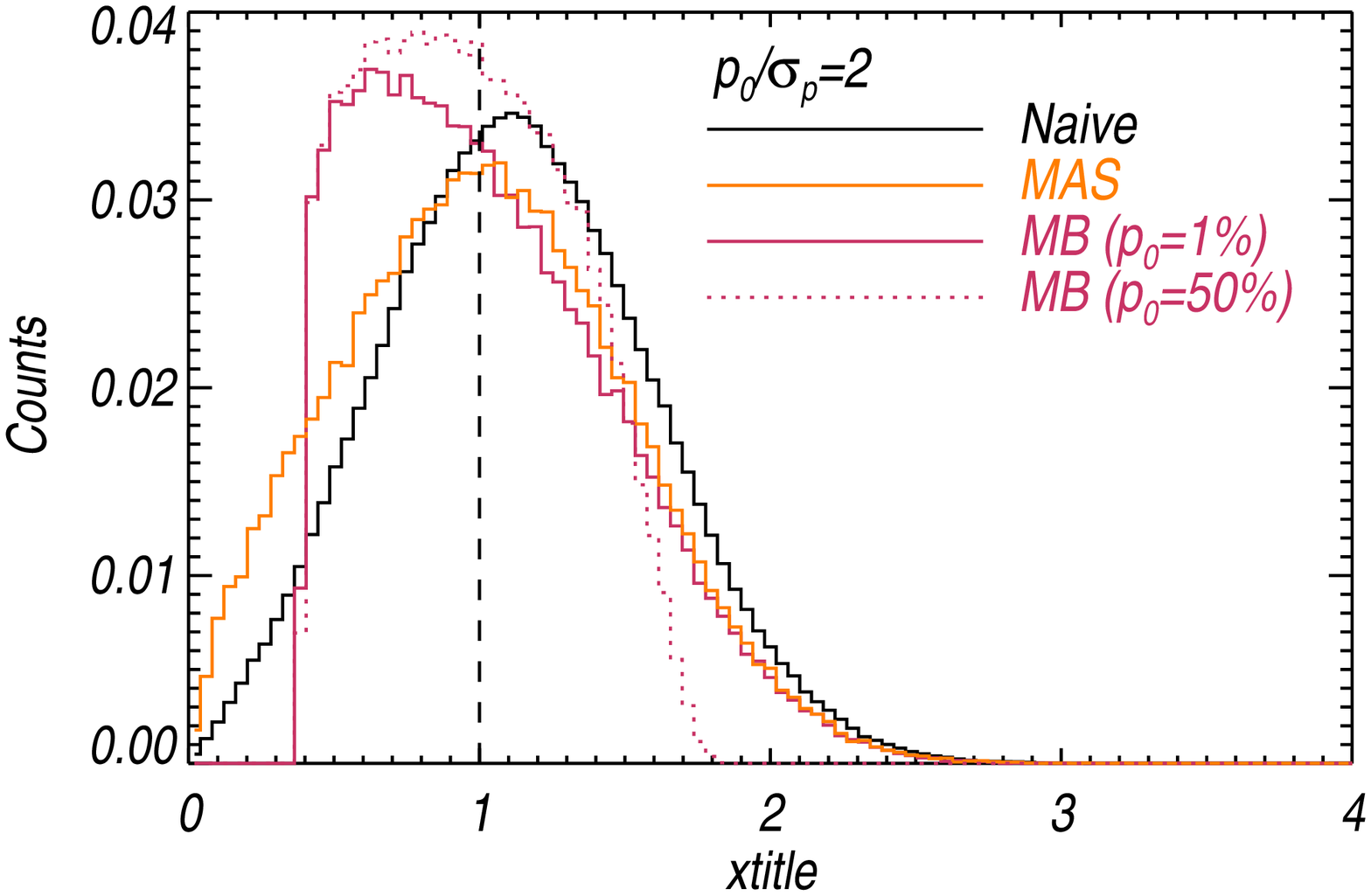}  \\
    \psfrag{xtitle}{$\hat{p}/p_0$}
  \includegraphics[width=.5\textwidth]{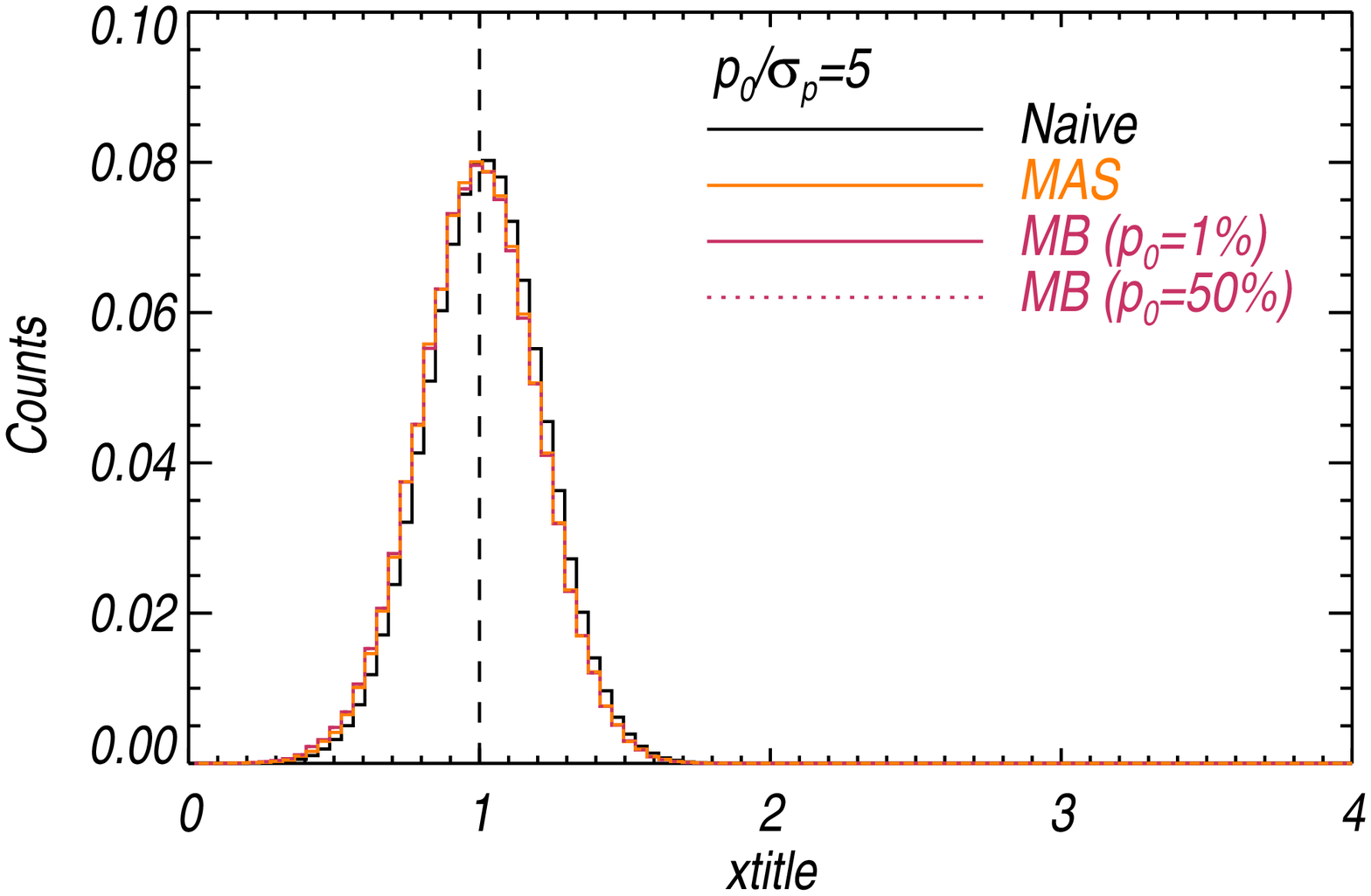}  \\
  \end{tabular}
\caption{Output distributions of the na\"ive (black),  MAS (orange) and the  MB (pink) $\hat{p}$ 
  estimators in the canonical case ($\varepsilon_\mathrm{eff}$=1), for three levels of the SNR
  $p_0/\sigma_p$=1,2 and 5 (from top to bottom). In the case of the MB estimator, 
  we show two setups of $p_0$=1\% and 50\% to illustrate the dependence of the output distribution on the $p_0$ value, 
  due to the prior used in the Bayesian approach ($\hat{p}_{\text{MB}} \in [0,1]$ so that $\hat{p}_{\text{MB}}/p_0 \in [0,1/p_0]$). 
  The other estimators are not sensitive to the true value $p_0$.}
\label{fig:estimator_comparison_histo}
\end{figure}

\section{$\hat{p}$ estimator performance}
\label{sec:comparison_p_estimators}

\subsection{Methodology}
\label{sec:comparison_p_methodoloy}

We investigate in this section the capability at providing polarization fraction estimates with low bias 
of the seven  $\hat{p}$ estimators introduced in the previous sections: 
the na\"ive measurements $p$, the Maximum Likelihood (ML), the Most Probable (MP and MP2), the Asymptotic (AS), 
the Modified Asymptotic (MAS) and the Mean Bayesian Posterior (MB) estimators.
Their performance is first quantified in terms of relative bias and risk  function of the
resulting estimates. Given true polarization parameters ($p_0$, $\psi_0$) and a covariance matrix $\tens{\Sigma}_p$, 
we build a sample of one million simulated measurements ($p$,$\psi$) by adding noise on the true Stokes parameters 
using the covariance matrix. We define the relative bias and risk  function  on $p$ as follows:
\begin{equation}
\mathrm{Bias}_p \equiv \frac{\left<\hat{p}\right> - p_0}{\sigma_{p,G}} \quad \mathrm{and} \quad \mathrm{Risk}_p \equiv \frac{\left<(\hat{p}-p_0)^2\right>}{\sigma_{p,G}^2} \, ,
\end{equation}
where $\hat{p}$ is the polarization fraction estimate computed on the simulated measurements $p$, 
$p_0$ is the true polarization fraction, $< >$ denotes the average computed over the simulated sample, 
and $\sigma_{p,G}$ is the estimate of the noise of the polarization fraction. 
The choice of $\sigma_{p,G}$ to scale the absolute bias and risk function, as a proxy of the 
$\hat{p}$ uncertainty, is motivated by the fact that it depends only on the effective ellipticity and not on $\psi_0$. 
Notice that this choice can lead to a relative risk function falling below 1 at low SNR, due to the fact that 
$\sigma_{p,G}^2$$>$$Var$ in this regime. 
The accuracy of the $p$ estimators is also quantified regarding the shape of their output distributions.
We use the Jarque-Bera estimator \citep{Jarque1980} as a test of normality of the output distribution, and defined by
\begin{equation}
JB = \frac{n}{6} \left(  \frac{\mu_3^2}{\mu_2^{3}} + \left(\frac{\mu_4}{\mu_2^2}-3\right)^2 / 4   \right) \, ,
\end{equation}
where $n$ is the number of samples and $\mu_i$ is the na\"ive estimate of the ith central moment of the distribution.
This test is based on the joint hypothesis of the skewness and the excess kurtosis being zero simultaneously. 
A value $JB$=0 means a perfect agreement with the {\it normality} to the 4th order, but does not prevent departure from the normality at higher orders. 
This $JB$ estimator tends to a $\chi^2$ test with two degrees of freedom when $n$ becomes large enough. Hence
the $JB$ has to satisfy the following condition $JB<\chi^2_{\alpha}$, once chosen a significance level $\alpha$.
For a significance level $\alpha$=5\% and 1\%, we get the conditions $JB<5.99$ and $JB<9.21$, respectively.

\subsection{Canonical case}
\label{sec:comparison_p_estimators_canonical}

We first assume the canonical simplification of the covariance matrix ($\varepsilon_\mathrm{eff}$=1). 
The  relative $\mathrm{Bias}_p$ and $\mathrm{Risk}_p$ quantities are shown on 
Fig.~\ref{fig:comparison_p_estimator} for the seven $\hat{p}$ estimators. 
We  recall  that the discontinuous estimators, shown in dashed line
(ML (blue), MP (light green), MP2 (green) and AS (red)), 
have an output distribution presenting a strong peak at zero, which leads to artificially lower  the statistical relative
$\mathrm{Bias}_p$ when simply including null values instead of using upper limits, 
as discussed in Sect.~\ref{sec:discontinuous_estimators}. Effectively
 these estimators show the lowest relative biases (top panel of Fig.~\ref{fig:comparison_p_estimator}) 
 compared to the MAS (orange) and MB (pink) estimators.
Hence the ML and MP2 estimators seem to statistically over-correct the data, below SNR=3. 
Consequently, the ML, MP and AS $\hat{p}$ estimators have to be used with an extreme care to 
deal with null estimates. We suggest here to focus on the  two continuous estimators: MAS and MB.

MAS provides the  better performances in terms of relative bias over the whole range of SNR, while 
MB appears less and less efficient at correcting the bias when the SNR tends to zero.
At larger SNR ($>$2), MB tends to slightly over-correct with a small negative  relative  bias (2\% of $\sigma_p$) 
up to SNR $\sim$5, while MAS converges quickly to a null  relative  bias for SNR $>$ 3.

The MB estimator clearly minimizes the risk function (in the range 0.7$<$SNR$<$3.2), 
as expected for this kind of posterior estimator.  
At larger SNR ($>$3.2) both MAS and MB have roughly the same behavior, even if the risk function associated to 
MAS appears slightly lower.

The resulting $\hat{p}_{\text{MB}}$ distribution is highly asymmetric at low SNR (see upper panels of Fig.~\ref{fig:estimator_comparison_histo}), 
with a sharp cutoff at 0.8$\sigma_p$. Moreover, we note that the output $\hat{p}_{\text{MB}}$ distribution depends not only on the SNR
$p_0/\sigma_p$, but also on the value of the true polarization fraction $p_0$. We report two cases, $p_0$=1\% (pink)
 and 50\% (dotted pink) in Fig.~\ref{fig:estimator_comparison_histo}. This comes from the prior of the Bayesian method, which 
 bounds the estimate $\hat{p}_{\text{MB}}$ between 0 and 1. As a consequence, the {\it normality} of the Bayesian distribution 
 is extremely poor, as pointed in the bottom panel of Fig.~\ref{fig:comparison_p_estimator}, where we show that the JB test of the MB estimator
 is larger than 9.21 (consistent with a $\chi^2_{0.01}$ test) over the whole range of SNR explored here (up to SNR$\sim$5).
On the contrary, the resulting $\hat{p}_{\text{MAS}}$ distribution of Fig.~\ref{fig:estimator_comparison_histo} looks much better, 
mimicking the Rayleigh distribution for low SNR and going neatly to the Gaussian regime, as pointed out by P14.
The JB of the MAS estimator is the lowest for SNR $>$3 (see bottom panel of 
Fig.~\ref{fig:comparison_p_estimator}), illustrating the consistency between the MAS distribution and the normal distribution.
Notice that all distributions, na\"ive,  MAS and MB, converge to a Gaussian distribution at higher SNR.

\begin{figure}[!tp]
    \psfrag{toto1}{$\kappa \in [0, 100 p_0]$}
    \psfrag{toto2}{$\kappa \in [0, 10 p_0]$}
    \psfrag{toto3}{$\kappa \in [0, 5 p_0]$}
    \psfrag{toto4}{$\kappa \in [0, 3 p_0]$}
    \psfrag{toto5}{$\kappa \in [0, 2 p_0]$}
  \includegraphics[width=.5\textwidth]{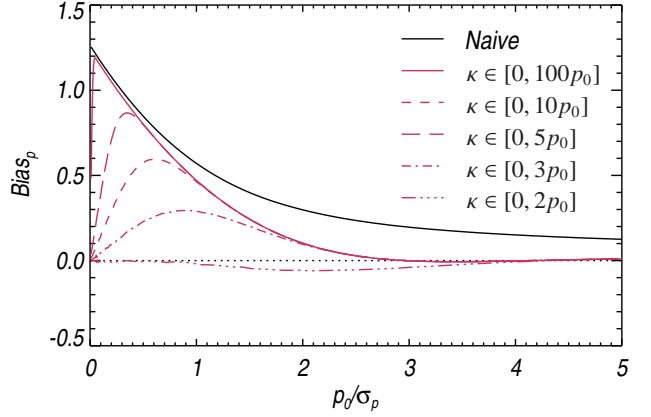}
  \caption{Impact of the flat prior interval upper limit on the  relative $\mathrm{Bias}_p$ performance of the MB estimator.}
\label{fig:impact_prior}
\end{figure}

\begin{figure}[!t]
    \psfrag{-----xtitle-----}{$<p_{0,i}>/\sigma_{p,G}$}
    \psfrag{----------ytitle----------}{$< \hat{p}_{i} - p_{0,i}> / \sigma_{p,G}$}
    \psfrag{MB toto2}{MB flat prior}
    \psfrag{MB toto3}{MB prior ($\hat{p}_i$)}
    \psfrag{MB toto4}{MB prior ($p_{0,i}$)}
  \includegraphics[width=.5\textwidth]{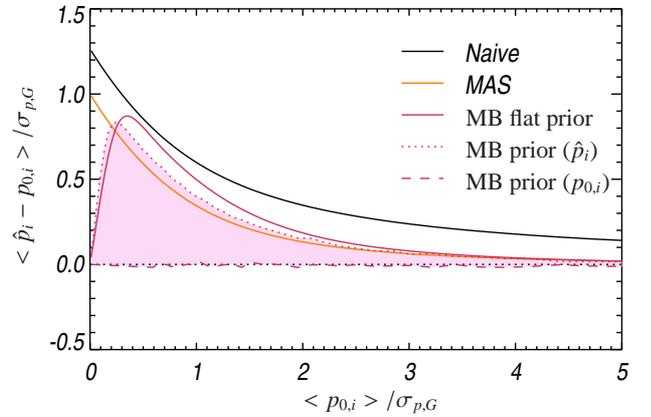} 
  \caption{Illustration of the improvement of the MB estimator performances when using evolved priors.
  Starting from an input distribution of true values ($p_{0,i}$), shown in Fig.~\ref{fig:impact_prior_histo_distrib}, the statistical relative bias is shown for four estimators: 
  na\"ive, MAS, and MB with three different priors.} 
\label{fig:impact_prior_histo}
\end{figure}

\subsection{Impact of the Bayesian prior}
\label{sec:priors}

\begin{figure}[!tph]
\vspace{1.1cm}
	\begin{tabular}{c}	
      \psfrag{toto1}{$<p_{0,i}>/\sigma_{p,G}$=1}
    \psfrag{MB toto2}{MB flat prior}
    \psfrag{MB toto3}{MB prior ($\hat{p}_i$)}
    \psfrag{MB toto4}{MB prior ($p_{0,i}$)}
  \includegraphics[width=.5\textwidth]{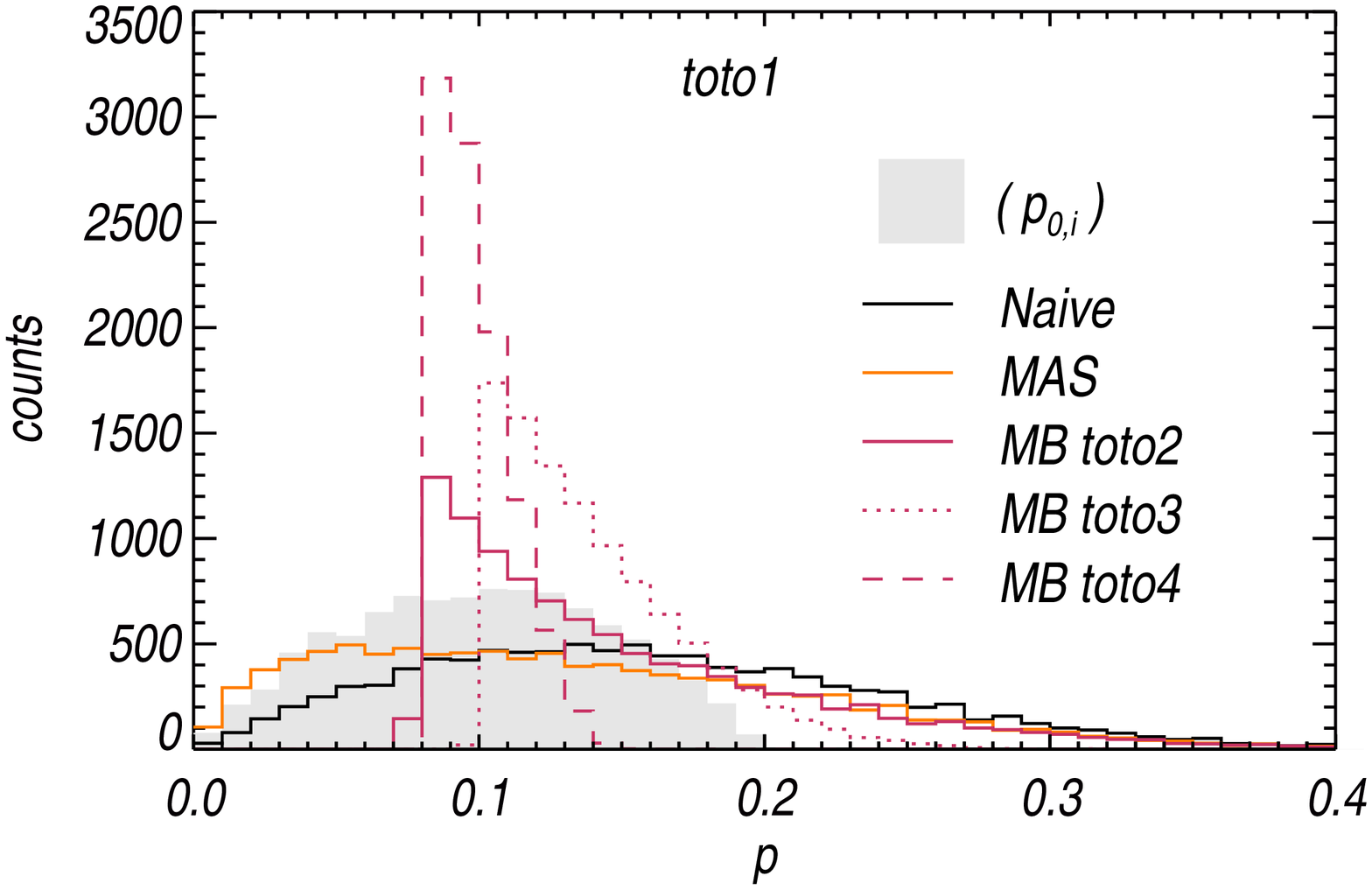} \\
    \psfrag{toto1}{$<p_{0,i}>/\sigma_{p,G}$=2}
      \psfrag{MB toto2}{MB flat prior}
    \psfrag{MB toto3}{MB prior ($\hat{p}_i$)}
    \psfrag{MB toto4}{MB prior ($p_{0,i}$)}
  \includegraphics[width=.5\textwidth]{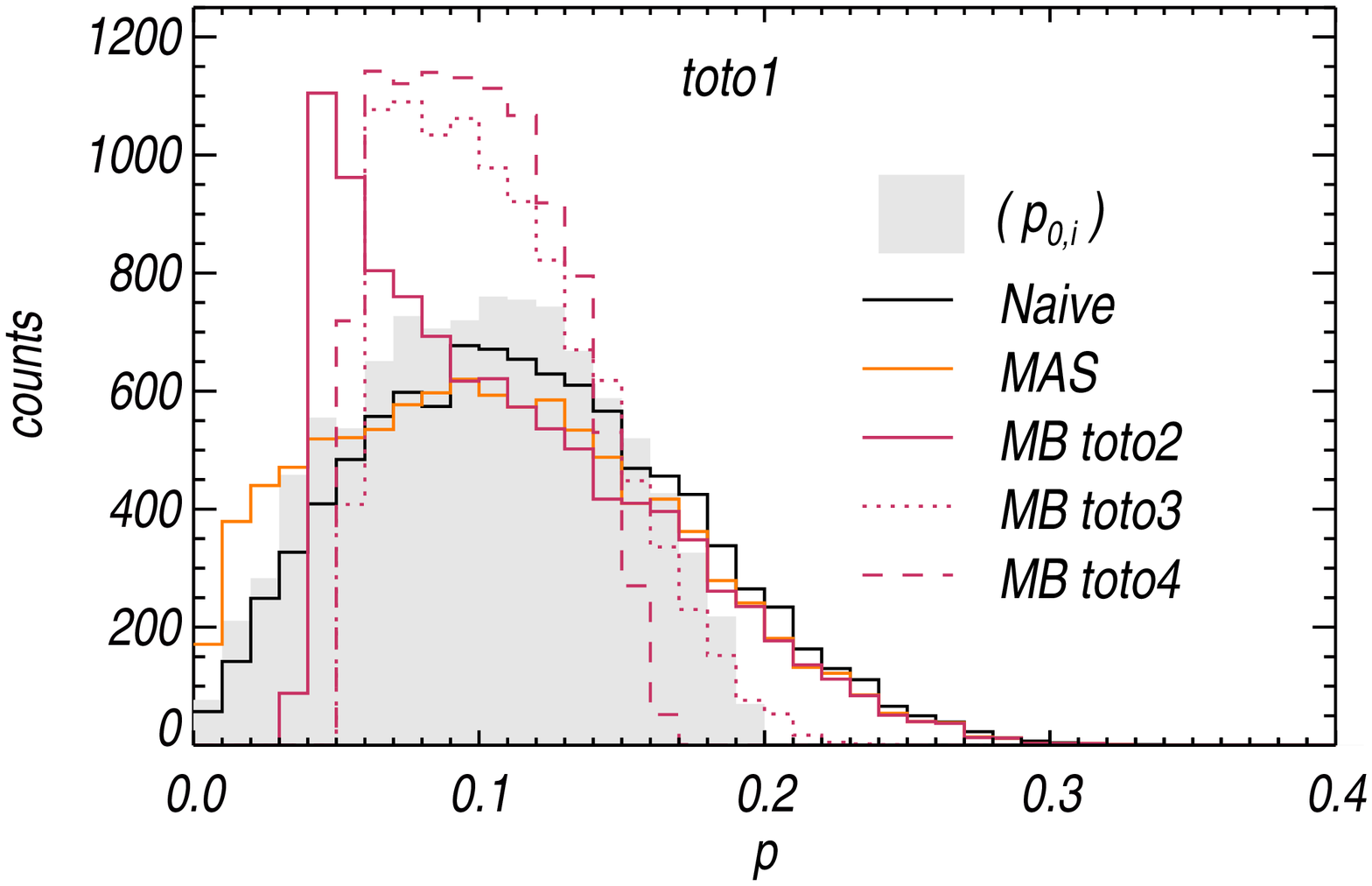} \\
    \psfrag{toto1}{$<p_{0,i}>/\sigma_{p,G}$=3}
    \psfrag{MB toto2}{MB flat prior}
    \psfrag{MB toto3}{MB prior ($\hat{p}_i$)}
    \psfrag{MB toto4}{MB prior ($p_{0,i}$)}
  \includegraphics[width=.5\textwidth]{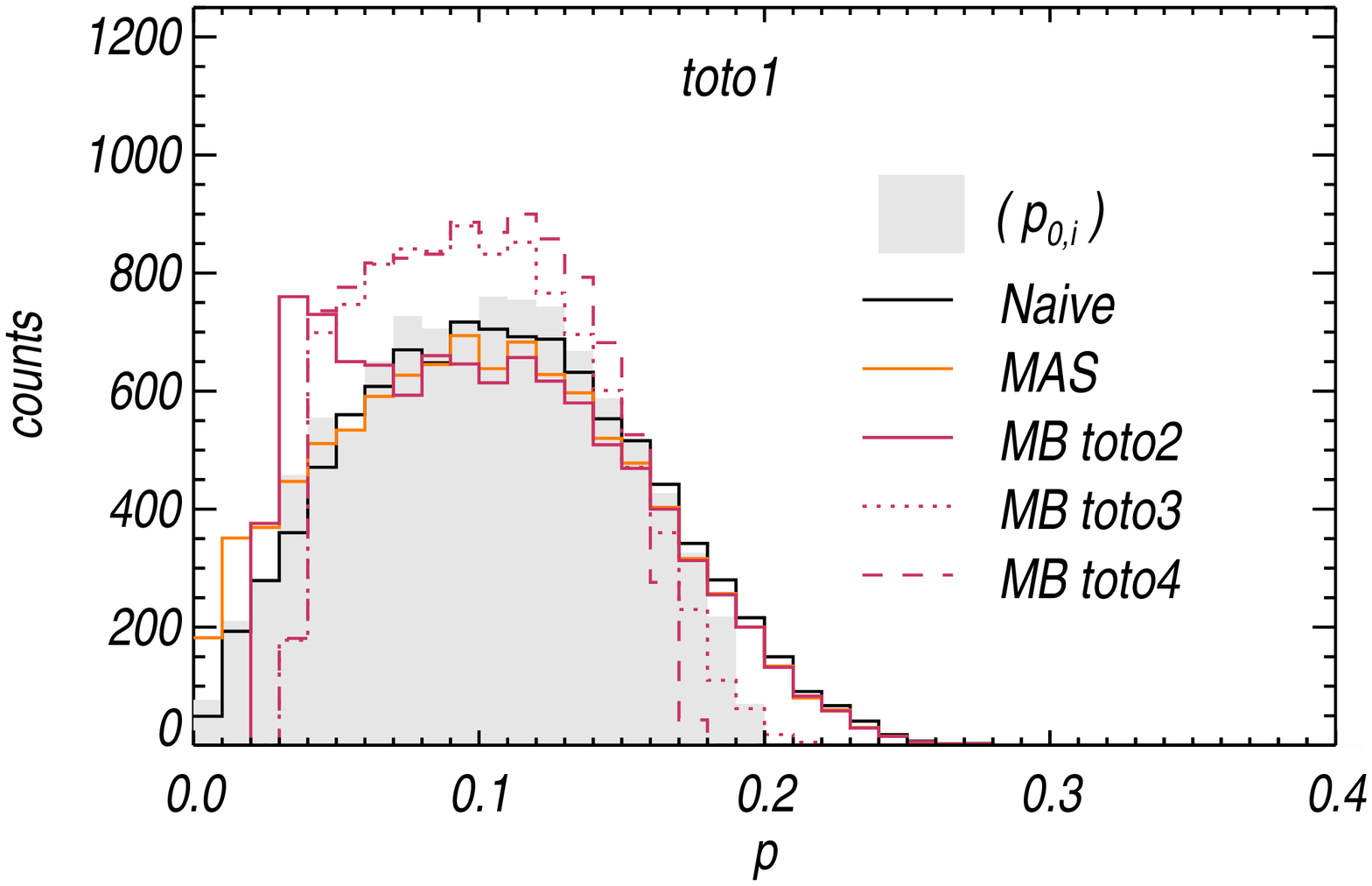}
  \end{tabular}
\caption{Output distributions  of the $\hat{p}$ estimates starting from a distribution of independent 
true values $(p_{0,i}$ centered around 10\% of polarization fraction (grey shaded region) shown at three
levels of noise characterized by the mean SNR $\langle{p_{0,i}}\rangle/\sigma_{p,G}$=1, 2 and 3 (top, middle and bottom, respectively).
The na\"ive (black) and MAS (orange) output distributions are compared to the MB output distributions obtained with three different priors:
flat prior between 0 and 1 (solid pink), set to the na\"ive output distribution (dotted pink) and set to the true input distribution (dashed pink).}
\label{fig:impact_prior_histo_distrib}
\end{figure}

\begin{figure}[!tph]
\vspace{1.1cm}
	\begin{tabular}{c}	
    \psfrag{toto1}{$<p_{0,i}>/\sigma_{p,G}$=1}
    \psfrag{MB toto2}{MB flat prior}
    \psfrag{MB toto3}{MB prior ($\hat{p}_i$)}
    \psfrag{MB toto4}{MB prior ($p_{0,i}$)}
  \includegraphics[width=.5\textwidth]{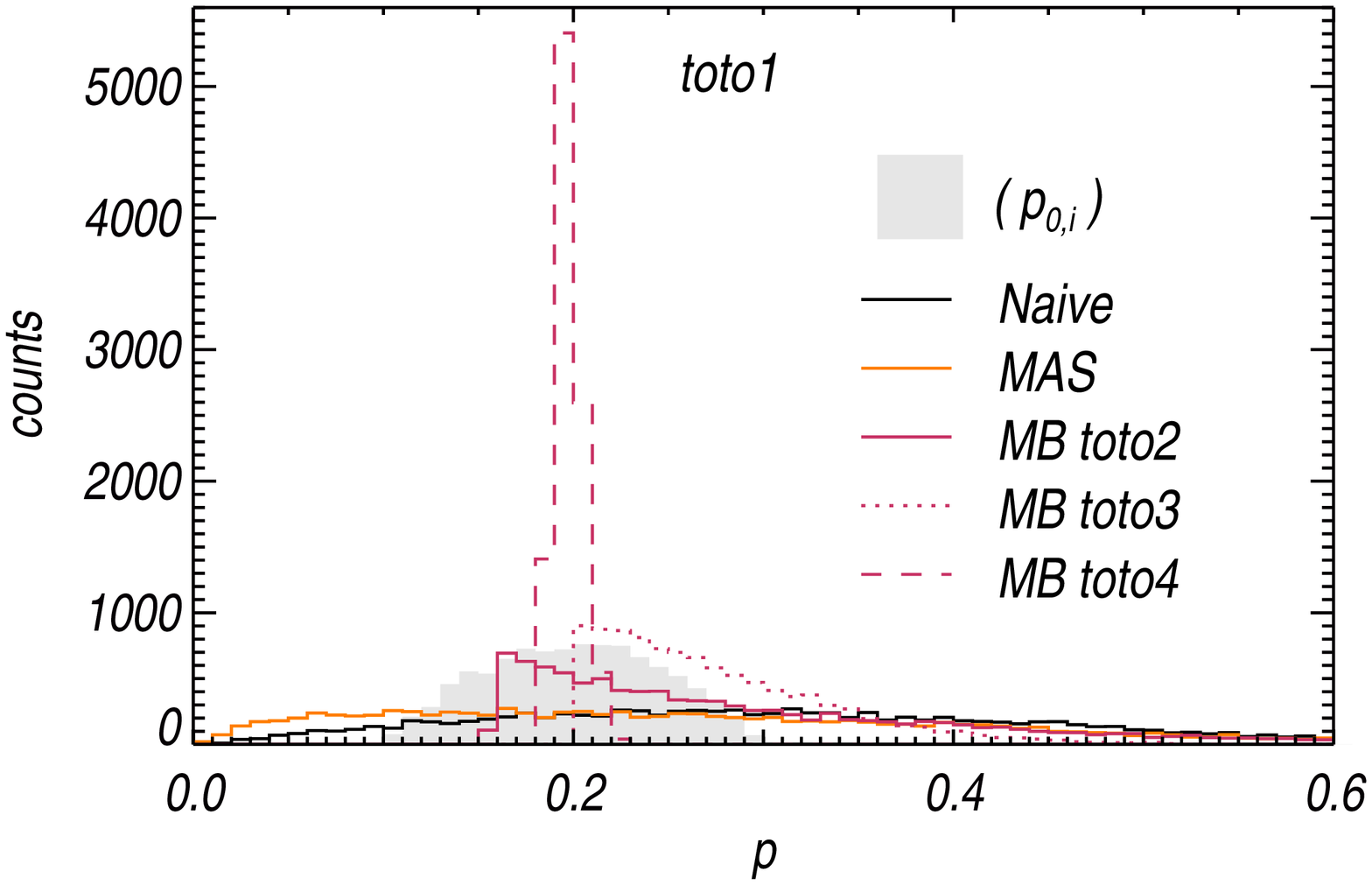} \\
    \psfrag{toto1}{$<p_{0,i}>/\sigma_{p,G}$=2}
      \psfrag{MB toto2}{MB flat prior}
    \psfrag{MB toto3}{MB prior ($\hat{p}_i$)}
    \psfrag{MB toto4}{MB prior ($p_{0,i}$)}
  \includegraphics[width=.5\textwidth]{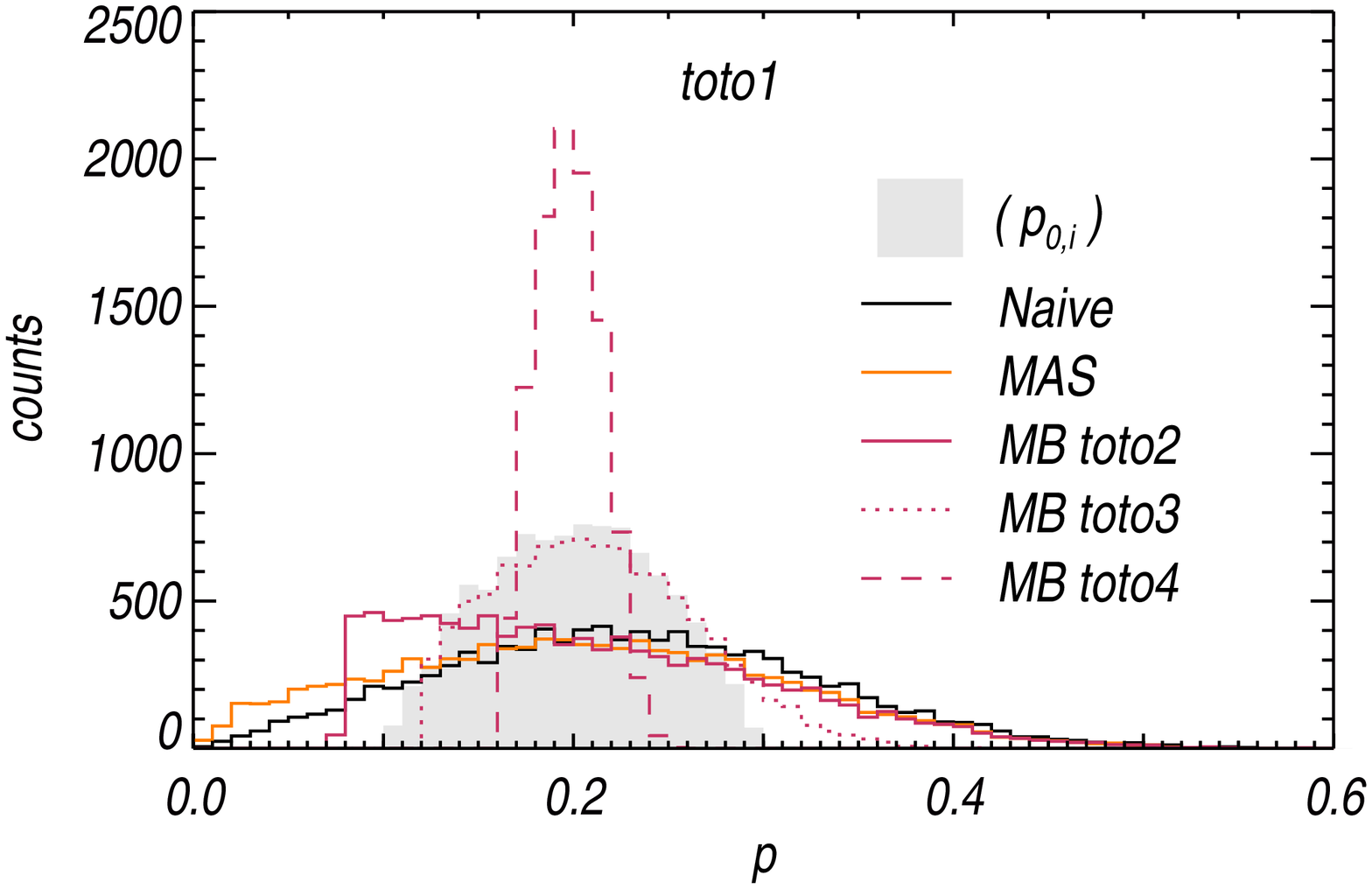} \\
    \psfrag{toto1}{$<p_{0,i}>/\sigma_{p,G}$=3}
    \psfrag{MB toto2}{MB flat prior}
    \psfrag{MB toto3}{MB prior ($\hat{p}_i$)}
    \psfrag{MB toto4}{MB prior ($p_{0,i}$)}
  \includegraphics[width=.5\textwidth]{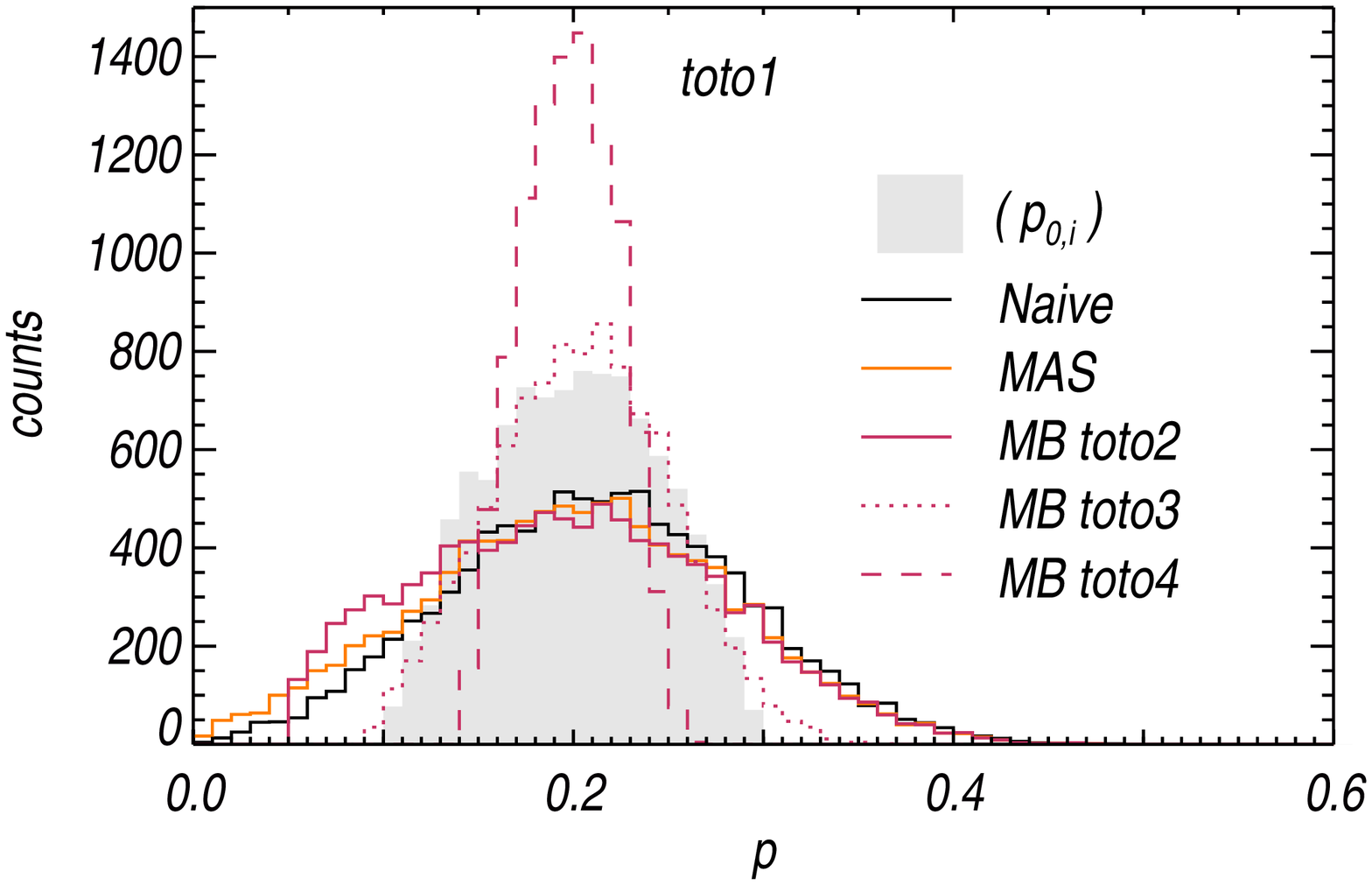}
  \end{tabular}
\caption{Same as Fig.~\ref{fig:impact_prior_histo_distrib} with a different initial distribution $(p_{0,i})$ centered on 20\% of polarization fraction.}
\label{fig:impact_prior_histo2_distrib}
\vspace{3cm}
\end{figure}

The choice of the prior is crucial in the Bayesian approach, and
we have seen how it is hard to define a non-informative prior in Sect.~\ref{sec:bayesian_estimators}.
The  MB estimator studied 
up to now assumes a flat prior in $p_0$ between 0 and 1, equivalent to no a priori knowledge. 
In practice when dealing with astrophysical data, we can bound the expected true values of the polarization fraction between much  tighter  limits. 
We know, for example, that the polarization fraction of the synchrotron signal peaks at  $\sim$75\%, but never reaches this maximum due 
to line-of-sight averaging. The maximum polarization fraction of the dust thermal emission is still a debated issue, 
but is unlikely to be larger than 20 to 30\% \citep{Benoit2004}. Appropriate priors can then be introduced to take into account this a priori physical knowledge into
the MB estimator.

We have already observed in Sect.~\ref{sec:comparison_p_estimators_canonical} how the output distribution of the $\hat{p}_{\text{MB}}$ estimates
is impacted by the value of the true $p_0$ (1\% or 50\%) due to the upper limit ($p_0$$<$1) of the prior, see Fig.~\ref{fig:estimator_comparison_histo}. 
We explore here a family of simple priors defined by $\kappa(p'_0) = 1/(kp_0)$ for $p'_0 \in[0, kp_0]$ and 0 otherwise, 
where we adjust the upper limit of the prior as a function of the expected true value.
We performed Monte Carlo simulations in the canonical case by setting the true value at $p_0$=1\% and varying the upper limit of the prior ($k=2,3,5,10,$ and 100).
The statistical  relative $\mathrm{Bias}_p$ of the MB estimators associated with each version of the priors are shown on Fig.~\ref{fig:impact_prior}.
The smaller the upper limit, the lower the relative $\mathrm{Bias}_p$, as expected. However the upper limit of the prior has to be very constraining ($k\le3$) to
observe a decrease of the relative bias in the range of SNR between 1.5 and 3. This requires very good a priori knowledge.
Using more relaxed priors ($k\ge5$) will not improve significantly the performances of the MB estimator at SNR$>$1.

When dealing with maps of polarized data, an interesting approach would be to start by estimating the histogram of $p$ values in the map and use it as a prior into
our MB estimators, even if this moves away from a strictly Bayesian approach again by introducing a data-dependent prior.
 As a first guess, the prior can be set to the histogram of the na\"ive estimates of $\hat{p}$, but 
a more sophisticated prior would be an histogram of $p$ deconvolved from the errors, using a Maximum Entropy method for example.

We illustrate the performance of the MB estimator with this kind of  prior on Figs.~\ref{fig:impact_prior_histo} and \ref{fig:impact_prior_histo_distrib}.
We start with a sample of 10~000 independent true values $(p_{0,i})$ ranging between 0 and 20\% of polarization fraction, 
with a distribution shown as the grey shaded histogram in Fig.~\ref{fig:impact_prior_histo_distrib} 
on which a random realization of the noise is added with the same noise level over the whole sample, leading
to varying SNRs through the sample.
We explore two extreme cases of the Bayesian prior, corresponding to i) an idealistic perfect knowledge of the input distribution 
and ii) its first guess provided by the na\"ive estimates. Hence the prior is chosen as the input 
distribution of the true $p_{0,i}$ values (dashed pink) and the output distribution of the na\"ive estimates (dotted pink). 
We compare the performance of these two new versions of the MB estimators 
with the na\"ive (black), MAS (orange) and flat prior MB (solid pink) estimators, in terms of relative bias in Fig.~\ref{fig:impact_prior_histo}.

We stress that the relative bias values are not defined as previously done in Sect.~\ref{sec:comparison_p_methodoloy}, 
but refer now to the mean of the difference
between each sample of true value $p_{0,i}$ and its associated estimate $\hat{p}_i$.
The pink shaded region provides the domain of the possible improvement of the
 MB estimators, by setting an appropriate prior as close as possible to the true distribution.
The improvements may seem spectacular, leading to a statistical relative bias close to zero at all SNRs in the best configuration (dashed line).
Caution is warranted, however, when looking at the output distributions associated with these new MB estimators on Fig.~\ref{fig:impact_prior_histo_distrib}, 
shown for three levels of the noise chosen so that the mean SNR is $\overline{p_0}/\sigma_{p,G}$=1, 2 and 3.
At low SNR ($\simeq$1), the output distribution of the MB estimator with a {\it perfect} prior (dashed line) is extremely peaked around 
the mean value of the sample $\overline{p_0}$, but does not match the input distribution at all. Even at higher SNR (2-3), 
the three MB output distributions suffer from the same feature already mentioned in Sect.~\ref{sec:comparison_p_estimators_canonical}, 
a sharp cutoff at low values of $p$.
Using a prior that is too constraining will yield dramatic cuts of the extremes values of the input distribution.
By contrast, the na\"ive prior is quite effective in that it allows the MB estimator 
 to recover the upper limit of the input distribution reasonably well at a SNR$\gtrsim$2, while the other estimators fail to do so at such low SNR.

The performance of the MB estimator with an evolved prior will also strongly 
depend on the initial true distribution of the polarization fraction. For example we
duplicated the analysis made above with a different initial distribution $(p_{0,i})$ centered on 20\% of
 polarization fraction instead of 10\% (see Fig.~\ref{fig:impact_prior_histo2_distrib}).
 In this configuration, the output distributions of the Bayesian estimators are not as much affected by the 
 cut-off at low $p$ as observed in Fig~\ref{fig:impact_prior_histo_distrib}.
 The MB estimator with the na\"ive prior appears extremely effective, even at low SNR ($\sim$2).

\subsection{Robustness to the covariance matrix}
\label{sec:robustness_covariance_matrix}

In PMA I we have discussed extensively the impact of the  asymmetry of the covariance matrix on the 
measurements of the polarization fraction. In particular, we have stressed that once the effective ellipticity
departs from the canonical case, the bias on the polarization fraction now depends on the true polarization angle $\psi_0$,
which remains unknown. We would like to explore in this section how the performance of the various $\hat{p}$ estimators are sensitive to
the  effective ellipticity of the covariance matrix.

\begin{figure*}[th!]
  \centering
  \begin{tabular}{cc}
  \psfrag{ytitle}{$\hat{p}$}
  \includegraphics[width=.5\textwidth]{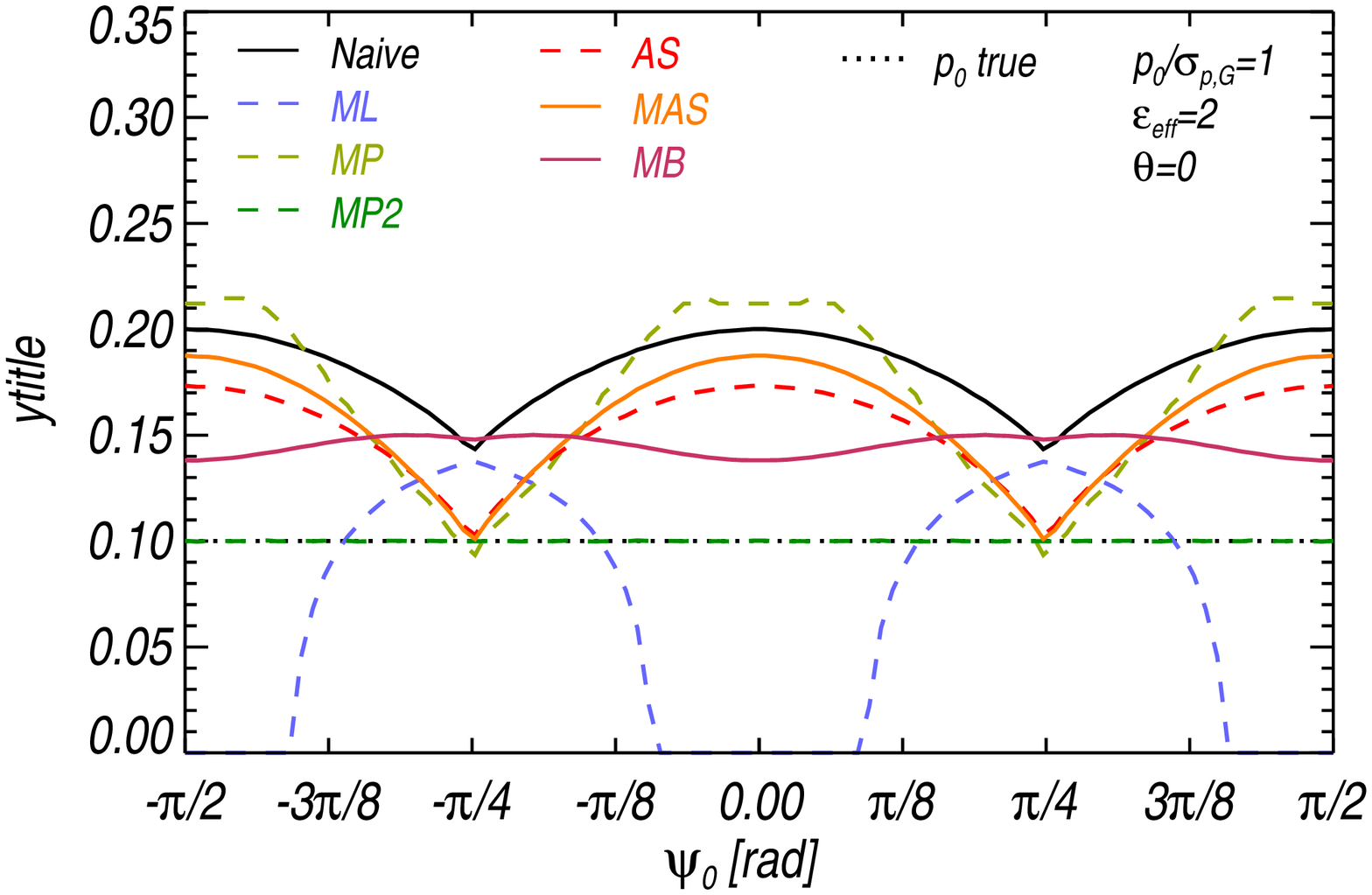} & 
  \psfrag{ytitle}{$\hat{p}$}
    \includegraphics[width=.5\textwidth]{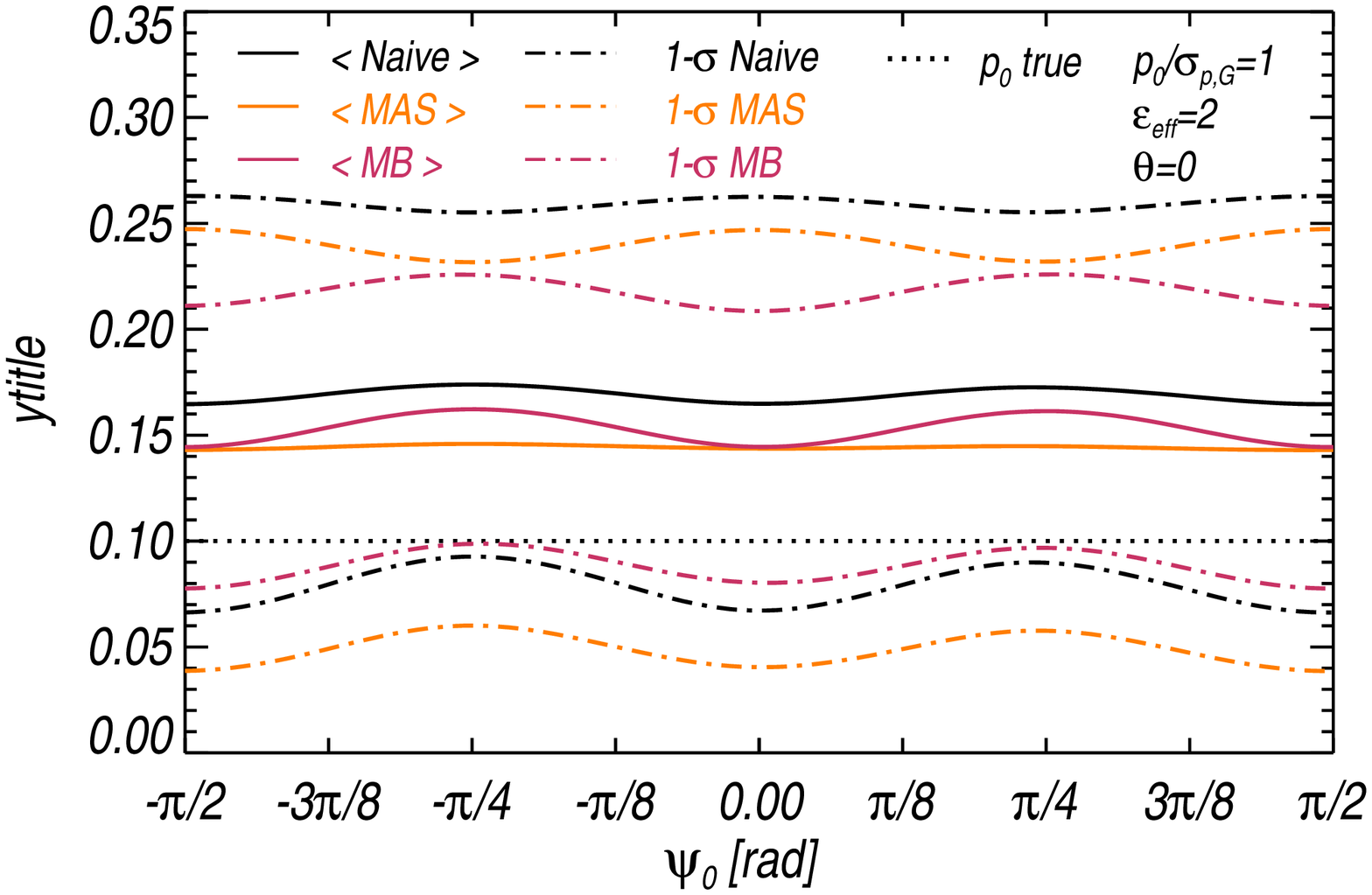}
    \end{tabular}
\caption{Illustration of the robustness of the $\hat{p}$ estimators against the unknown $\psi_0$ parameter when
the covariance matrix departs from the canonical value. The covariance matrix is setup with $\varepsilon_\mathrm{eff}$=2 and a SNR $p_0/\sigma_{p,G}$= 1, and 
a true polarization fraction $p_0$=0.1. For each value of $\psi_0$, we first illustrate (on the left panel) the performance of the 7 estimators on
 one particular measurement set to the maximum of the pdf. We focus then on the statistical average estimates $\hat{p}$
computed over 10~000 Monte-Carlo realizations for the na\"ive, MAS and MB estimators (right panel). The full lines stand for the mean, and the dot-dash lines for the 
1-$\sigma$ dispersion.}
\label{fig:impact_cov_epsilon}
\end{figure*}

\begin{figure*}[bh!]
  \centering
  \begin{tabular}{cc}
  \includegraphics[width=.5\textwidth]{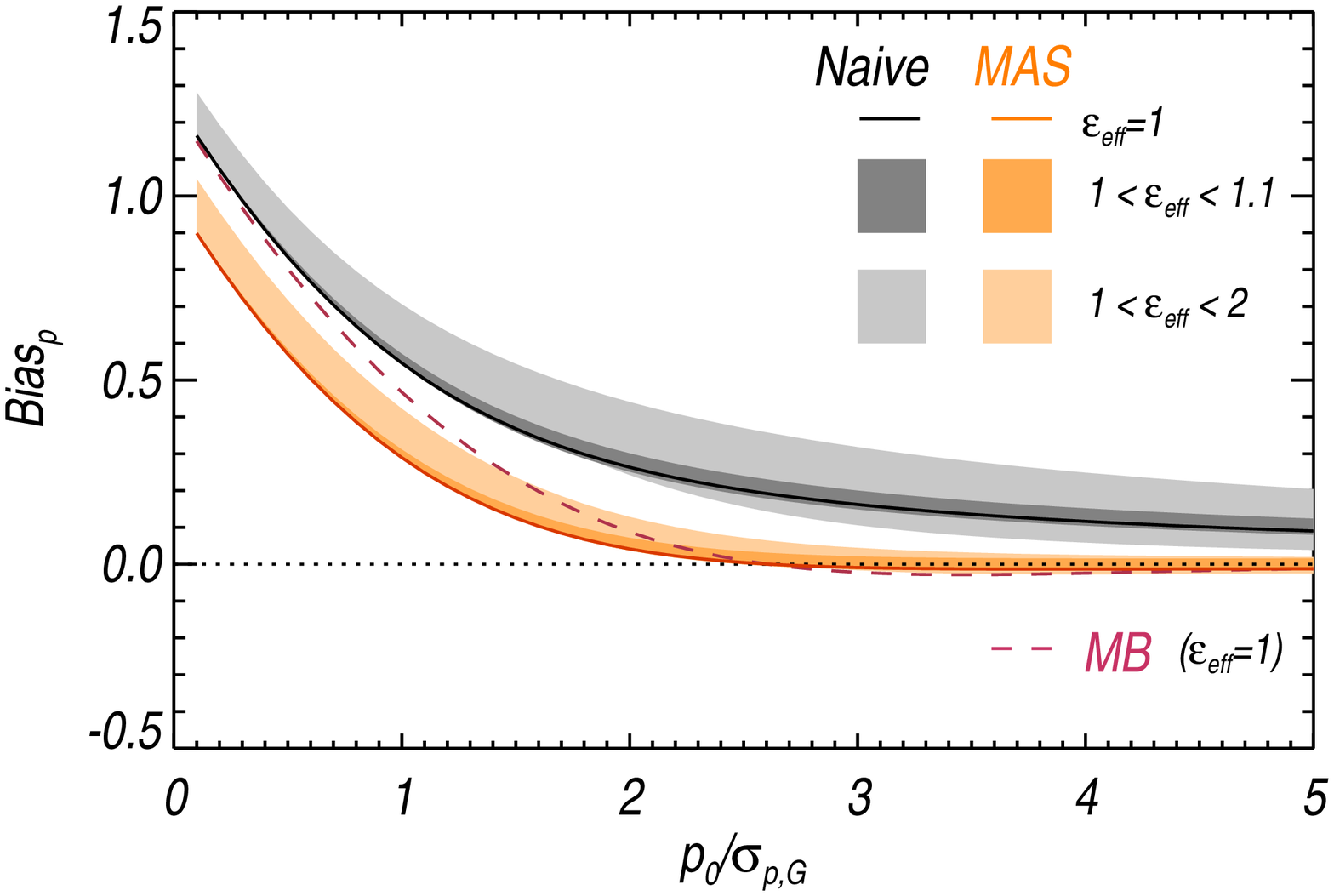} &
  \includegraphics[width=.5\textwidth]{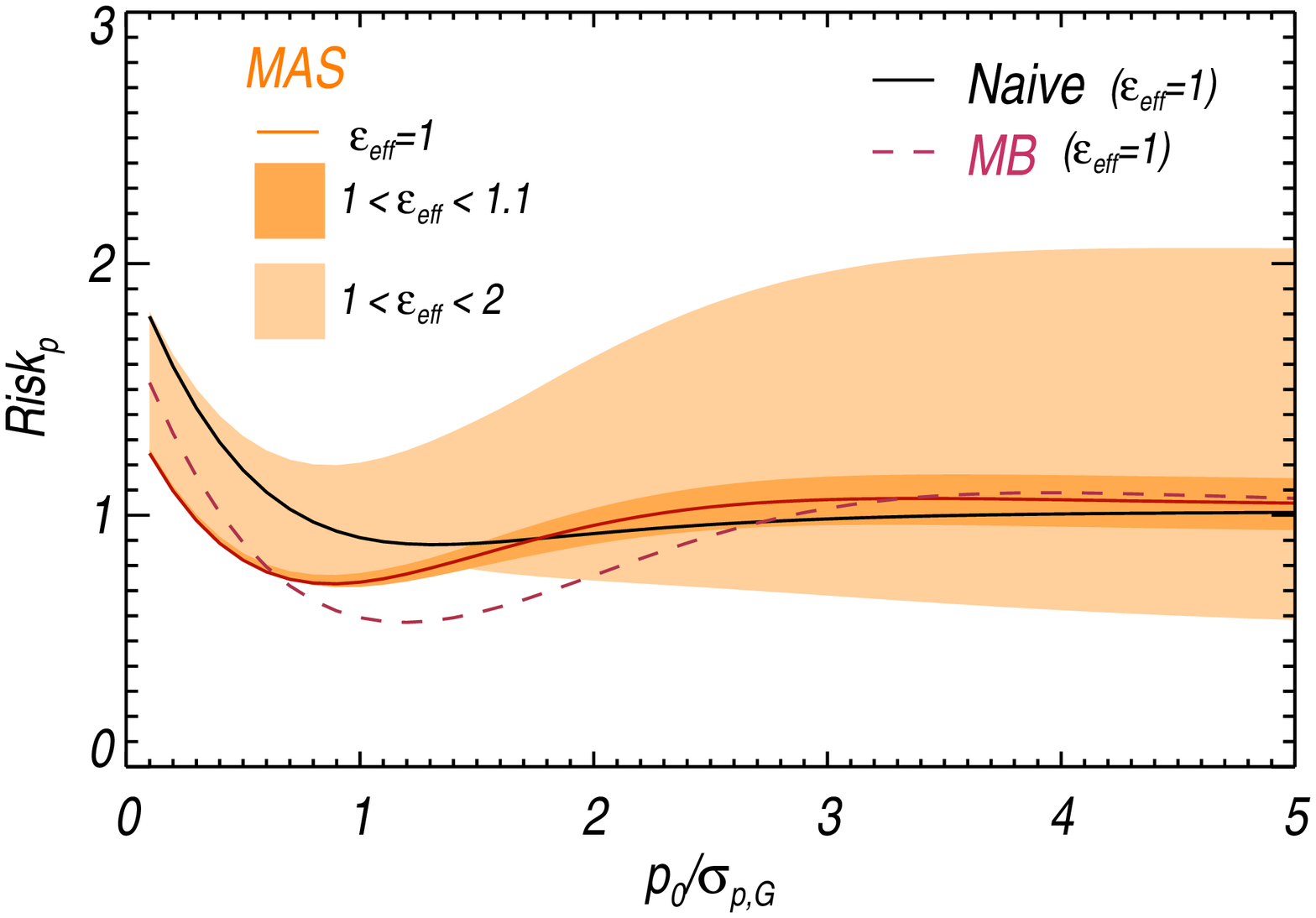} \\
  \includegraphics[width=.5\textwidth]{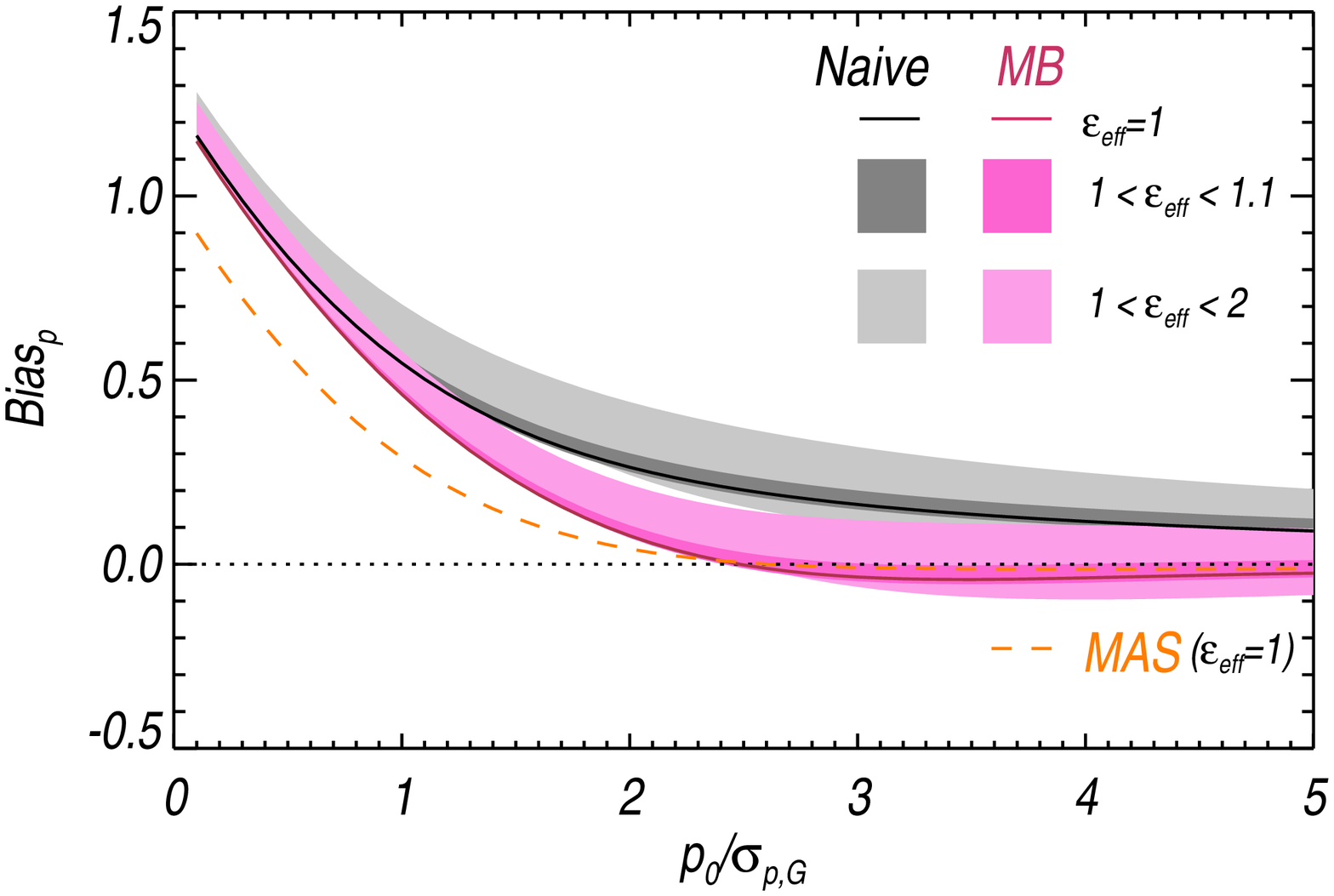} &
  \includegraphics[width=.5\textwidth]{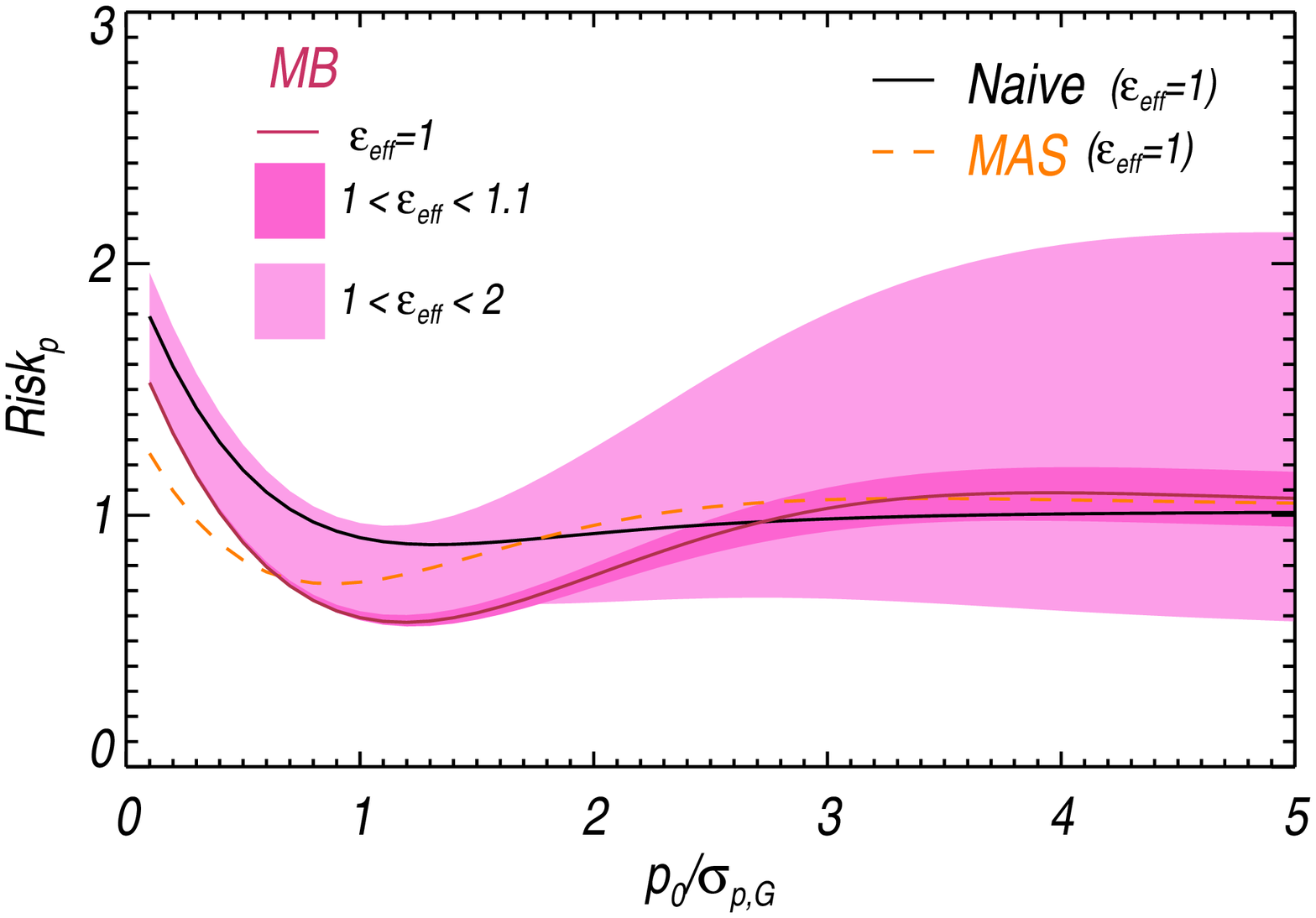} \\
\end{tabular}
\caption{Impact of the effective ellipticity of the covariance matrix on the statistical relative $\mathrm{Bias}_p$ (left column) and $\mathrm{Risk}_p$ (right column) 
quantities in the {\it extreme} (light shaded region) and {\it low} (dark shaded regions) regimes, for 
both MAS (orange, top) and MB (pink, bottom) $\hat{p}$ estimators.
The domain of the na\"ive measurements is repeated in grey shaded regions on both plots. 
The canonical case of the MAS (and MB) is also repeated on each panel in dashed orange (and pink) lines.}
\label{fig:estimator_comparison_fullcov}
\end{figure*}

We illustrate the dependence of the $\hat{p}$ estimators on the true polarization angle $\psi_0$ in Fig.~\ref{fig:impact_cov_epsilon}.
Given true polarization parameters ($p_0$=0.1 and $\psi_0$ ranging between -$\pi/2$ and $\pi/2$)
and a covariance matrix characterized by $\varepsilon_\mathrm{eff}$=2 and $\theta$=0 (left panel),  and a SNR
$p_0/\sigma_{p,G}$=1, we first set the polarization 
measurements ($p$, $\psi$) to the maximum of the pdf $f_{2D}$ (left panel).
We apply then the six estimators on these measurements to get  the $\hat{p}$ estimates for each $\psi_0$ between -$\pi/2$ and $\pi/2$. 
With this particular setting, the MP2 (green) estimator gives back 
the true polarization fraction $p_0$ whatever the polarization angle $\psi_0$, by definition of this estimator 
and the choice of the measurement in this example. On the contrary, the MP (light green) and the ML (blue) 
estimators are extremely sensitive to the true polarization angle $\psi_0$, yielding estimates spanning between 0 and 1.4$p_0$, while
the AS (red) and MAS (orange) estimators yield results spanning between 1 to 1.8$p_0$ when $\psi_0$ varies.
The MB (pink) estimator provides stable estimates in the range 1.4 to 1.5\,$p_0$,
which is consistent with the fact that the posterior estimators minimize the risk function.  This of course has a cost, and the MB estimator 
provides here the largest averaged  relative bias compared to the other methods, with the exception of the na\"ive (black) one.

More generally,  for each value of the true polarization angle $\psi_0$ between $-\pi/2$ and $\pi/2$,
we build a sample of 10\,000 simulated measurements using the same setup of the covariance matrix as above. 
Then we compute the statistical average of the na\"ive, MAS and MB estimates (black, orange and pink lines, respectively) 
obtained on this simulated sample, 
with their associated 1-$\sigma$ dispersion (black, orange and pink dot-dash lines, respectively), as shown in the right panel of Fig.~\ref{fig:impact_cov_epsilon}.
 The averaged MB estimates present the same characteristic as shown on the left panel. 
 By contrast, the averaged MAS estimates are independent from the unknown $\psi_0$ true polarization angle. 
 The MAS 1-$\sigma$ dispersion is, however, slightly larger than the MB 1-$\sigma$ dispersion.

The impact of the  effective ellipticity of the covariance matrix is then analysed statistically
 for the MAS and MB estimators only in Fig.~\ref{fig:estimator_comparison_fullcov}. 
 Instead of looking at the accuracy of the $\hat{p}$ estimators 
 around one particular measurement (the most probable one) as done in Fig.~\ref{fig:impact_cov_epsilon}, 
 for each set of true polarization parameters ($p_0$=0.1, $\psi_0$), with $\psi_0$ ranging between
 -$\pi/2$ and $\pi/2$, we perform Monte Carlo simulations. For each set of true polarization parameters, 
 we build a sample of 100~000 simulated measurements on which we apply the MAS and MB estimators to finally compute
 the statistical relative $\mathrm{Bias}_p$ and $\mathrm{Risk}_p$, as defined in Sect.~\ref{sec:comparison_p_methodoloy}.
 This is done for various setups of the covariance matrix chosen to cover the whole range of the {\it extreme} and {\it low} regimes.
 The minimum and maximum relative $\mathrm{Bias}_p$ and $\mathrm{Risk}_p$ are then computed over the whole range of $\psi_0$ and effective ellipticity $\varepsilon_\mathrm{eff}$
  in each regime of the covariance matrix  to build the shaded regions of Fig.~\ref{fig:estimator_comparison_fullcov}
 for the MAS (top panels) and MB (bottom panels) $\hat{p}$ estimators.
The domain of the na\"ive measurements in each regime is repeated in grey shaded regions, while 
we show the result in orange shaded regions for the MAS and pink shaded regions for the MB estimators. 
 It appears that the relative $\mathrm{Bias}_p$ of the MAS estimator is less impacted by a change of ellipticity for SNR$>$2 than the MB estimator, 
even in the {\it extreme} regime of the covariance matrix.
The dependance of the risk function on the ellipticity is almost identical for the two estimators around their respective canonical curve.
The thickness of the risk function  region is slightly smaller for the MB estimator than for the MAS estimator at low SNR ($<$3), 
while it is the opposite for larger SNR ($>$3), as already observed in the canonical case. 

\begin{figure}[t]
\begin{tabular}{c}
  \psfrag{-----------------ytitle-----------------}{$\mathcal{P} \left( p_0 \in \left[ \hat{p} - \sigma^{\rm low}_{\hat{p}} , \hat{p} + \sigma^{\rm up}_{\hat{p}}  \right] \right)\, [\%]$}
 \includegraphics[width=9cm]{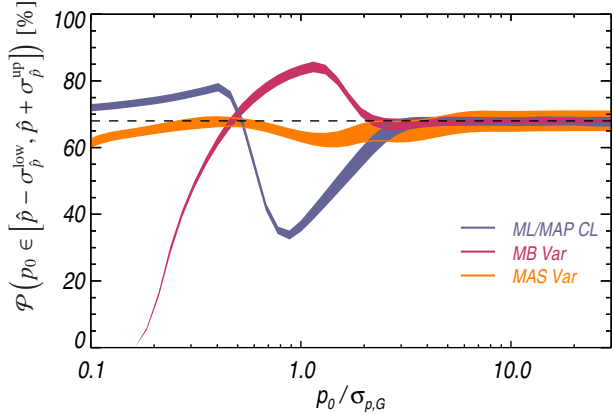} 
 \end{tabular}
 \caption{Probability of finding the true polarization angle $p_0$ inside the interval  $[\hat{p}-\sigma^{\rm low}_{\hat{p}} , \hat{p}+\sigma^{\rm up}_{\hat{p}}]$, where
 $\sigma^{\rm low}_{\hat{p}}$ and $\sigma^{\rm up}_{\hat{p}}$ are the lower and upper limits of each estimator: 
 credible intervals ML/MAP (blue), a posteriori variance MB (pink) and MAS variance (orange). It is  plotted as a function of the SNR $p_0/\sigma_{p,G}$.
Monte-Carlo simulations have been carried on in {\it low} regime of the covariance matrix.
The Gaussian level at 68\% is shown as a dashed line. }
 \label{fig:uncertainties_cl_mb_ml_mas}
\end{figure}

\begin{figure}
\begin{tabular}{c}
  \psfrag{-----------------ytitle-----------------}{$\mathcal{P} \left( p_0 \in \left[ \hat{p} - \sigma^{\rm low}_{\hat{p}} , \hat{p} + \sigma^{\rm up}_{\hat{p}}  \right] \right)\, [\%]$}
  \psfrag{xtitle}{$\hat{p} / \sigma_{\hat{p}}$}
 \includegraphics[width=9cm]{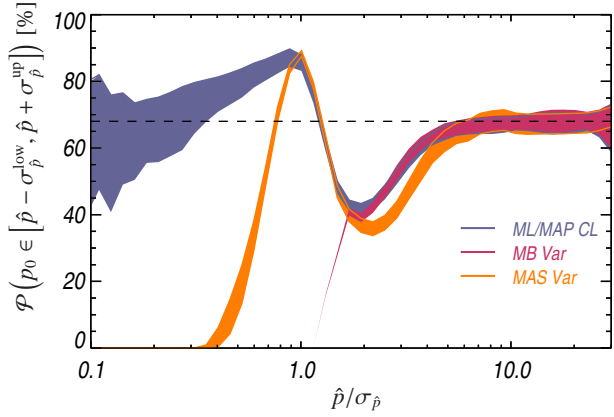} 
 \end{tabular}
 \caption{Same as Fig.~\ref{fig:uncertainties_cl_mb_ml_mas} but plotted as a function of the measured SNR $\hat{p}/\sigma_{\hat{p}}$. }
 \label{fig:uncertainties_cl_mb_ml_mas_measured_snr}
\end{figure}


\subsection{Polarization fraction uncertainty estimates}
\label{sec:uncertainty_comparison}

The questions of estimating the polarization uncertainties and how uncertainties are propagated are
essential in reliable polarization analysis. 
The best approach consists of building the confidence intervals to retrieve
robust estimates of the lower and upper limits of the 68, 95 or 99.5\% intervals, which is valid
even when the distribution is not Gaussian. As already mentioned in sect.~\ref{sec:confidence_intervals}, building optimized confidence intervals 
including the full knowledge of the covariance matrix may represent a challenge for large samples of data.
Hence P14 provides analytic approximations of such  confidence intervals for the MAS  estimator, 
which can be extremely useful.

A commonly used approach, however, is to provide the 1-$\sigma$ dispersion, 
assuming the Gaussian distribution of the $\hat{p}$ estimates as a  first approximation. 
We have already stressed the difference between the  risk  function  and the variance, and the limitations of the latter to derive robust uncertainties in the presence of bias.
We compare below the performance of the usual uncertainty estimates introduced in Sect.~\ref{sec:uncertainties} to provide robust 
68\% tolerance intervals: MAS variance, credible intervals MAP and 1-$\sigma$ a posteriori dispersion MB.

Starting with a true $p_0$ value, we have performed Monte-Carlo simulations in the {\it low} regime of the covariance matrix, 
by exploring the whole range of the true polarization angle $\psi_0$, with an SNR spanning from 0 to 30. 
For each simulated measurement ($p$,$\psi$), we compute the $\hat{p}$ estimates with their uncertainty 
estimators $\sigma_{\hat{p}}$.  We then compute the a posteriori probability
to find the true $p_0$ inside the interval $[\hat{p}-\sigma^{\rm low}_{\hat{p}} , \hat{p}+\sigma^{\rm up}_{\hat{p}}]$. 
In the case of the MAP estimator,  the lower and upper limits of the interval, $\hat{p}_{\text{MAP}}- \sigma^{\rm low}_{\hat{p}_{\text{MAP}}}$
and $\hat{p}_{\text{MAP}} + \sigma^{\rm up}_{\hat{p}_{\text{MAP}}}$, are set to $p^{\rm low}_{\text{MAP}}$ 
and $p^{\rm up}_{\text{MAP}}$, respectively, (with $\lambda$=68 as defined in Sect.~\ref{sec:credible_intervals}), which can be asymmetric.
We report the results compared to the expected 68\% level in Fig.~\ref{fig:uncertainties_cl_mb_ml_mas}.
We recall that this comparison approach is frequentist again, while anything derived from the Bayesian pdf
is used to build single estimates and to be compared with the confidence intervals.

As pointed  out in Sect.~\ref{sec:variance_risk}, the theoretical variance associated with the MAS estimator
still tends to provide slightly lower probabilities than the expected 68\% at low SNR, mainly due to the asymmetry of the distribution. 
The variance  associated to the MB estimator, which is more biased at low SNR, gives extremely low probability to recover the true
$p_0$ value at low SNR ($<0.5$). By contrast, it provides probabilities greater than 68\% (as high as 90\%) for SNR between 0.5 and 2.
This comes from the fact that the MB variance statistically over-estimates by a 
factor of 2 the exact variance of the a posteriori $\hat{p}_{\text{MB}}$ distribution at low SNR ($<$2). 
Thus the MB uncertainty estimator yields conservative estimates of the uncertainty for SNR $>$0.5.
At high SNR ($>$3) all these uncertainty estimators provide compatible estimates of the probability close to 68\%.

\begin{figure}
\begin{tabular}{c}
  \psfrag{ytitle}{$\hat{p} / \sigma_{\hat{p}}$}
  \psfrag{xtitle}{$p_0 / \sigma_{p,0}$}
 \includegraphics[width=9cm]{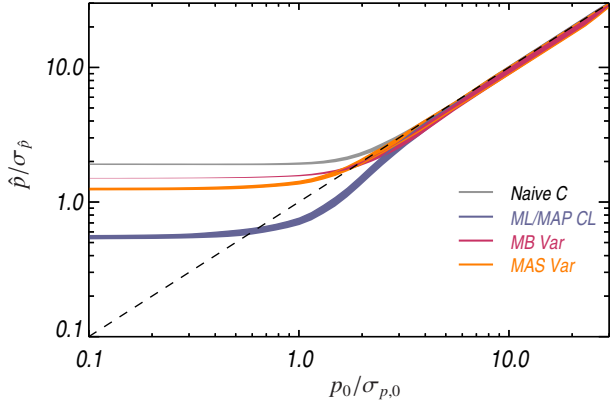} 
 \end{tabular}
 \caption{Average Measured SNR computed over 10\,000 Monte-Carlo simualtions
 as a function of the true SNR for four methods : Naive $\hat{p}/\sigma_{p,C}$ (dark), 
 MAP confidence intervals $\hat{p}_{\text{ML}}/\sigma_{\hat{p},\text{MAP}}$ (blue), 
 MB  $\hat{p}_{\text{MB}} / \sigma_{\hat{p},\text{MB}}$ (pink) and 
 MAS variance $\hat{p}_{\text{MAS}} / \sigma_{\hat{p},\text{MAS}}$(orange).
 The covariance matrix is taken in its {\it low} regime.}
 \label{fig:uncertainties_snr}
\end{figure}

Because the true SNR is always unknown (see Sect.~\ref{sec:snr}), the probability to 
find the true $p_0$ value in the confidence interval is also shown as a function of the measured SNR 
in Fig.~\ref{fig:uncertainties_cl_mb_ml_mas_measured_snr}. 
This much more realistic picture shows that the variance estimates provide reliable probability for measured SNR larger than $\sim$6. 

\subsection{Polarization signal-to-noise ratio}
\label{sec:snr}

In any real measurement, the true SNR $p_0/\sigma_{p,G}$ remains unknown.
From observations, we only have access to the measured SNR, which can be obtained by 
the ratio $\hat{p} / \sigma_{\hat{p}}$ associated with each estimator, or by a confidence interval approach (see P14), 
which is much more robust at a low true SNR. We show in Fig.~\ref{fig:uncertainties_snr} the accuracy of the 
measured SNR compared to the true SNR for the four following methods: the na\"ive estimate plus Classical estimate of the uncertainty, 
the MAS estimate with the associated variance, the MB estimate and its variance, and the ML estimate with the MAP credible  intervals.
We observe that all methods agree only for a true SNR larger than 3, giving back the true SNR in this regime. 
Below this true SNR, the measured SNR becomes extremely biased whatever the method used, due to the bias of the measurement $\hat{p}$
itself, but also due to the bias introduced by the variance as an estimate of the uncertainty when the output distribution departs from the Gaussian regime.

\begin{figure*}[]
  \begin{tabular}{lcc}
\begin{minipage}[c]{.06\linewidth}
 \begin{tabular}{l}
$ p_0/\sigma_{p,G}=0.5$ 
\end{tabular} 
\end{minipage} &
\begin{minipage}[c]{.4\linewidth}   \includegraphics[width=1.2\textwidth]{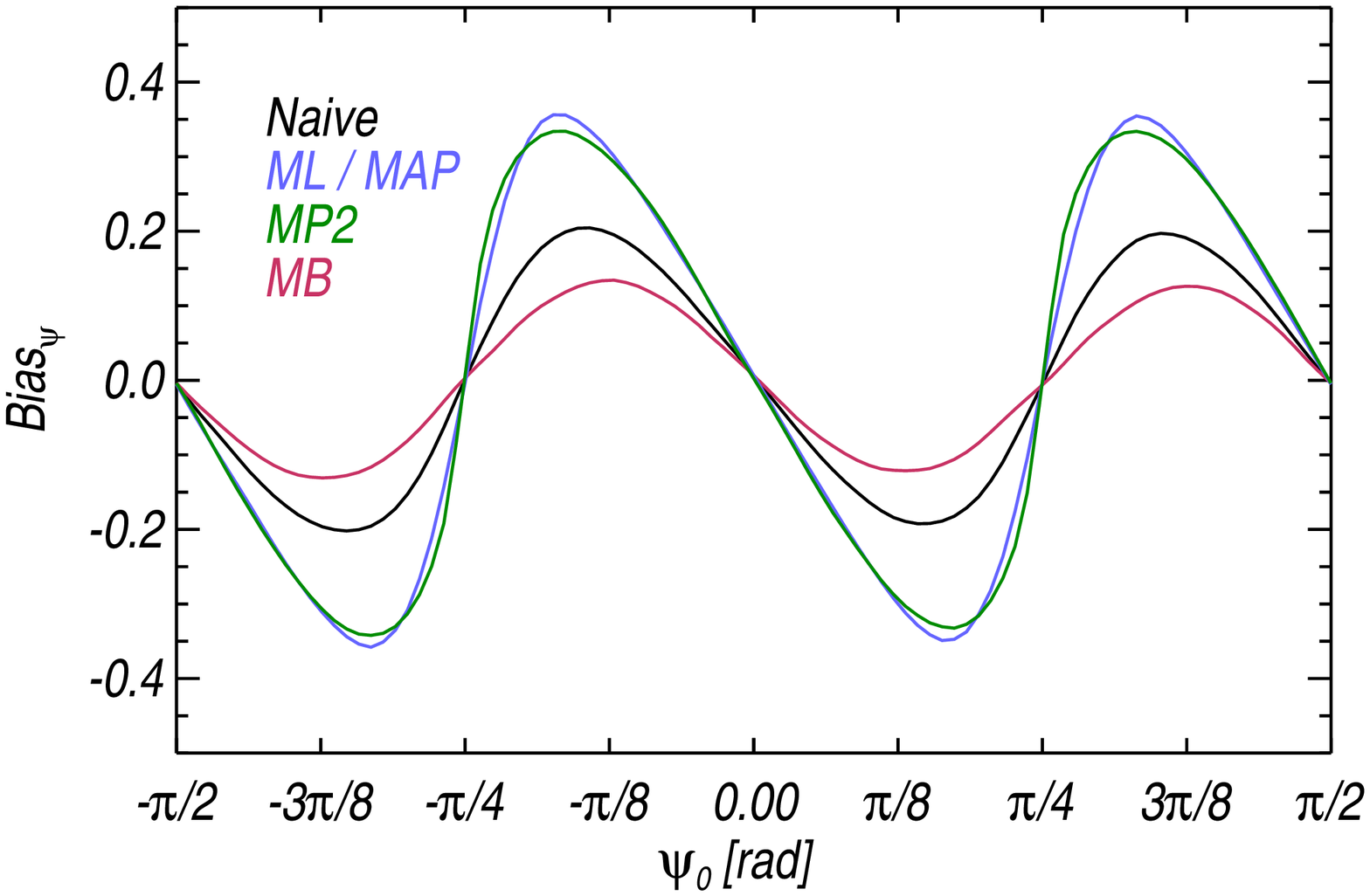} \end{minipage}&
\begin{minipage}[c]{.4\linewidth}   \includegraphics[width=1.2\textwidth]{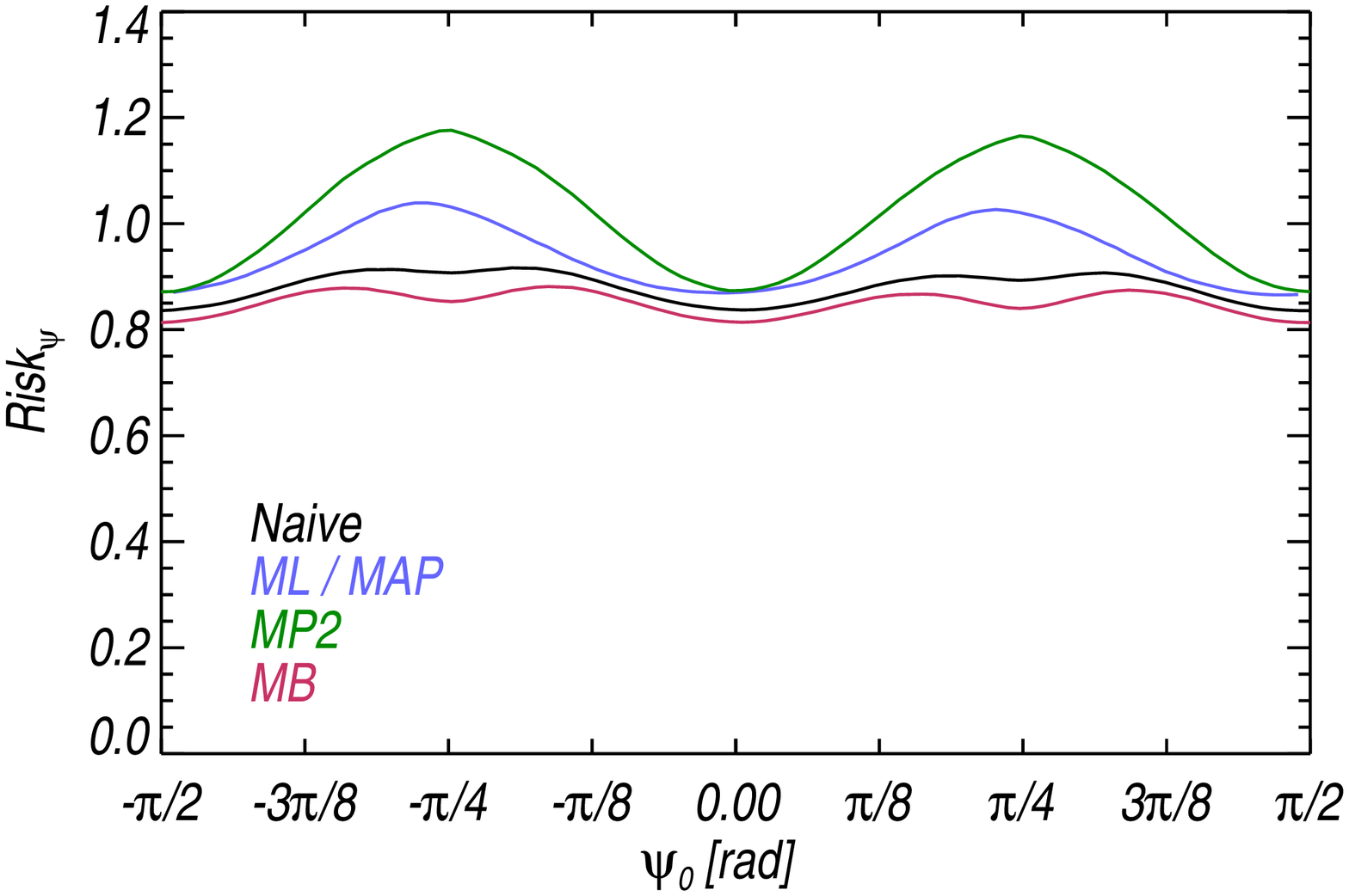} \end{minipage}\\

\begin{minipage}[c]{.06\linewidth}
 \begin{tabular}{l}
$ p_0/\sigma_{p,G}=1$ 
\end{tabular} 
\end{minipage} &
\begin{minipage}[c]{.4\linewidth}   \includegraphics[width=1.2\textwidth]{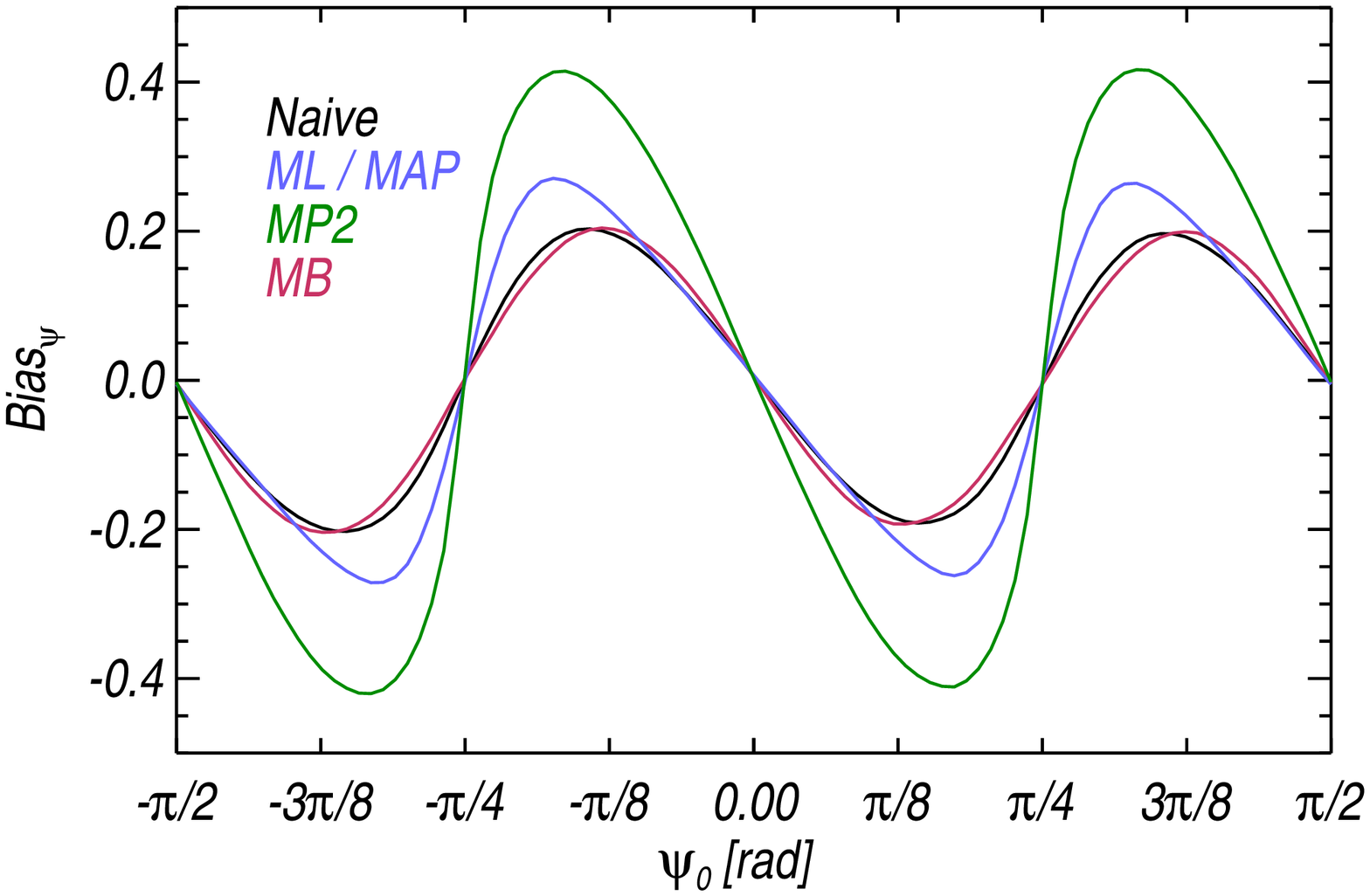} \end{minipage}&
\begin{minipage}[c]{.4\linewidth}   \includegraphics[width=1.2\textwidth]{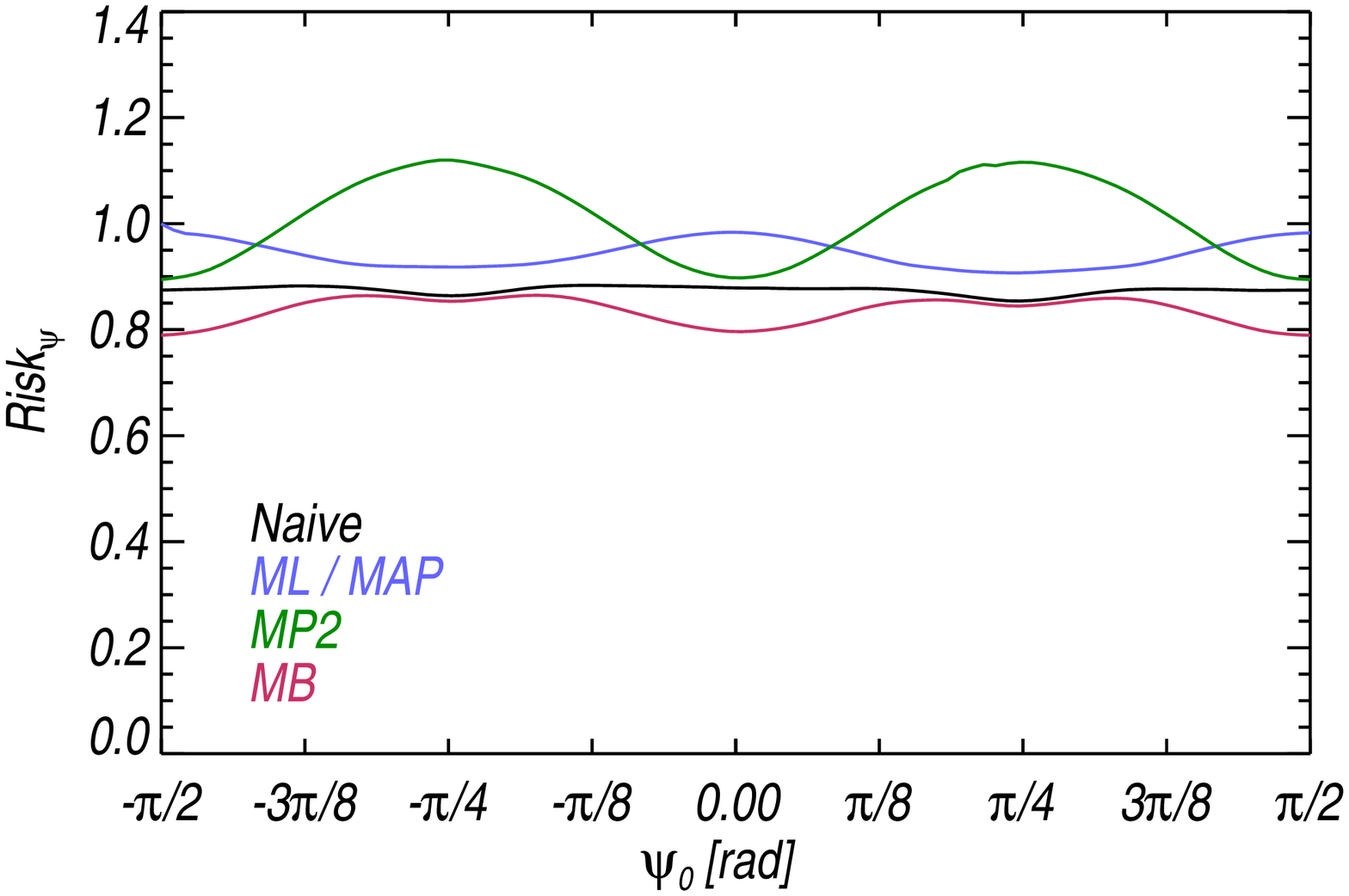}\end{minipage} \\
  
\begin{minipage}[c]{.06\linewidth}
 \begin{tabular}{l}
$p_0/\sigma_{p,G}=2$ 
\end{tabular} 
\end{minipage} &
\begin{minipage}[c]{.4\linewidth}   \includegraphics[width=1.2\textwidth]{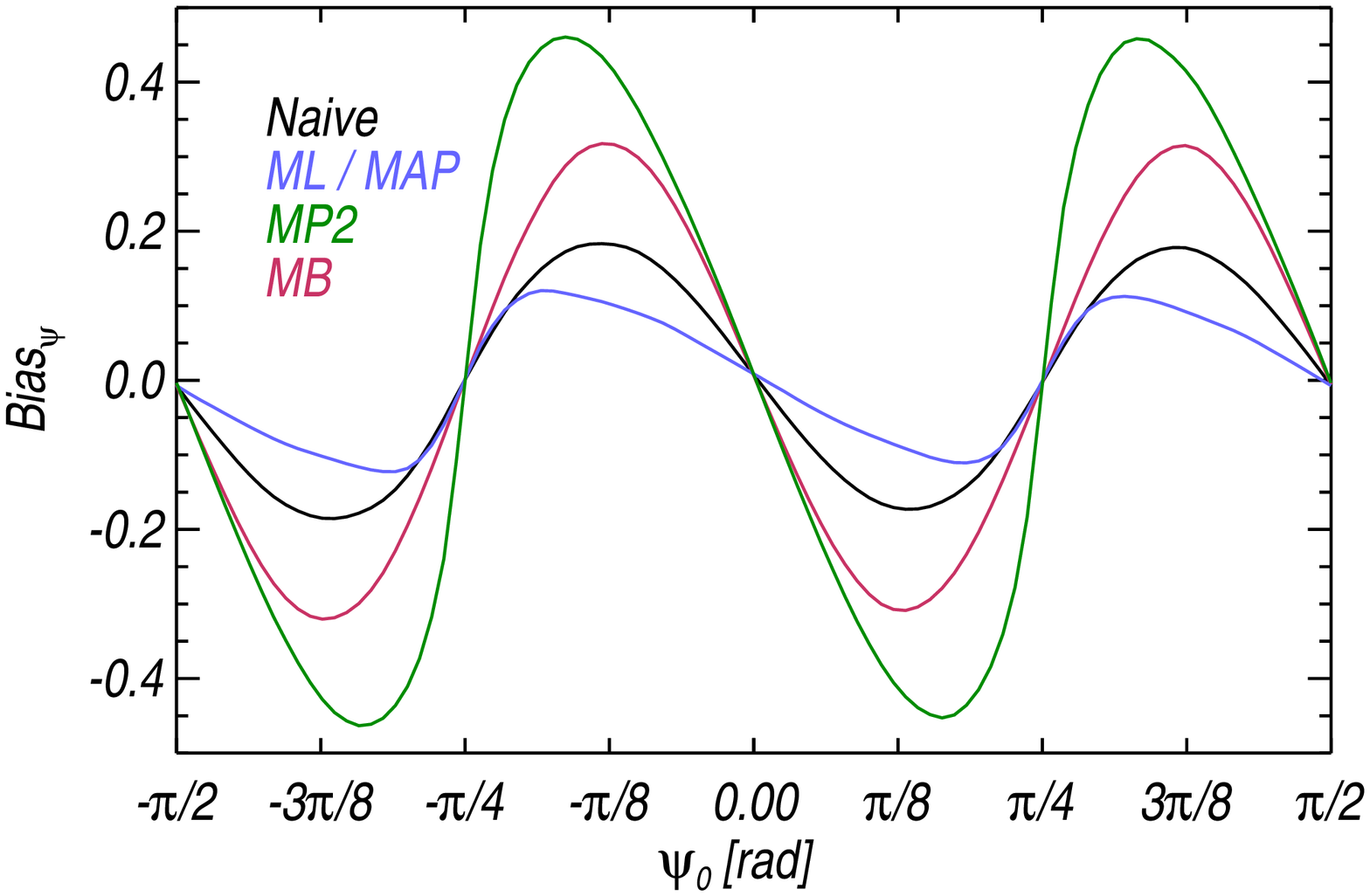}\end{minipage} &
\begin{minipage}[c]{.4\linewidth}   \includegraphics[width=1.2\textwidth]{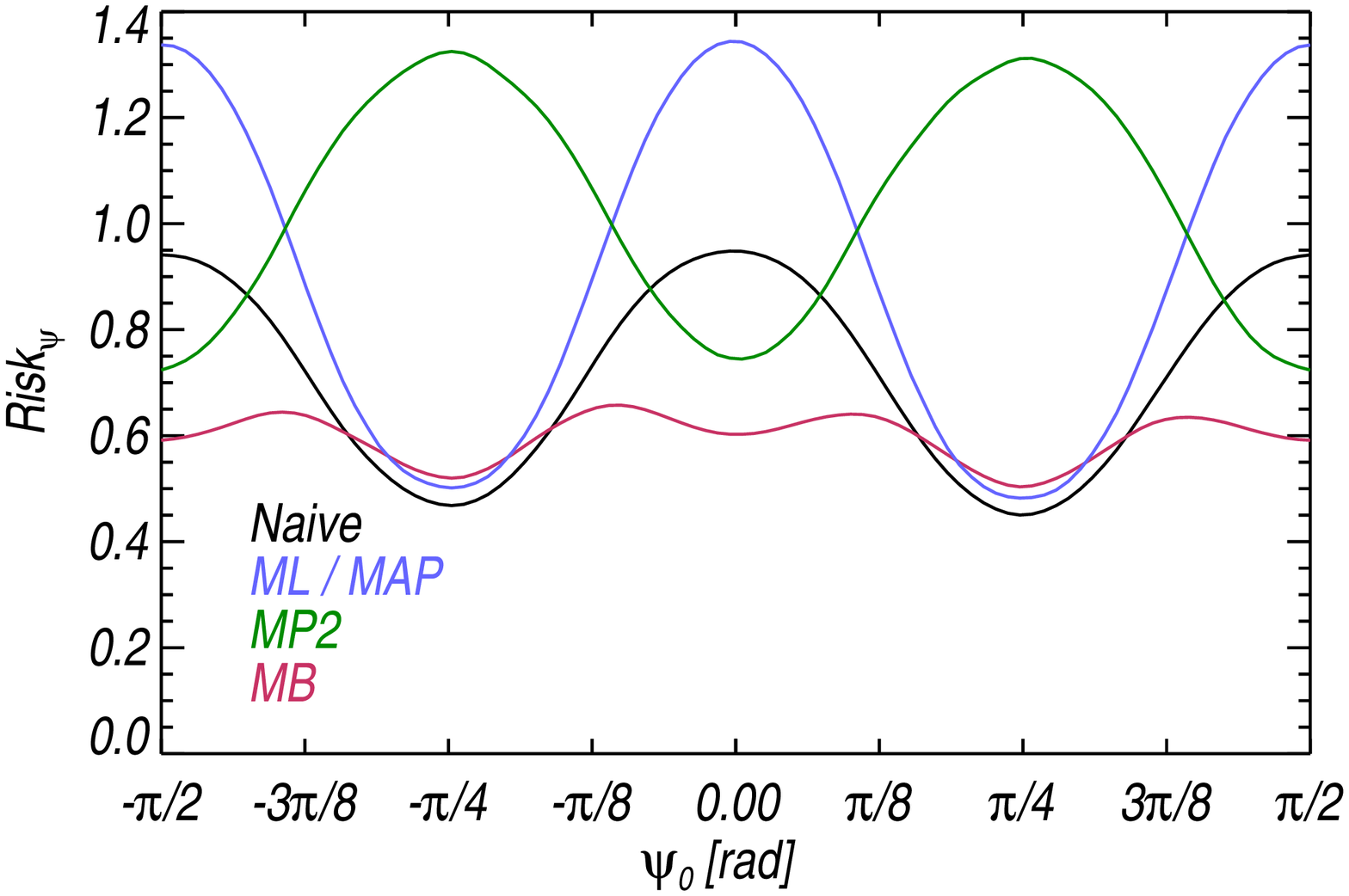} \end{minipage}\\
  
\begin{minipage}[c]{.06\linewidth}
 \begin{tabular}{l}
$ p_0/\sigma_{p,G}=5$ 
\end{tabular} 
\end{minipage} &
\begin{minipage}[c]{.4\linewidth}   \includegraphics[width=1.2\textwidth]{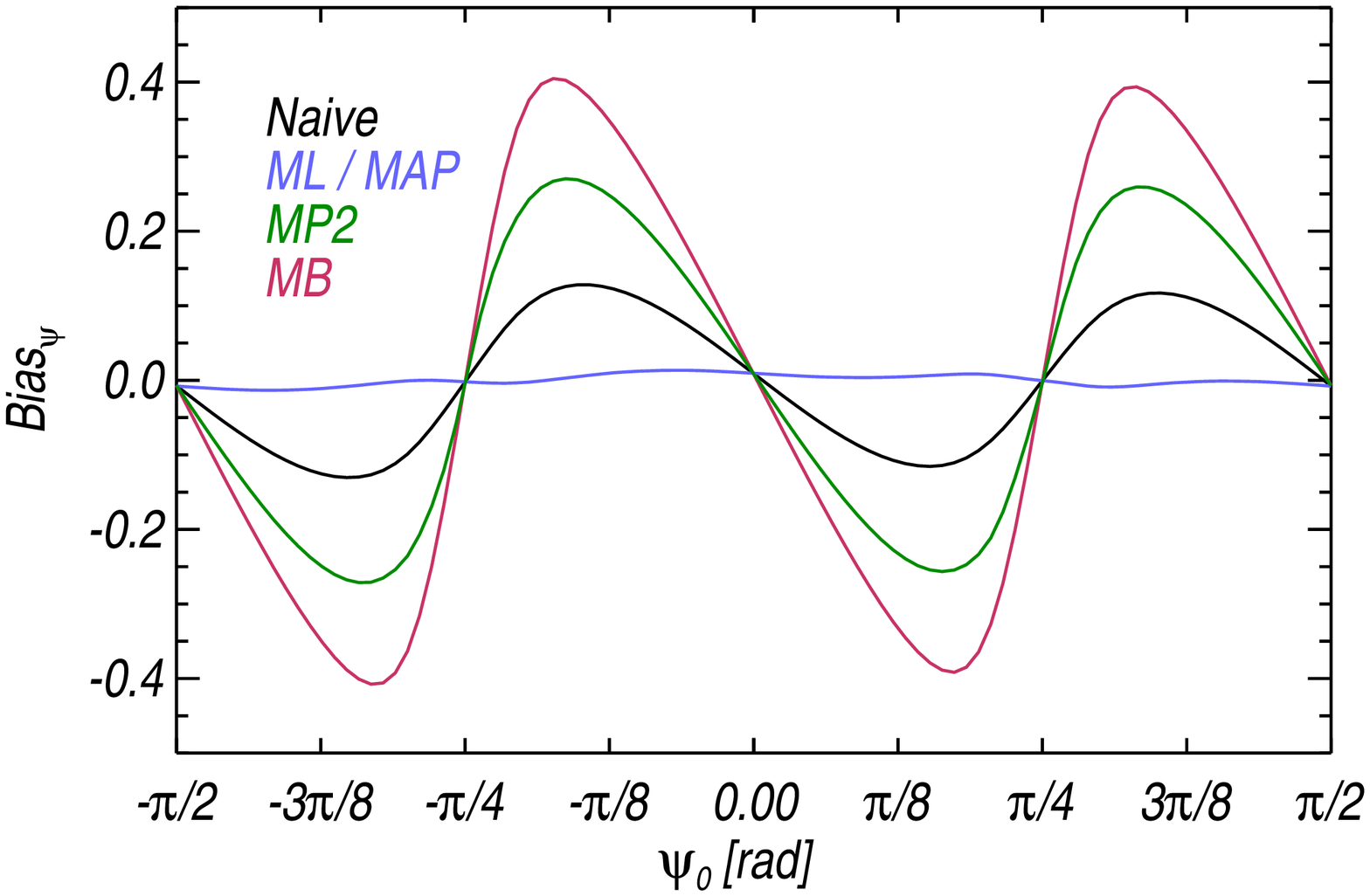} \end{minipage}&
\begin{minipage}[c]{.4\linewidth}   \includegraphics[width=1.2\textwidth]{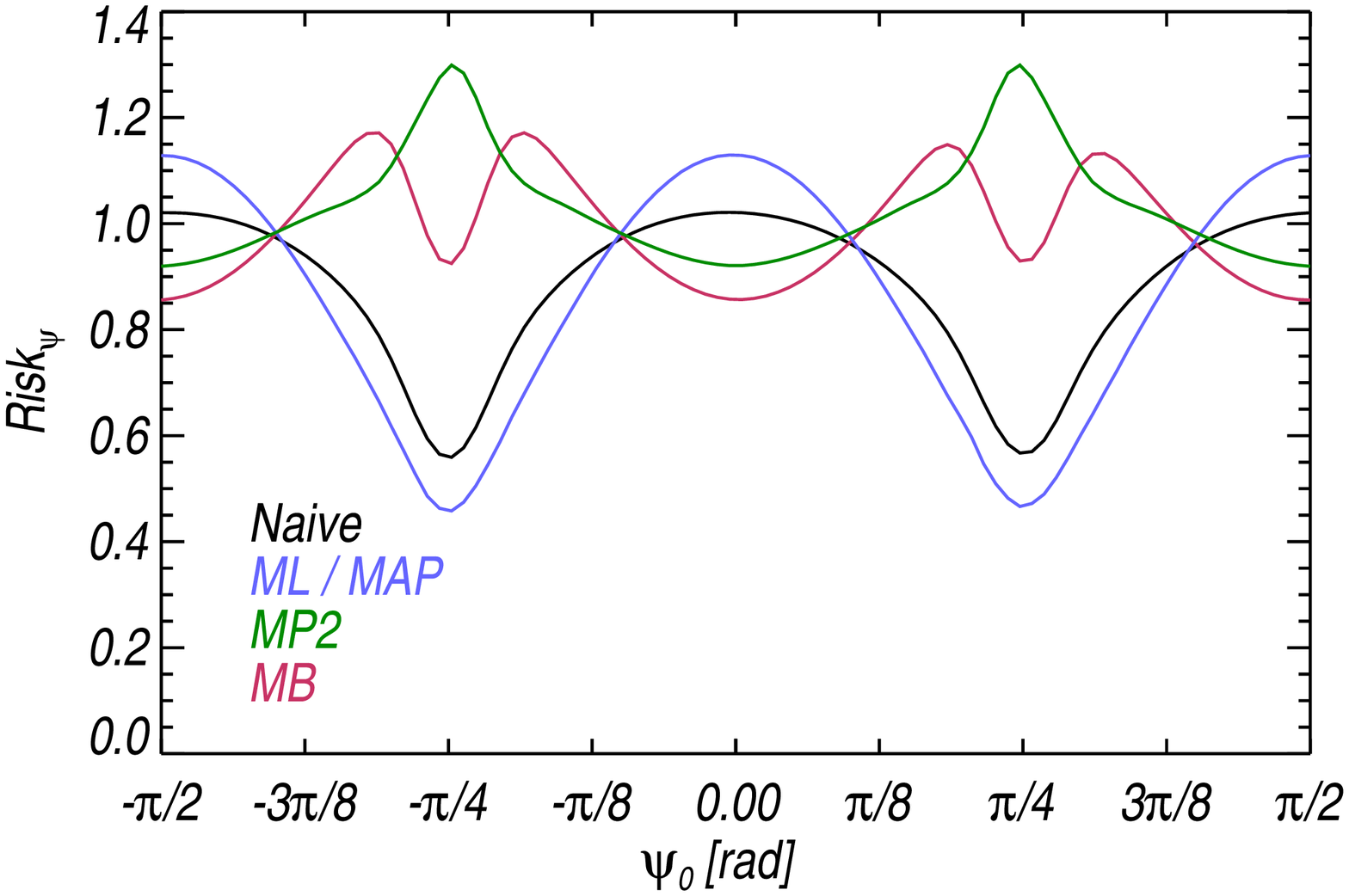} \end{minipage}\\
  
\end{tabular}
\caption{Comparison of the relative $\mathrm{Bias}_{\psi}$ (left) and $\mathrm{Risk}_{\psi}$ (right) quantities of the four $\hat{\psi}$ estimators: 
Naive (black), ML (blue), MP2 (green) and MB (pink) plotted as a function of the true  polarization angle $\psi_0$ 
and computed at four SNR $p_0/\sigma_{p,G}$=0.5, 1, 2 and 5. 
The covariance matrix is set to $\varepsilon=2$ and $\rho=0$ ($\varepsilon_\mathrm{eff}=2$).}
\label{fig:estimator_comparison_psi}
\end{figure*}

\begin{figure*}
\begin{tabular}{cc}
 \includegraphics[width=9cm]{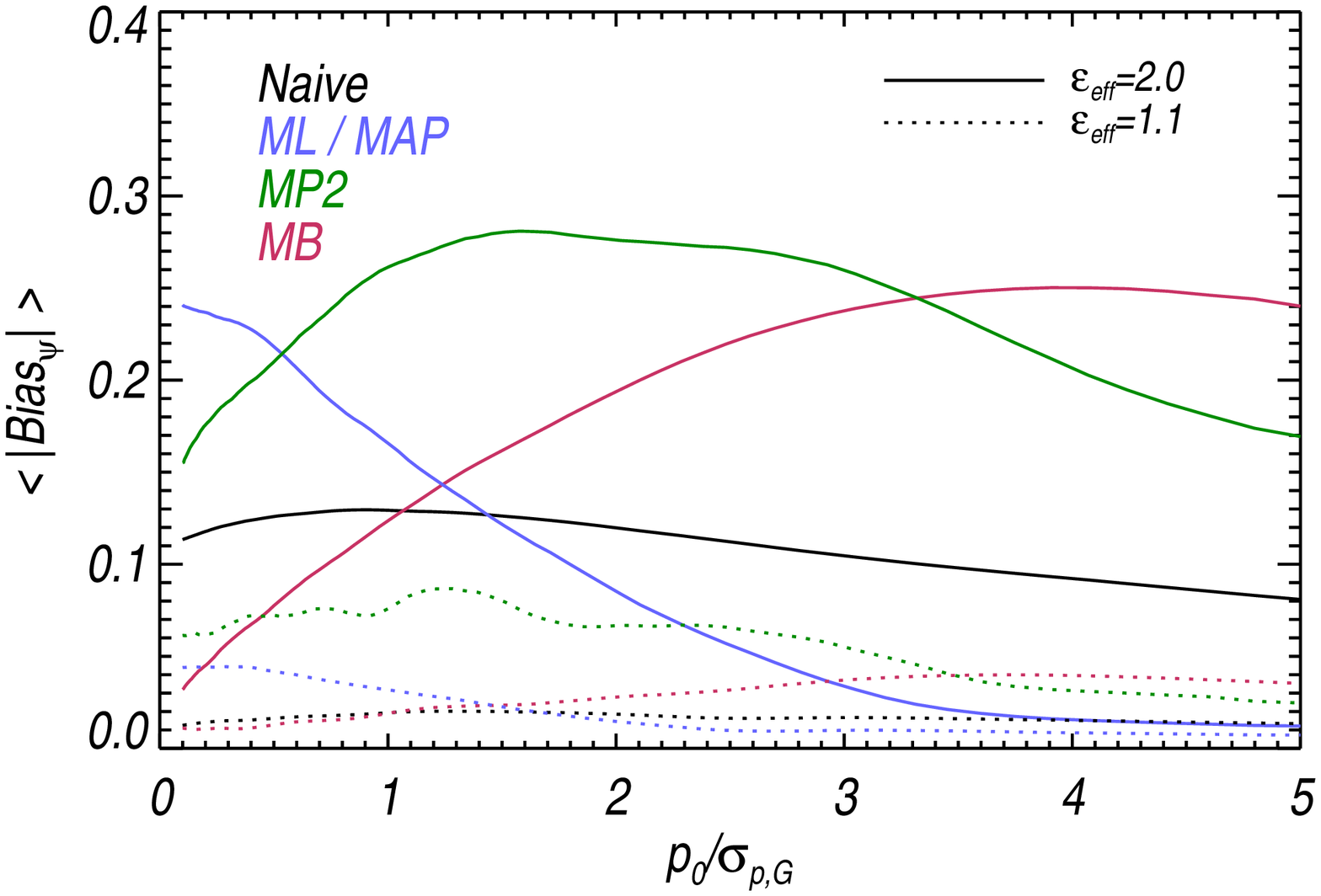}  & 
 \includegraphics[width=9cm]{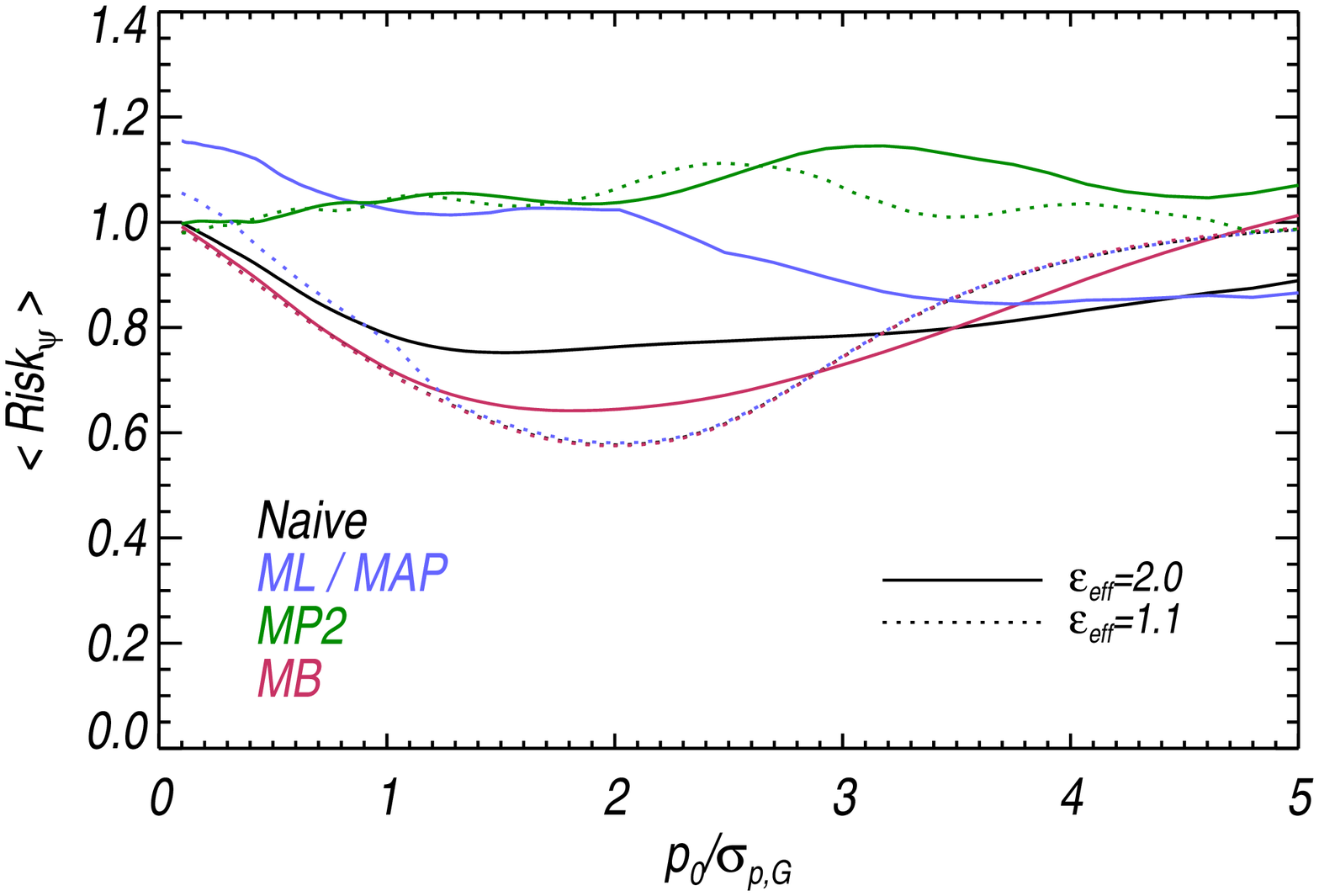}  
 \end{tabular}
 \caption{Statistical relative $\left|\mathrm{Bias}_{\psi}\right|$ (left panel) and $\mathrm{Risk}_{\psi}$ (right panel) averaged over $\psi_0$ between $-\pi/2$ and $\pi/2$, 
 as a function of the SNR on $p_0/\sigma_{p,G}$,  for the four $\hat{\psi}$ estimators:
 na\"ive (black), ML / MAP (blue), MP2 (green) and MB (pink). We consider two setups of the covariance matrix here: $\varepsilon_\mathrm{eff}$=2 (solid line) and 
 and $\varepsilon_\mathrm{eff}$=1.1 (dotted line).}
 \label{fig:psi_mean_bias_risk}
\end{figure*}

\section{$\hat{\psi}$ estimator performance}
\label{sec:comparison_estimators_psi}

\subsection{Methodology}

As pointed out by PMA I, once the covariance matrix is not canonical ($\varepsilon_\mathrm{eft}>1$), a bias of the 
polarization angle measurements $\psi$ appears with respect to the true polarization angle $\psi_0$. This bias may be positive or negative.
We propose to compare the accuracy at correcting the bias  of the polarization angle of the four following $\hat{\psi}$ estimators:
na\"ive measurements $\psi$, the ML $\hat{\psi}_{\text{ML}}$ 
(which is equivalent to the MAP $\hat{\psi}_{\text{MAP}}$), the MP2 $\hat{\psi}_{\text{MP2}}$ and
the MB $\hat{\psi}_{\text{MB}}$.

Similarly to the $\hat{p}$ estimators, we define the  relative  bias and  risk  function on $\hat{\psi}$ as follows:
\begin{equation}
\mathrm{Bias}_{\psi} \equiv \frac{\left< \hat{\psi} - \psi_0  \right>}{\sigma_{\psi,0}}  \quad \mathrm{and} 
\quad  \mathrm{Risk}_{\psi} \equiv  \frac{\left< (\hat{\psi} - \psi_0)^2  \right> }{\sigma_{\psi,0}^2} \, ,
\end{equation}
where $\hat{\psi}$ is the polarization angle estimate computed on the simulated measurements $\psi$, 
 $\psi_0$ is the true polarization fraction and angle, $< >$ denotes the average computed over the simulated sample, and
 $\sigma_{\psi,0}$ is the standard deviation of the  simulated measurements.

\subsection{Performance Comparison}
\label{sec:efficiency_comparison_psi}

We explore the performance of the four $\hat{\psi}$ estimators at four SNR=0.5, 1, 2 and 5 (from top to bottom) 
and a covariance matrix with an effective ellipticity $\varepsilon_\mathrm{eff}$=2, on Fig.~\ref{fig:estimator_comparison_psi}.
The relative $\mathrm{Bias}_{\psi}$ (left panels) and $\mathrm{Risk}_{\psi}$ (right panels) are plotted as a function of the true polarization angle $\psi_0$.
While the MB (pink) estimator seems to provide the least biased estimates with the lowest risk  function at low SNR ($<$1),
it becomes the least efficient at higher SNR. On the contrary, the ML
 (or MAP too) presents poor performances at low SNR, but
provides impressive results at high SNR, reducing the relative bias close to zero at a SNR of 5.
The MP2 estimator does not present any satisfactory properties: strong relative bias and risk  function  in almost all cases. Hence
this $\hat{\psi}_{\text{MP2}}$ estimator can be ruled out.

An overview of the performance of the four $\hat{\psi}$ estimators as a function of the SNR is shown 
on Fig.~\ref{fig:psi_mean_bias_risk}, after marginalization over all the possible values of the $\psi_0$ parameter. As
the  relative $\mathrm{Bias}_{\psi}$ can be positive or negative depending on $\psi_0$, we compute the average of the absolute value of the relative bias, 
$< |\mathrm{Bias}_{\psi} | >$ as an indicator of the statistical performance of the estimators whatever the true polarization angle is.
We observe again on the left panel of Fig.~\ref{fig:psi_mean_bias_risk} that the MB (pink) estimator provides the lowest  relative  bias for SNR$<$1.2, while the
ML is especially powerful for SNR$>$2. All estimators provide almost the same results for the average $\mathrm{Risk}_{\psi}$ (left panel), 
even if MB appears slightly better than the others, including the na\"ive measurements.

The examples provided above have been computed with an {\it extreme} effective ellipticity ($\varepsilon_\mathrm{eff}$=2) to emphasize the observations, 
but the same conclusions can be reached for lower values of the ellipticity. See, for example, the case with $\varepsilon_\mathrm{eff}$=1.1 
shown in dotted line in Fig.~\ref{fig:psi_mean_bias_risk}. 
In the {\it low} regime of the covariance matrix, however, the statistical  relative
bias on $\psi$ is very small, typically smaller than 5\% of the dispersion, so that the need to correct the bias on $\psi$ remains extremely limited.

\subsection{Polarization angle uncertainty estimates}
\label{sec:uncertainty_psi_comparison}

Once a reliable estimate of $\hat{\psi}$ based on the MB and ML (MAP) estimators has been obtained, 
we would like to build a robust estimate of the associated uncertainties $\sigma_{\hat{\psi}}$, 
which should be done by building confidence intervals. Because this last step could represent important 
efforts in some cases, for example when dealing with the full covariance matrix,  
we detail other methods below. 

One option is to use the uncertainty associated with the MB estimator, $\sigma_{\hat{\psi},\text{MB}}$ (see Eq.~\ref{eq:sigma_psi_mb}). 
Another is to use the 
credible intervals built around the MAP estimates on the posterior pdf.
We can keep the lower and upper limits, $\psi^{\rm low}_{\text{MAP}}$ and  $\psi^{\rm up}_{\text{MAP}}$ 
computed for a 68\% credible interval, 
or build a symmetrized uncertainty:
\begin{equation}
\sigma_{\hat{\psi},{MAP}} = \frac{1}{2} \left( \psi^{\rm up}_{\text{MAP}} - \psi^{\rm low}_{\text{MAP}} \right) \, .
\end{equation}
A third option consists in taking the classical uncertainty given in PMA I, derived from the derivatives of the polarization parameters.
PMA I has already shown that this $\hat{\psi}$ uncertainty estimator, associated with the na\"ive measurements, tends to systematically
underestimate the true dispersion of the $\psi$ distribution. 

\begin{figure}
 \includegraphics[width=9cm]{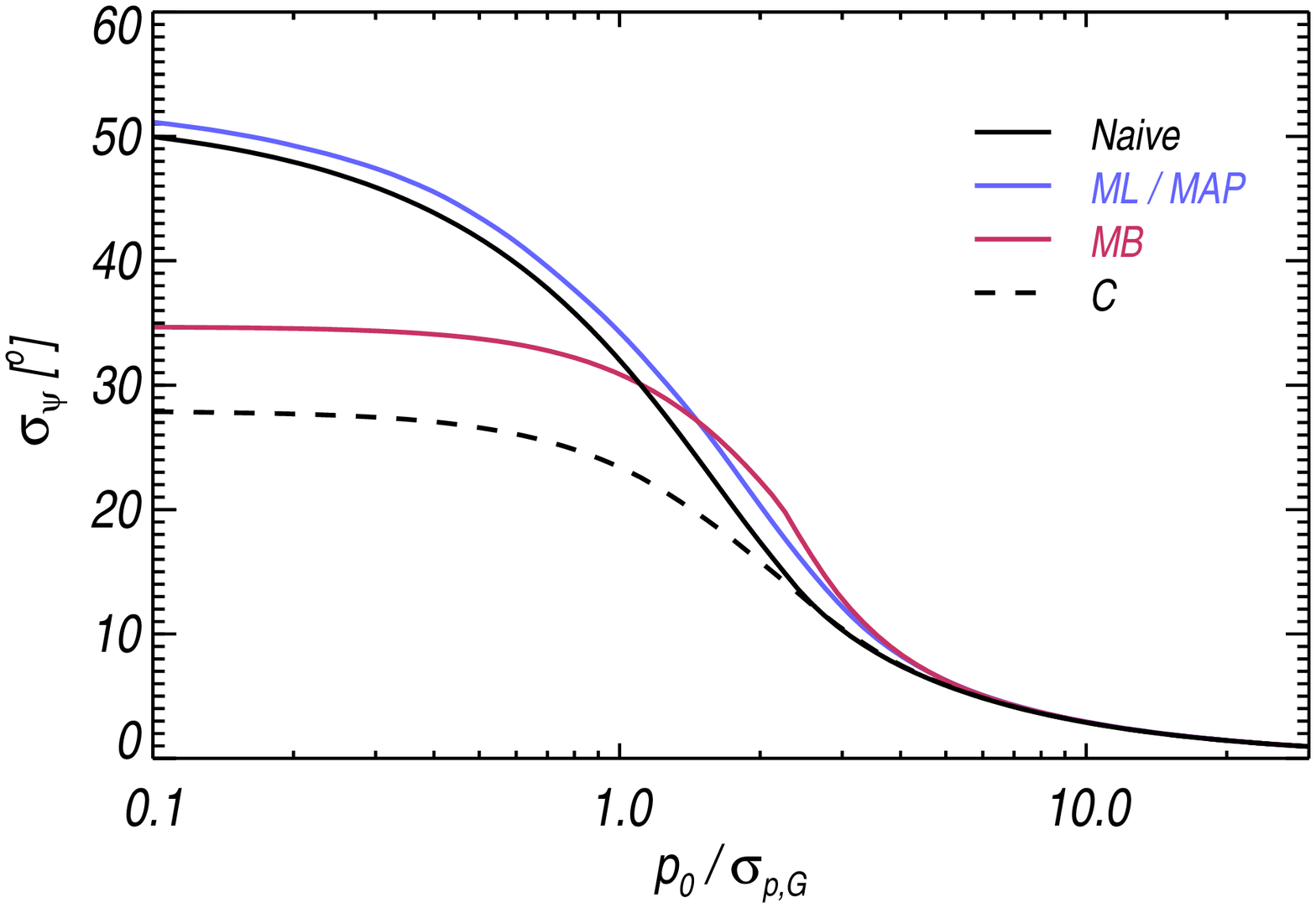}
 \caption{Average polarization angle uncertainty as a function of the SNR in the canonical case: true uncertainty $\sigma_{\psi,0}$ (black), 
 Classical estimate $\sigma_{\psi,C}$ (C, dashed dark), ML $\sigma_{\hat{\psi},\text{MAP}}$ (blue) 
 and MB  $\sigma_{\hat{\psi},\text{MB}}$ (pink) estimators.
 The covariance matrix is assumed to be canonical.}
 \label{fig:uncertainties_sigphi_all}
\end{figure}

We first assume the canonical simplification of the covariance matrix, which implies that the $\psi$ measurements are not statistically biased.
We also recall  that under such assumptions the ML (MAP) and MB $\hat{\psi}$ estimators will give back the measurements $\psi$. 
We study, however, how the uncertainties associated with these two estimators can be used to get a reliable estimate of the uncertainty $\sigma_{\hat{\psi}}$.
Starting from a true ($p_0$, $\psi_0$), we simulate a sample of 50~000 simulated
 measurements $p$, $\psi$ at  a given SNR $p_0/\sigma_{p}$, on which we apply the two 
 ML (MAP) and MB $\hat{\psi}$ estimators and their associated uncertainty $\sigma_{\hat{\psi},\text{MAP}}$ and $\sigma_{\hat{\psi},\text{MB}}$, respectively.
 From this simulated set we can derive the averaged $\sigma_{\hat{\psi}}$ for both methods. 
 Because all estimators give back the measurements in the canonical case, we compare 
 the MAP (blue) and MB (pink) polarization angle uncertainties 
 estimators directly to the true dispersion (black) of the $\psi$ measurements in Fig.~\ref{fig:uncertainties_sigphi_all}. 
 We also repeat the average of the classical estimates (dashed line) of the polarization uncertainty estimate, which has been shown by  PMA I (see their Fig.~7) to underestimate by a factor of two the true uncertainty at low SNR ($<$2).
We observe that the MAP estimator $\sigma_{\hat{\psi},\text{MAP}}$ provides an extremely good estimate of the polarization angle 
uncertainty compared to the true one over the whole range of SNR, even if slightly conservative up to a SNR of 5.
The MB estimator $\sigma_{\hat{\psi},\text{MB}}$ provides consistent estimates of the uncertainty 
from intermediate SNR$\sim$1, but still underestimates at lower SNR ($<$1).

\begin{figure}
\begin{tabular}{c}
  \psfrag{-----------------ytitle-----------------}{$\mathcal{P} \left( \psi_0 \in \left[ \hat{\psi} - \sigma^{\rm low}_{\hat{\psi}} , \hat{\psi} + \sigma^{\rm up}_{\hat{\psi}}  \right] \right)\, [\%]$}
 \includegraphics[width=9cm]{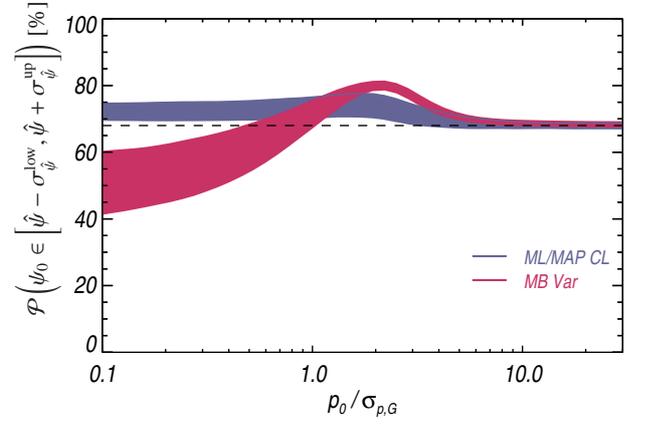} 
 \end{tabular}
 \caption{Probability to find the true polarization angle $\psi_0$ inside the interval  $[\hat{\psi}-\sigma^{\rm low}_{\hat{\psi}} , \hat{\psi}+\sigma^{\rm up}_{\hat{\psi}}]$, where
 $\sigma^{\rm low}_{\hat{\psi}}$ and $\sigma^{\rm up}_{\hat{\psi}}$ are the lower and upper uncertainties for each estimator, ML/MAP (blue) and MB (pink), and
 plotted as a function of the SNR $p_0/\sigma_{p,G}$.
Monte-Carlo simulations have been carried out in the {\it low} regime of the covariance matrix.
The expected level at 68\% is shown as a dashed line. }
 \label{fig:uncertainties_psi_cl_mb_ml}
\end{figure}

\begin{figure}
\begin{tabular}{c}
  \psfrag{-----------------ytitle-----------------}{$\mathcal{P} \left( \psi_0 \in \left[ \hat{\psi} - \sigma^{\rm low}_{\hat{\psi}} , \hat{\psi} + \sigma^{\rm up}_{\hat{\psi}}  \right] \right)\, [\%]$}
  \psfrag{xtitle}{$\hat{p} / \sigma_{\hat{p}}$}
 \includegraphics[width=9cm]{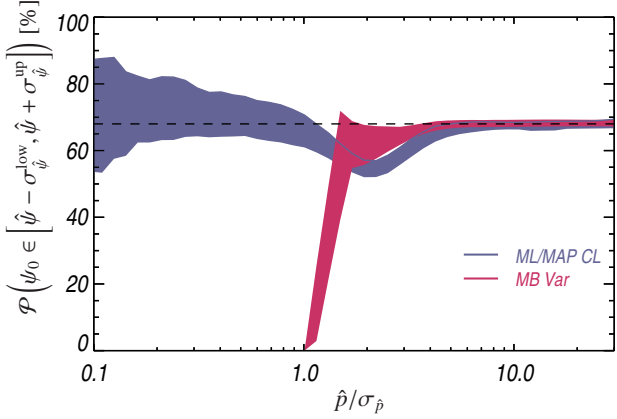} 
 \end{tabular}
 \caption{Same as Fig.~\ref{fig:uncertainties_psi_cl_mb_ml}, but plotted as a function of the measured SNR $\hat{p}/\sigma_{\hat{p}}$.}
 \label{fig:uncertainties_psi_cl_mb_ml_measured_snr}
\end{figure}

In the non-canonical case a statistical bias on $\psi$ appears, which can be 
partially corrected using the appropriate $\hat{\psi}$ estimators (see Sect.~\ref{sec:efficiency_comparison_psi}), leading
to an output distribution of the $\hat{\psi}$ estimates. We quantify the performance of the $\psi$ uncertainty estimators via Monte-Carlo simulations, as done for the $\hat{p}$ uncertainties. Starting from a set of polarization parameters 
($p_0$=0.1, -$\pi$/2$<$$\psi_0$$<$$\pi$/2), 
we build a sample of simulated measurements ($p$, $\psi$) using various
setups of the covariance matrix in the {\it low} regime, and various SNRs ranging from 0 to 30.
We then compute the a posteriori probability to find the true polarization angle $\psi_0$ in the interval 
 $[\hat{\psi}-\sigma^{\rm low}_{\hat{\psi}} , \hat{\psi}+\sigma^{\rm up}_{\hat{\psi}}]$,  where $\sigma^{\rm low}_{\hat{\psi}}$
  and $\sigma^{\rm up}_{\hat{\psi}}$ are symmetrized.  The results are shown as a function of the 
 true SNR $p_0/\sigma_{p,G}$ in Fig.~\ref{fig:uncertainties_psi_cl_mb_ml}  and of the measured SNR
 $\hat{p}/\sigma_{\hat{p}}$ in Fig.~\ref{fig:uncertainties_psi_cl_mb_ml_measured_snr}. 
 We observe that the MAP estimator provides slightly conservative probabilities over the whole range of SNR. 
The MB estimator gives low probabilities to recover the true polarization angle $\psi_0$ for a true SNR $<$1,  and a measured SNR$<$2.

\section{Three-dimensional case}
\label{sec:3Dcase}

In all of the preceding sections, the total intensity $I$ was assumed to be perfectly known, $I=I_0$. 
in some cases, however, this assumption is not valid as discussed by PMA1. 
For instance, one needs to subtract from the observed intensity signal any unpolarized 
component, leading to three main issues:  i) the derived polarization fraction may be grossly underestimated if this is not done properly, 
ii) this subtraction may be subject to a relatively large uncertainty, larger than the noise on the total intensity, and could lead to diverging estimates of the polarization fraction when intensity crosses null values ; iii) this uncertainty on this unpolarized component intensity level 
should be included in the 3D noise covariance matrix, and propagated to the uncertainty estimates of the polarization fraction.
This happens for instance when dealing with the polarization fraction of the Galactic dust component  at high latitude, 
where the total intensity of the signal is strongly contaminated by the unpolarized signal of the Cosmic Infrared Background (CIB).

The Bayesian approach has the definite advantage over other estimators discussed here
 in that it can deal fairly easily with three-dimensional $(I,Q,U)$ noise. However, 
 an uncertain total intensity still poses problems, which are most acute in low brightness regions, 
 since the noisy $I$ may become null or negative, leading to infinite or negative polarization fractions. 
 With this in mind, it is possible that the choice of the prior in $p_0$ and $I_0$ may have a strong impact 
 on the $\hat{p}_\mathrm{MB}$ estimate. One may for instance choose to allow for negative $I_0$ in low-brightness regions, 
 which implies extending the definition range of the polarization fraction to the negative part, leading to a prior defined on [-1,1].
 Another possibility in this case, and possible development of the present paper, is to extend the dimensionality 
 of the problem to include the unpolarized intensity component $I_\mathrm{offset}$, e.g., 
 with a flat prior between $I_\mathrm{offset,min}$ and $I_\mathrm{offset,max}$, and still imposing $I_0>0$. 

Let us stress that the Bayesian approach is also currently the only one that can deal with correlation between total intensity $I$ to Stokes $Q$ and $U$. We note, however, (i) new and forthcoming polarization data sets have a much better control of these systematics, and (ii) the impact of these correlations between noise components on the polarization fraction and angle bias is quite limited, as shown by PMA1.

\section{Conclusion}
\label{sec:conclusion}

We have presented in this work an extensive comparison of the performance of the polarization fraction and angle estimators.
While \citet{Simmons1985} focused on the common estimators of  the polarization fraction, such as 
the Maximum Likelihood (ML), the Most Probable (MP) and the Asymptotic (AS), and \citet{Quinn2012} 
suggested to use a Bayesian approach to estimate the polarization fraction, 
we have generalized all these methods to take into account the full covariance matrix of the Stokes parameters. 
We have also included in this comparison a novel estimator of the polarization fraction, the Modified Asymptotic  \citep[MAS, ][]{Plaszczynski2014}.
In addition, we have performed for the first time a comparison of the performance of the polarization angle estimators, 
since a statistical bias of $\psi$ is expected when the covariance matrix departs from its canonical form.
We have followed a frequentist methodology to investigate the properties of the polarization estimators, 
even when dealing with the frequentist estimators inspired by the Bayesian approach.

The question of the performance of a $\hat{p}$ or $\hat{\psi}$ estimator depends intrinsically on the analysis we would like to carry out with these quantities.
Including or not the full covariance matrix is one of the first questions that must be handled, 
but the more important aspect relies on the  properties of the output distribution of each estimator.
In practice, a compromise between three frequentist 
criteria has to be found: a minimum bias, a minimum risk  function, and the shape of  the output distribution, in terms of non-Gaussianity.
We present below a few recipes associated to typical use cases.

{\it - Build a mask.} It is usually recommended to build a mask on the intensity map, instead of using the SNR of the polarization fraction, 
so that  no values of the polarization fraction (especially low values of $p$) are discarded in the further analysis.
It can be useful, however, to build a mask based on the SNR of a polarization fraction map when
we are interested in strong values of the polarization fraction only, and we try to reject $p$ estimates artificially boosted by the noise.
This is the case when we look for the maximum value of $p$, for example. In this context 
we suggest following the prescription of P14, using a combination of the 
MAS estimator with confidence intervals. This method allows building conservative domains where the SNR is ensured to be greater than a given threshold. 
P14 provide numerical approximations in the canonical case. If one wants to take into account the specificity of the noise properties in each pixel, 
confidence intervals can be built for any covariance matrix (including ellipticity and correlation), but it could require intensive computing.
Another alternative in that case is to build credible intervals using the posterior distribution (MAP).

{\it - Large maps of the polarization fraction with high SNR on the intensity. } 
Another typical use case is to provide large maps of the polarization fraction with the associated uncertainty,  when
the intensity is assumed to be perfectly known.
Because of their discontinuous distributions presenting a peak at $\hat{p}$=0 and their strong 
dependance to the unknown true polarization angle $\psi_0$,  the common estimators of $p$ ML, MP and AS 
are not well designed for this purpose. These estimators could produce highly discontinuous patterns with  zero values over the output $\hat{p}$ map  when 
the SNR goes below 4, 
 which may imply complicated analysis including upper limits values. 
In order to avoid such issues, we first suggest using the MAS estimator which has been shown to produce the lowest relative bias, with a continuous output distribution which 
becomes close to a Gaussian for SNR larger than 2. Moreover, the  relative risk function  associated with the MAS estimator becomes competitive for SNR$>$3,
while the MB estimator minimizes the  relative risk function  for an intermediate SNR, between 1 and 3.
The uncertainties can then be derived again from the confidence or credible intervals, depending on the ellipticity of the covariance matrix. 
A second option, especially suited for intermediate SNR (2-3), consists in performing a preliminary analysis on the data 
to build a prior from the $\hat{p}$ distribution,  which can then be injected into the MB estimator. 
The performance of this method strongly rely of the properties of the initial true distribution. 
It is particularly efficient for true polarization fractions largely greater than zero, to avoid the major drawback of the 
MB estimator presenting a lower-limit proportional to the noise level.
Hence the MB (with flat prior) estimator presents a cut-off at 0.8$\sigma_p$, so that it can never provide null estimates of $\hat{p}$.
We stress that above a SNR of 4, all methods (except MP2) fall in agreement.

{\it - Combined polarization fraction and angle analysis. }The Bayesian estimators of $\hat{p}_{\text{MB}}$ and $\hat{\psi}_{\text{MB}}$ may be used to build estimates of the polarization
 fraction and angle simultaneously, by taking into account the full covariance matrix, including the ellipticity and correlation between
 $Q$ and $U$, and the correlation between total and polarized intensity. 
 This could be useful when performing an analysis over large areas with inhomogeneous noise properties,
 when the SNR on the intensity becomes problematic, or when an important  correlation between $I$ and ($Q$, $U$) exists. 
 Nevertheless we stress that the output distributions of the MB estimates are strongly asymmetric at low SNR ($<$3), and 
 that the Bayesian uncertainty estimates can not be used as typical Gaussian 68\% tolerance intervals. 
 
{\it - Low SNR on the intensity. } 
We recommend in this case to use the Bayesian estimators which allow simultaneous estimates 
of the intensity and the polarization parameters, taking into account the full covariance matrix, and include the impact of the uncertainty of the intensity
on the polarization fraction estimate.

 {\it - Very low SNR studies. } Very low SNRs studies may require different approaches. We have seen that at low SNR, all estimators 
 provide biased estimates of the polarization fraction, with highly asymmetrical distributions. The more conservative option in this case is to use the 
 confidence or credible intervals. 
 Similarly the question of assessing the unpolarized level of a set of data (i.e. SNR$\sim$0) 
 has been first raised by \citet{Clarke1993}. They suggested to used Kolmogorov test to compare the measurement distributions with the expectation
derived from the Rice distribution with $p_0$=0. Another option is to build the likelihood in two dimensions ($Q$,$U$) to perform a $\chi^2$ test with $Q_0$=$U_0$=0.
A last method is to use the Bayesian posterior probability $B(p_0|p,\sigma_p)$ to assess the probability to have $p_0$=0 
for a given measurement or a series of measurements by convolving all individual pdfs.
 
{\it - Polarization angle.} Concerning the polarization angle estimates $\hat{\psi}$, we have shown that the ML provides the best performance in terms of
 relative  bias and  risk function for SNR$>$1. It corrects a potential bias of $\psi$ when the covariance matrix is not under its canonical form.
Because the ML and MAP estimators give equivalent results, the MAP can be used to 
efficiently build credible intervals and symmetric uncertainties, which have been shown to be in a very good agreement with the output distributions.
 Nevertheless we stress that the level of the  absolute bias of $\psi$ remains extremely limited compared to the dispersion of the polarization angle 
 in most cases (i.e. in the {\it low} and {\it tiny} regime of the covariance matrix), so that it can be usually neglected.

\begin{acknowledgements}
This paper was developed to support the analysis of data from the
\Planck\ satellite.  The development of \Planck\ has been supported by: ESA; CNES and
CNRS/INSU-IN2P3-INP (France); ASI, CNR, and INAF (Italy); NASA and DoE
(USA); STFC and UKSA (UK); CSIC, MICINN, JA, and RES (Spain); Tekes,
AoF, and CSC (Finland); DLR and MPG (Germany); CSA (Canada); DTU Space
(Denmark); SER/SSO (Switzerland); RCN (Norway); SFI (Ireland);
FCT/MCTES (Portugal); and PRACE (EU). A description of the Planck
Collaboration and a list of its members, including the technical or
scientific activities in which they have been involved, can be found
at \url{http://www.sciops.esa.int/index.php?project=planck&page=}\\ \url{Planck_Collaboration}.
We acknowledge the use of the Legacy Archive for Microwave Background
Data Analysis (LAMBDA), part of the High Energy Astrophysics Science
Archive Center (HEASARC). HEASARC/LAMBDA is a service of the
Astrophysics Science Division at the NASA Goddard Space Flight Center.
Some of the results in this paper have been derived using the
{\healpix} package.   
We would also like to thank P. Leahy, S. Prunet and M. Seiffert for their very useful comments.
\end{acknowledgements}

\appendix

\section{Most Probable in general case}
\label{sec:most_probable_detail}

The MP2 estimators, $\hat{p}_{\text{MP2}}$  and $\hat{\psi}_{\text{MP2}}$, 
have to satisfy the Eqs.~\ref{eq:mp_1}  and \ref{eq:mp_2} simultaneously.
 These relations can be solved, using the fully developed expression of $f_{2D}$ including the terms of the inverse matrix $\tens{\Sigma}_p^{-1}$:
\begin{equation}
\tens{\Sigma}_p^{-1}= \left(\begin{array}{cc}
 v_{11} & v_{12} \\
 v_{12} & v_{22} \\
\end{array}\right) \, ,
\end{equation}
leading to 

\begin{eqnarray}
\hat{\psi}_{\text{MP2}} & = & \frac{1}{2} \arctan \left (\frac{ \left( \left( v_{11}v_{22} - v_{12}^2 \right)p^2 - v_{11} \right) \sin2\psi  +
 v_{12} \cos2\psi}{\left( \left( v_{11}v_{22} - v_{12}^2 \right)p^2 - v_{22} \right) \cos2\psi  + v_{12} \sin2\psi} \right) \, , \nonumber \\
\hat{p}_{\text{MP2}} & = & \frac{A_1}{ \left( A_2 \cos2\hat{\psi}_{\text{MP2}} + A_3 \sin2\hat{\psi}_{\text{MP2}} \right) } \, ,
\end{eqnarray}
with
\begin{eqnarray}
A_1 & \equiv & p \left( v_{11} \cos^22\psi+ v_{22} \sin^22\psi + 2 v_{12} \cos2\psi\sin2\psi \right) -1/p \, , \nonumber \\
A_2 &  \equiv & v_{11}  \cos2\psi + v_{12}\sin2\psi \, , \nonumber \\ 
A_3 &  \equiv & v_{22}  \sin2\psi + v_{12}\cos2\psi \, .
\end{eqnarray}
This analytical solution only depends on the input measurements ($p$, $\psi$) and the covariance matrix $\tens{\Sigma}_p$.
Because the polarization fraction must be positive, there exists a lower 
limit of the SNR so that $\hat{p}_{\text{MP2}}=0$. In
that case $\hat{\psi}_{\text{MP2}} $ is not constrained any more and can be chosen to any possible value, we will set it equal to the measurement $\psi$.
Moreover, this expression can be simplified when $\rho=0$, which implies that $v_{12}=0$, leading to:
\begin{eqnarray}
\hat{\psi}_{\text{MP2}} & = & \frac{1}{2} \arctan \left (\frac{ \left( p^2 - 1/v_{22} \right)}{\left( p^2 - 1/v_{11} \right)} \tan2\psi \right) \, , \\
\hat{p}_{\text{MP2}} & = & \frac{  p \left( v_{11} \cos^2 2\psi + v_{22} \sin^22\psi \right) -1/p }{\left(  v_{11}  \cos2\psi\cos2\hat{\psi}_{\text{MP2}}  + v_{22}  \sin2\psi \sin2\hat{\psi}_{\text{MP2}} \right) } \, . \nonumber
\end{eqnarray}
In the canonical case ($v_{12}$=0, $v_{11}$=$v_{22}$=1/$\sigma_p^2$), we recover the expression derived by \citet{Quinn2012}:
\begin{eqnarray}
\hat{\psi}_{\text{MP2}} & = & \psi \, , \nonumber \\
\hat{p}_{\text{MP2}} & = & \Bigg \{  \begin{array}{ll}
(p - \sigma_p^2 / p)  & \, \, \, \mathrm{for} \, \, p > \sigma_p \\
0 &  \, \, \, \mathrm{for} \, \,  p \le \sigma_p
\end{array} \, .
\end{eqnarray}

\section{Bayesian Posterior pdf}
\label{sec:pdf_posterior}

\begin{figure*}[tp]
\begin{tabular}{cccccc}
&  \begin{minipage}[c]{.15\linewidth}
$\quad$
 \begin{tabular}{l}
$ \varepsilon = 1$ \\
$ \rho  = 0$ \\
\end{tabular} \\
$\Bigg($
 \begin{tabular}{l}
$ \varepsilon_\mathrm{eff} = 1$ \\
$ \theta  = 0$ \\
\end{tabular} 
$\Bigg)$
\end{minipage} & 
 \begin{minipage}[c]{.15\linewidth}
$\quad$
 \begin{tabular}{l}
$ \varepsilon = 1/2$ \\
$ \rho  = 0$ \\
\end{tabular} \\ 
$\Bigg($
 \begin{tabular}{l}
$ \varepsilon_\mathrm{eff} = 1/2$ \\
$ \theta  = 0$ \\
\end{tabular} 
$\Bigg)$
\end{minipage} & 
\begin{minipage}[c]{.15\linewidth}
$\quad$
 \begin{tabular}{l}
$ \varepsilon = 2$ \\
$ \rho  = 0$ 
\end{tabular}  \\ 
$\Bigg($
 \begin{tabular}{l}
$ \varepsilon_\mathrm{eff} = 2$ \\
$ \theta  = 0$ \\
\end{tabular} 
$\Bigg)$
\end{minipage} &
\begin{minipage}[c]{.15\linewidth}
$\quad$
 \begin{tabular}{l}
$ \varepsilon = 1$ \\
$ \rho  = -1/2$ 
\end{tabular}  \\ 
$\Bigg($
 \begin{tabular}{l}
$ \varepsilon_\mathrm{eff} \sim 1.73$ \\
$ \theta  = -\pi/4$ \\
\end{tabular} 
$\Bigg)$
\end{minipage} &
\begin{minipage}[c]{.15\linewidth}
$\quad$
 \begin{tabular}{l}
$ \varepsilon = 1$ \\
$ \rho  = 1/2$
\end{tabular}  \\ 
$\Bigg($
 \begin{tabular}{l}
$ \varepsilon_\mathrm{eff} \sim 1.73$ \\
$ \theta  = \pi/4$ \\
\end{tabular} 
$\Bigg)$
\end{minipage} 
\\
\begin{minipage}[c]{.13\linewidth}
 \begin{tabular}{l}
$p_0/\sigma_{p,G}=0.1$ 
\end{tabular} 
\end{minipage} &
\begin{minipage}[c]{.15\linewidth} \includegraphics[width=3cm, viewport=200 0 600 400]{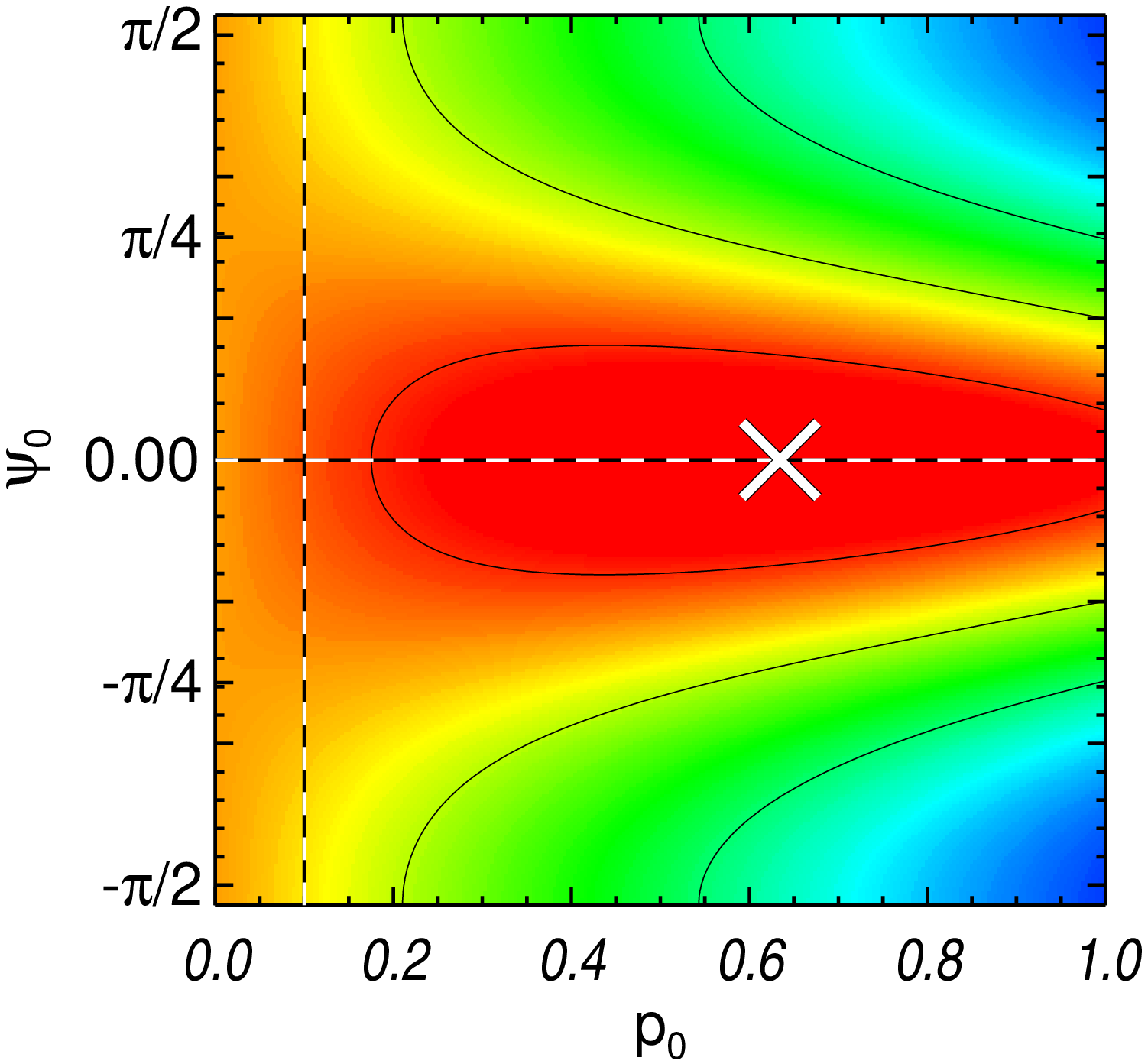} \end{minipage} & 
\begin{minipage}[c]{.15\linewidth} \includegraphics[width=3cm, viewport=200 0 600 400]{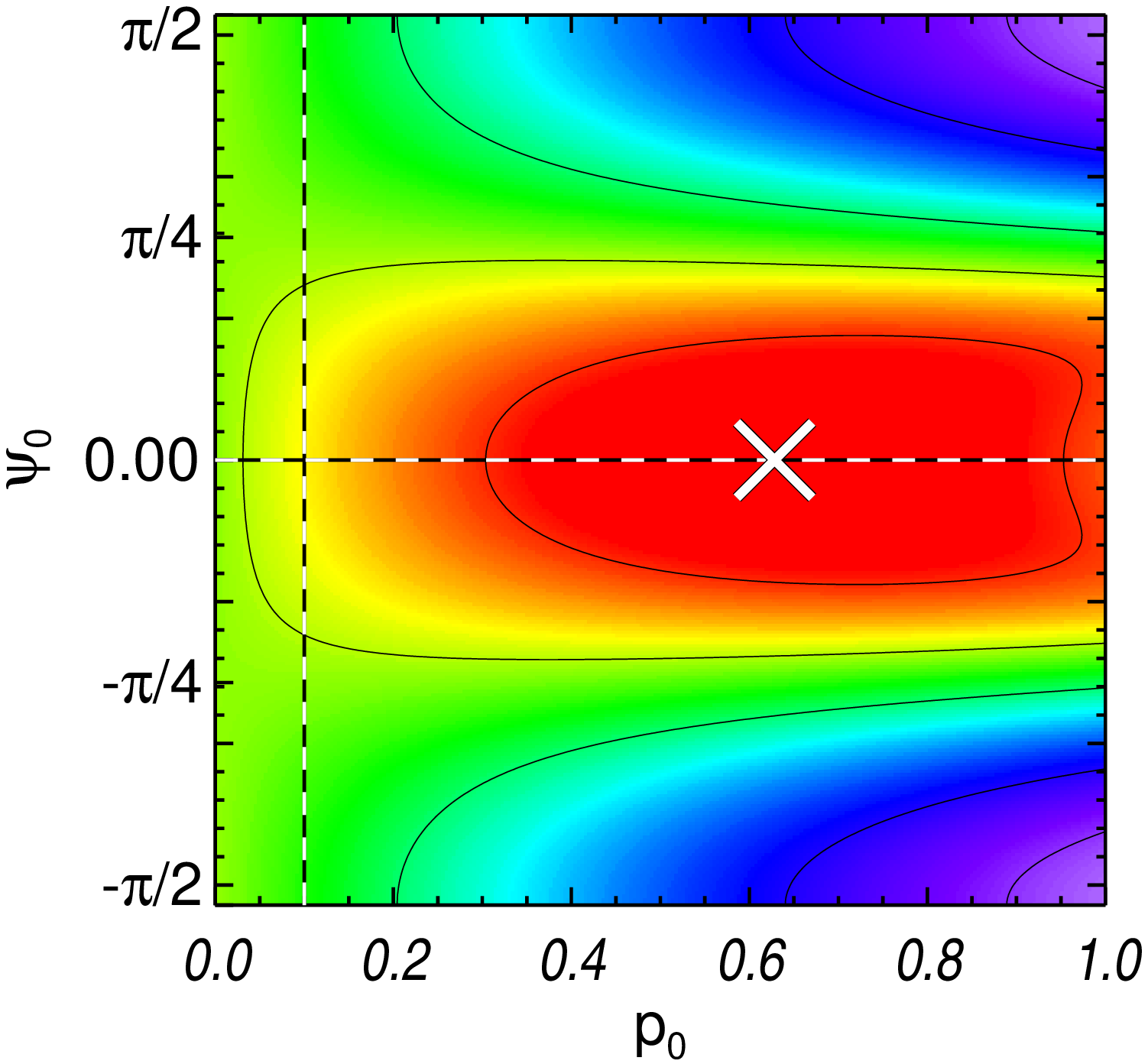} \end{minipage} &
\begin{minipage}[c]{.15\linewidth} \includegraphics[width=3cm, viewport=200 0 600 400]{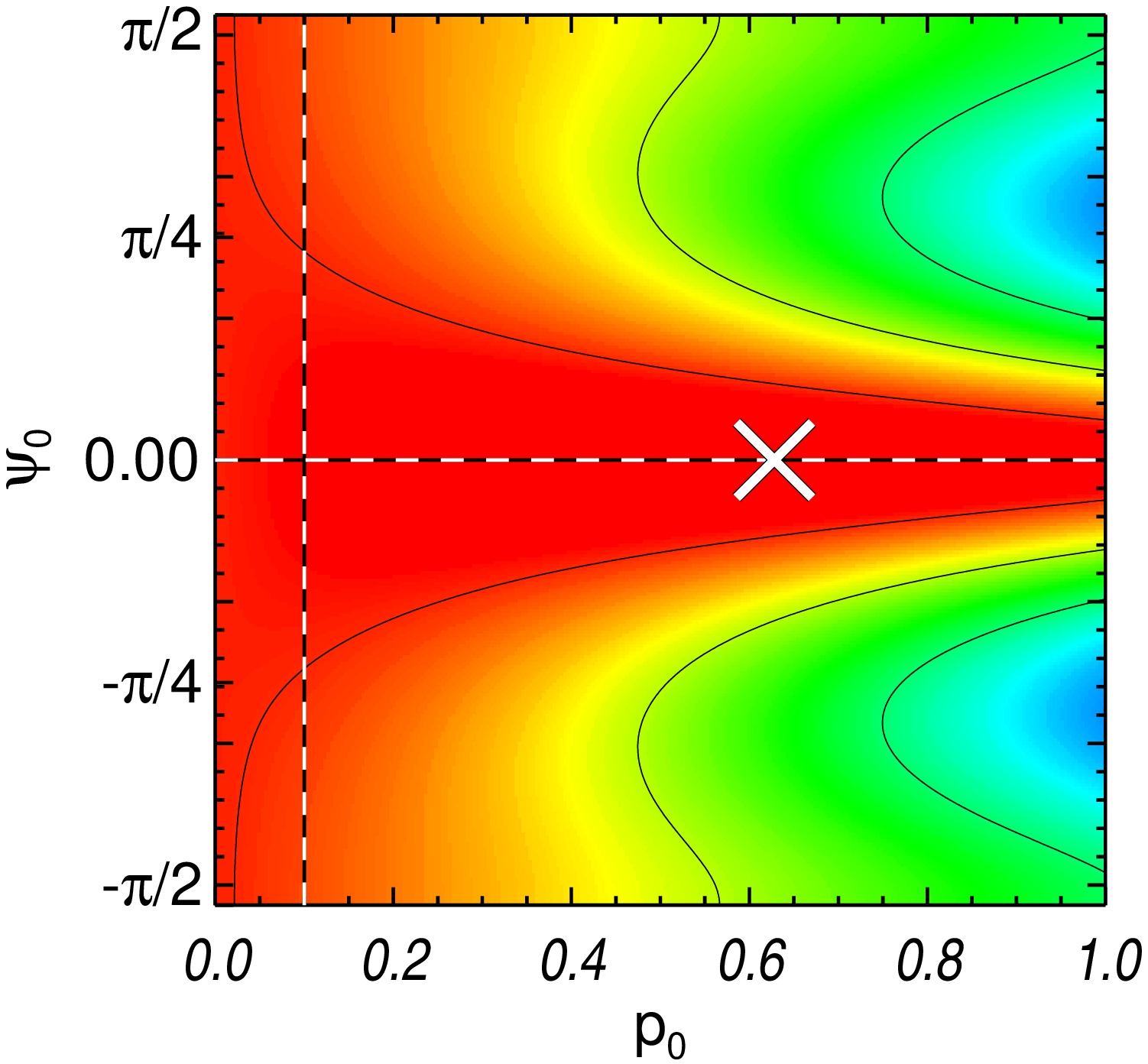} \end{minipage} &
\begin{minipage}[c]{.15\linewidth} \includegraphics[width=3cm, viewport=200 0 600 400]{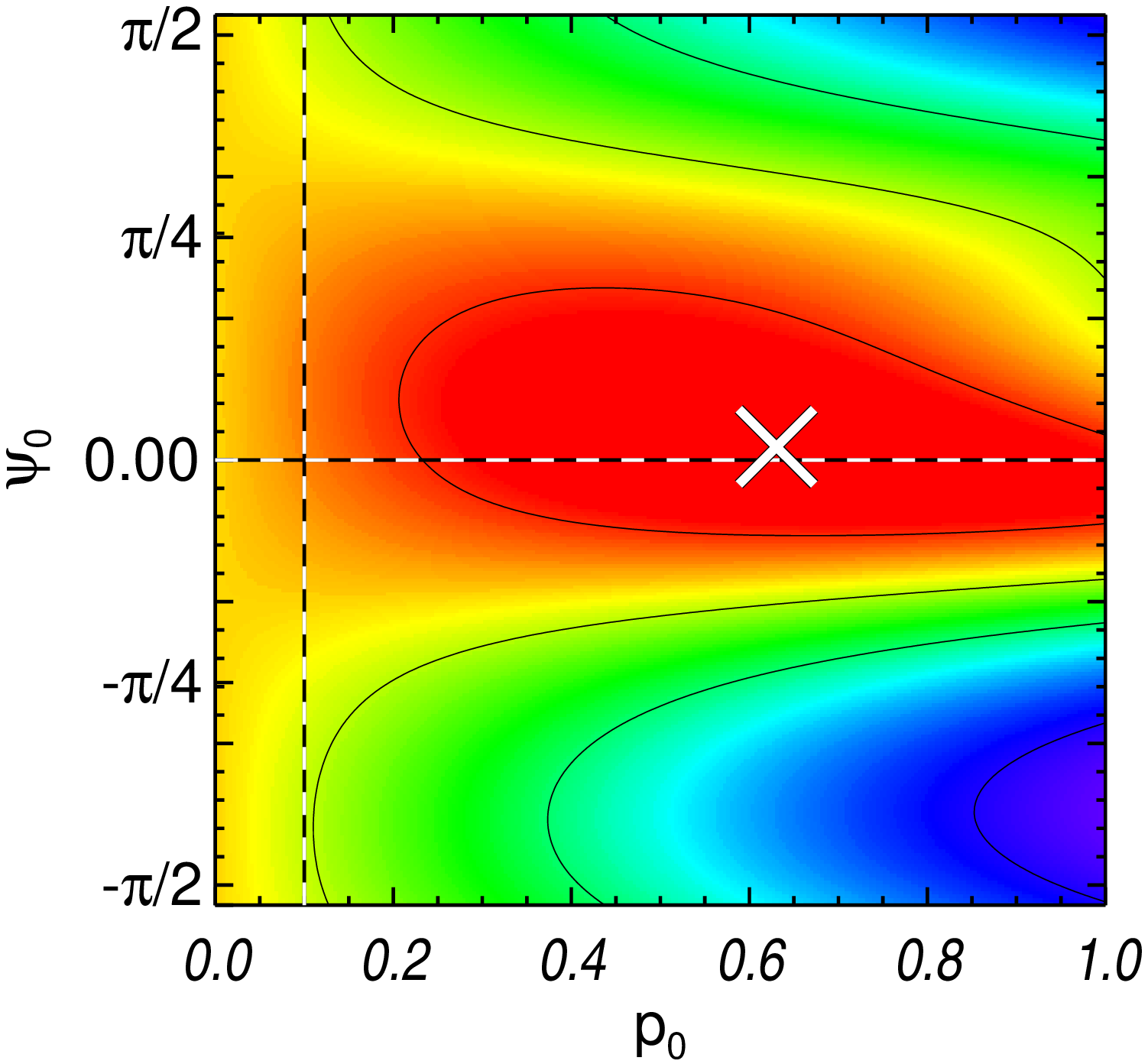} \end{minipage} &
\begin{minipage}[c]{.15\linewidth} \includegraphics[width=3cm, viewport=200 0 600 400]{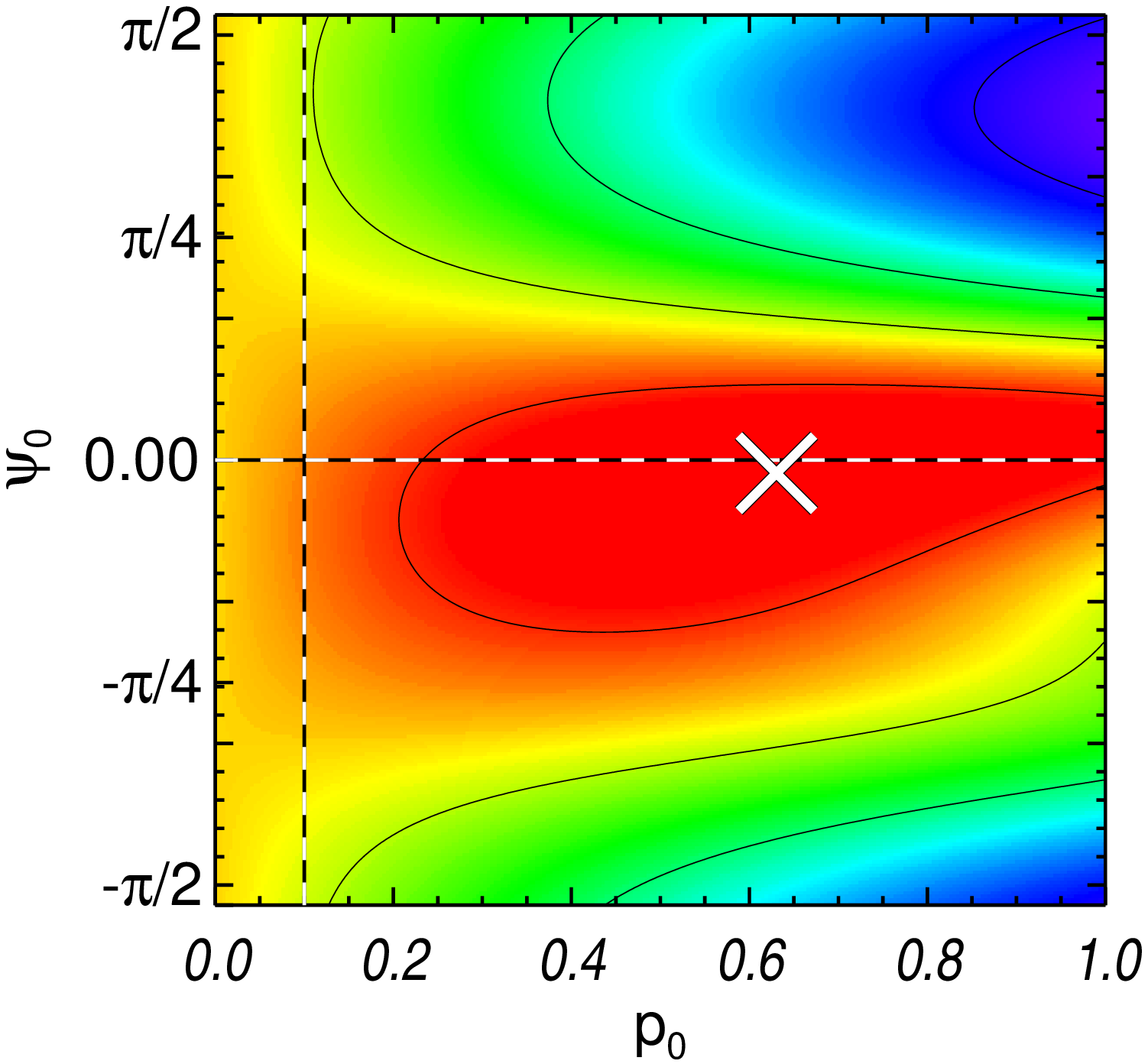} \end{minipage} \\

\begin{minipage}[c]{.13\linewidth}
 \begin{tabular}{l}
$p_0/\sigma_{p,G}=0.5$ 
\end{tabular} 
\end{minipage} &
\begin{minipage}[c]{.15\linewidth} \includegraphics[width=3cm, viewport=200 0 600 400]{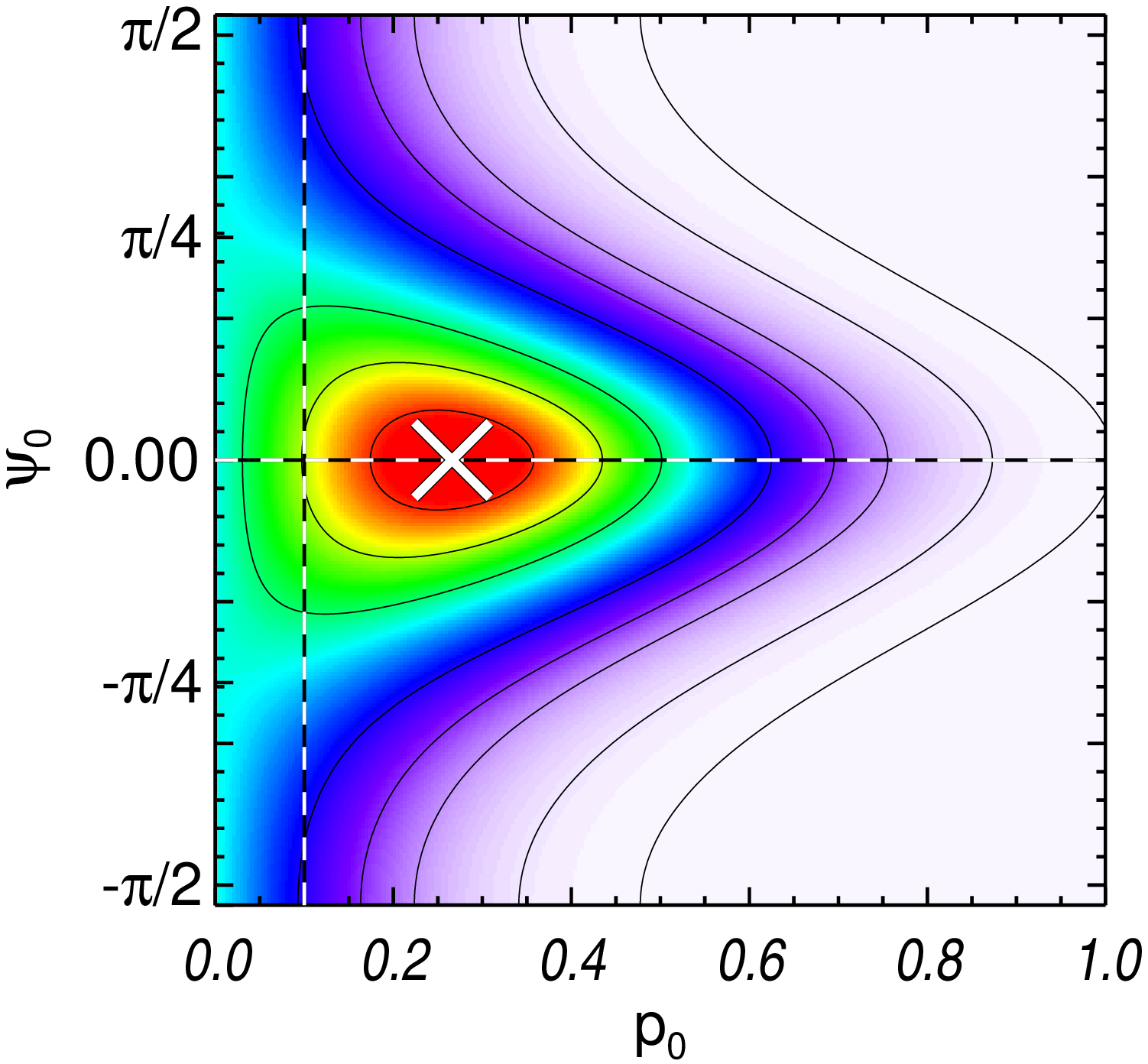} \end{minipage} &
\begin{minipage}[c]{.15\linewidth} \includegraphics[width=3cm, viewport=200 0 600 400]{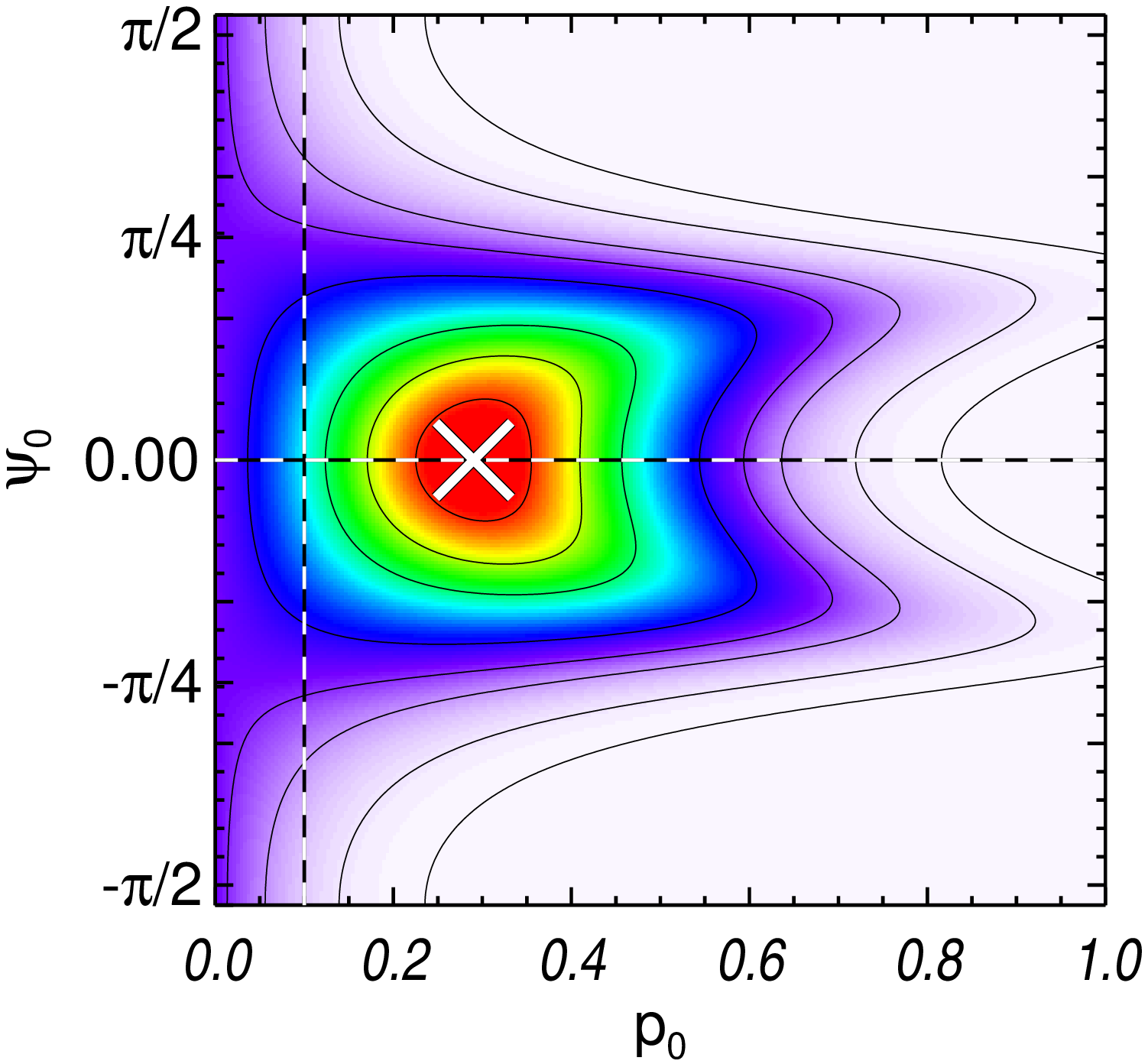} \end{minipage} &
\begin{minipage}[c]{.15\linewidth} \includegraphics[width=3cm, viewport=200 0 600 400]{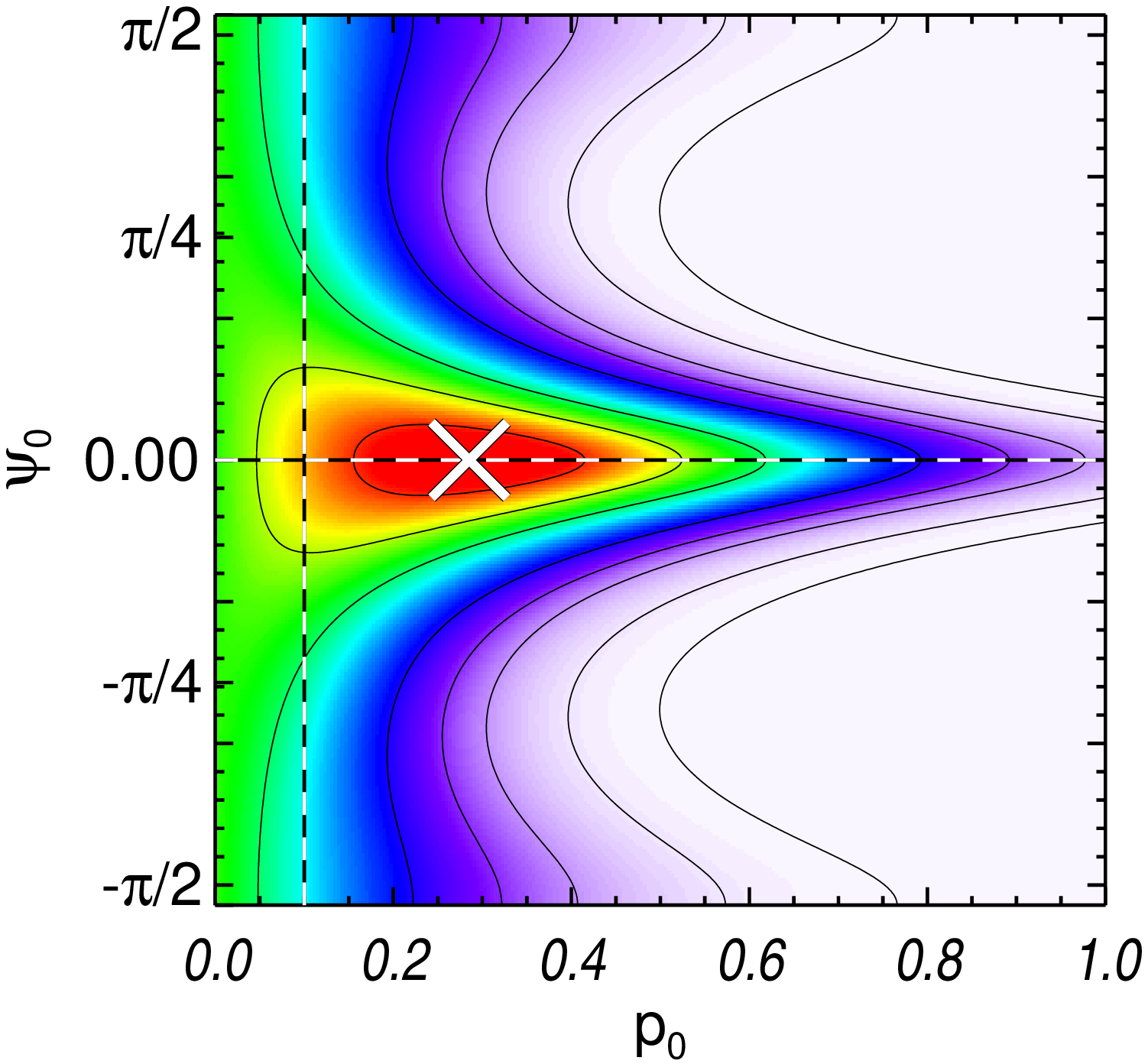} \end{minipage} &
\begin{minipage}[c]{.15\linewidth} \includegraphics[width=3cm, viewport=200 0 600 400]{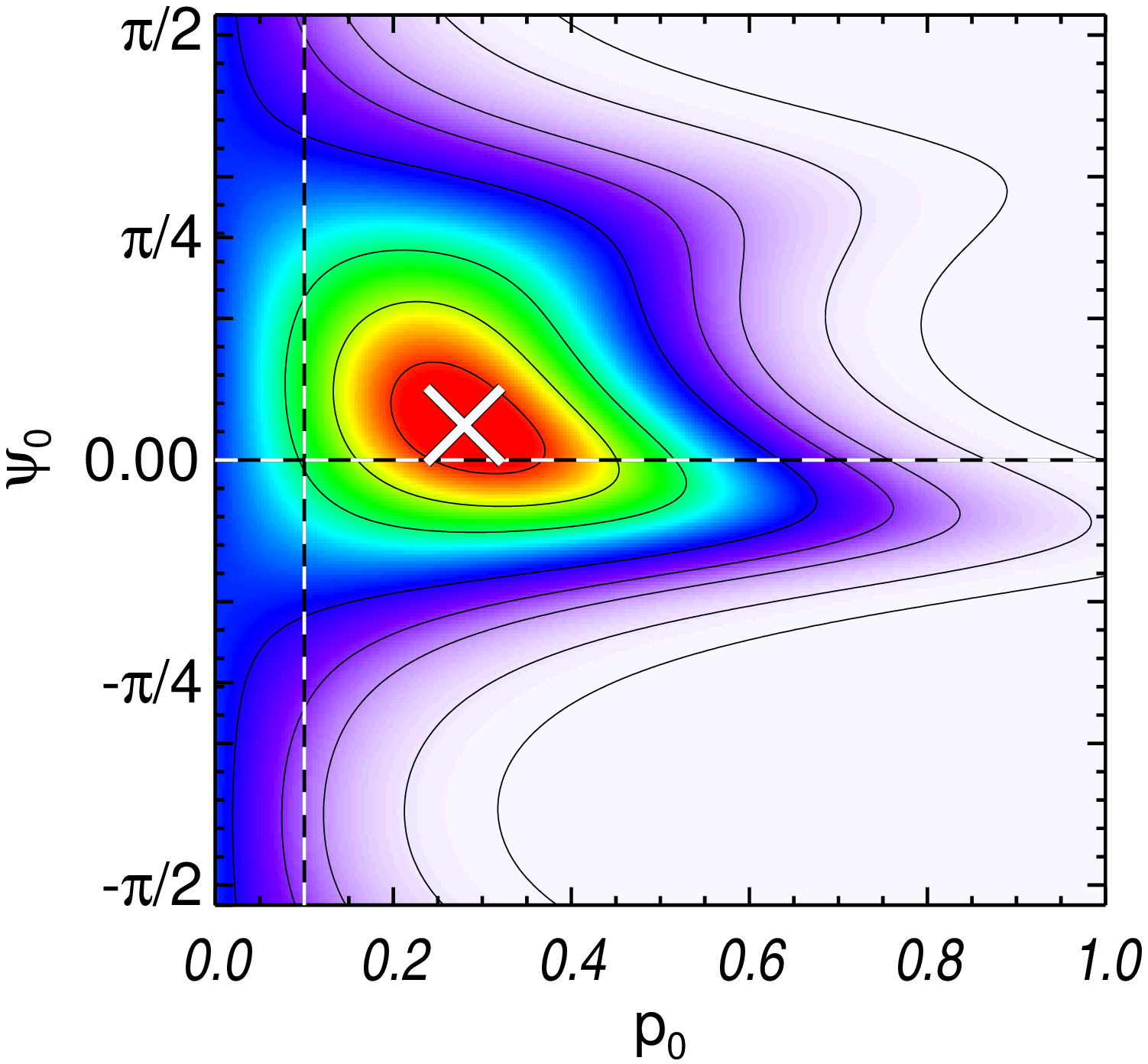} \end{minipage} &
\begin{minipage}[c]{.15\linewidth} \includegraphics[width=3cm, viewport=200 0 600 400]{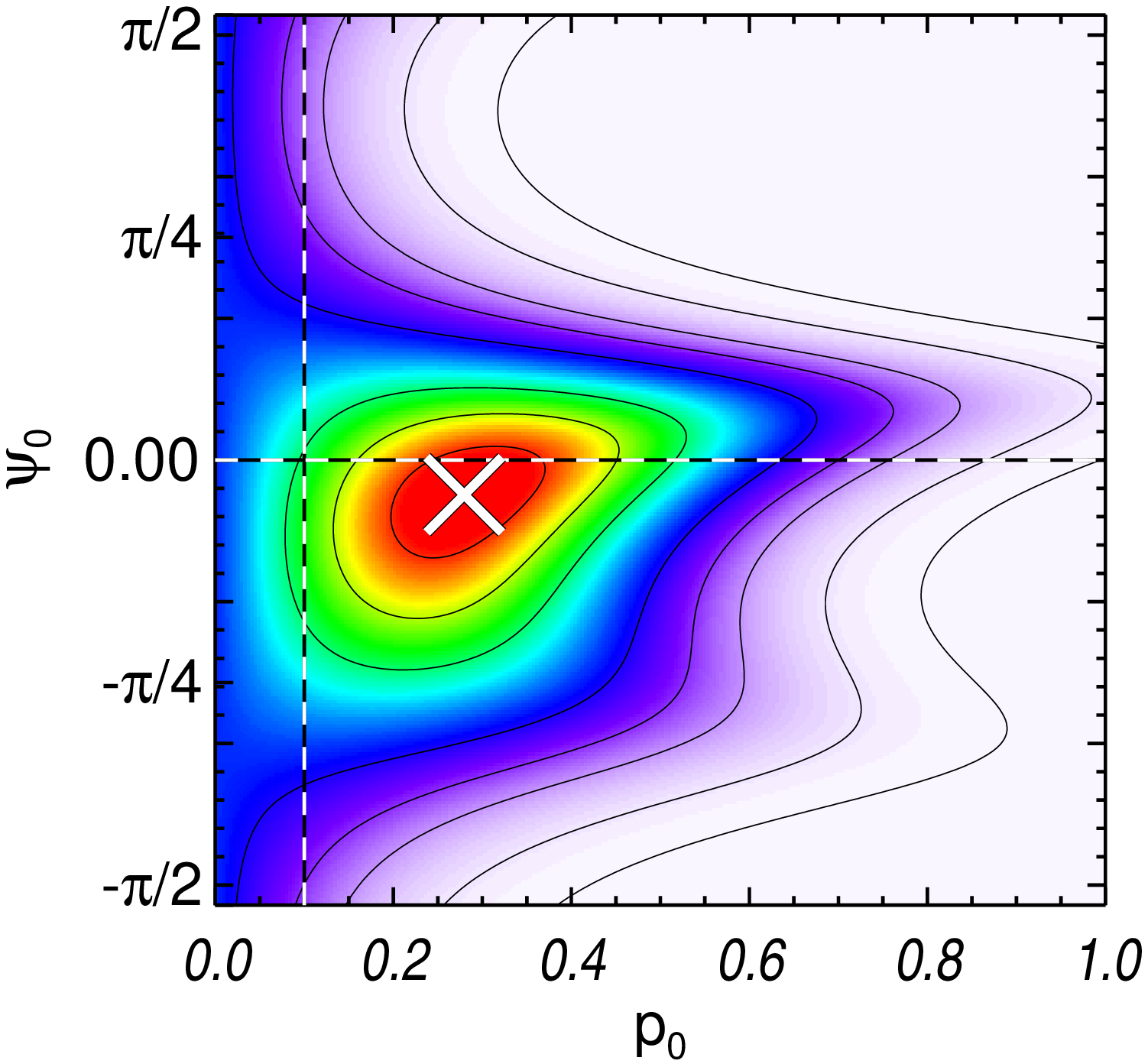} \end{minipage} \\

\begin{minipage}[c]{.13\linewidth}
 \begin{tabular}{l}
$p_0/\sigma_{p,G}=1$ 
\end{tabular} 
\end{minipage} &
\begin{minipage}[c]{.15\linewidth} \includegraphics[width=3cm, viewport=200 0 600 400]{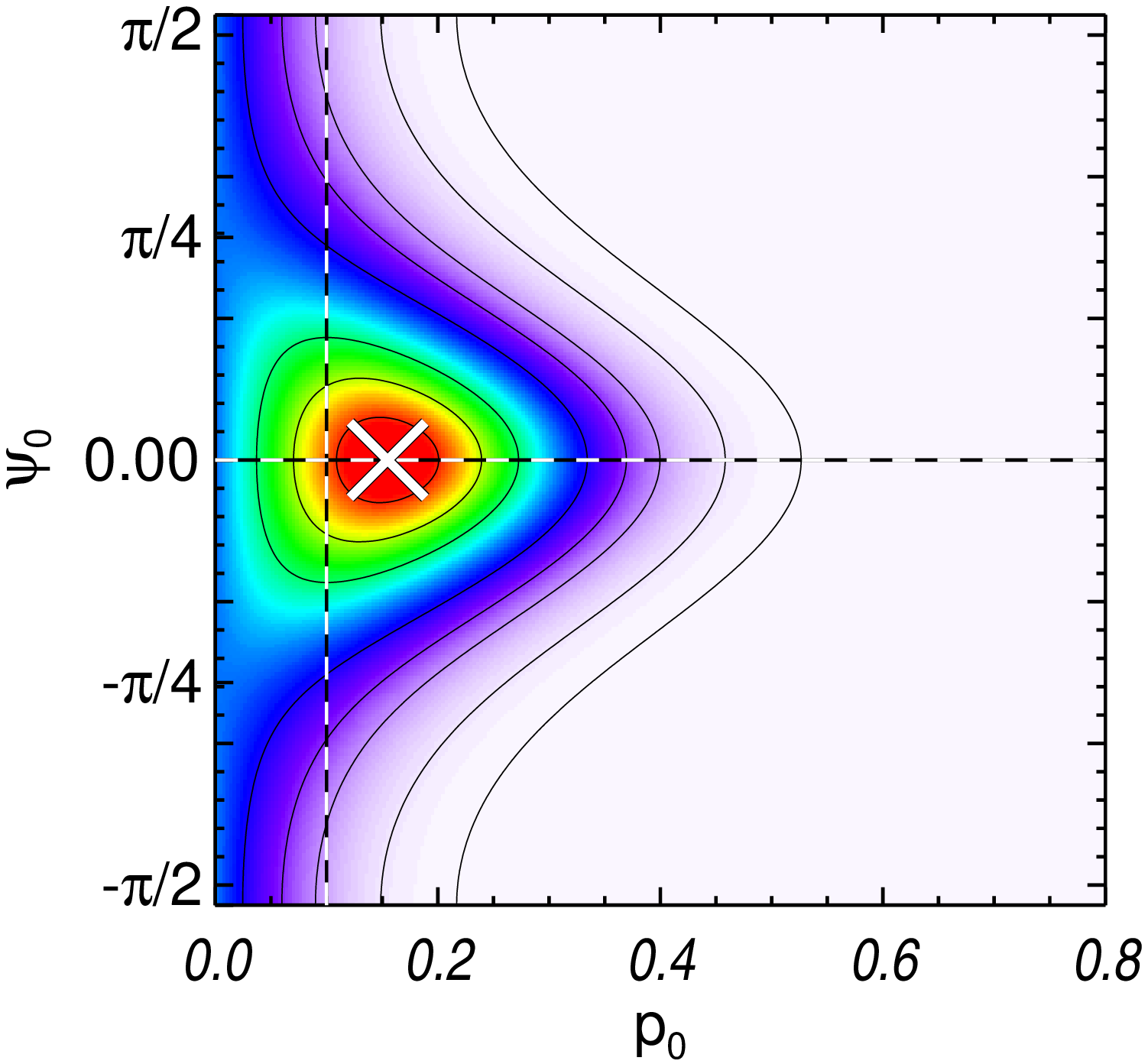} \end{minipage} &
\begin{minipage}[c]{.15\linewidth} \includegraphics[width=3cm, viewport=200 0 600 400]{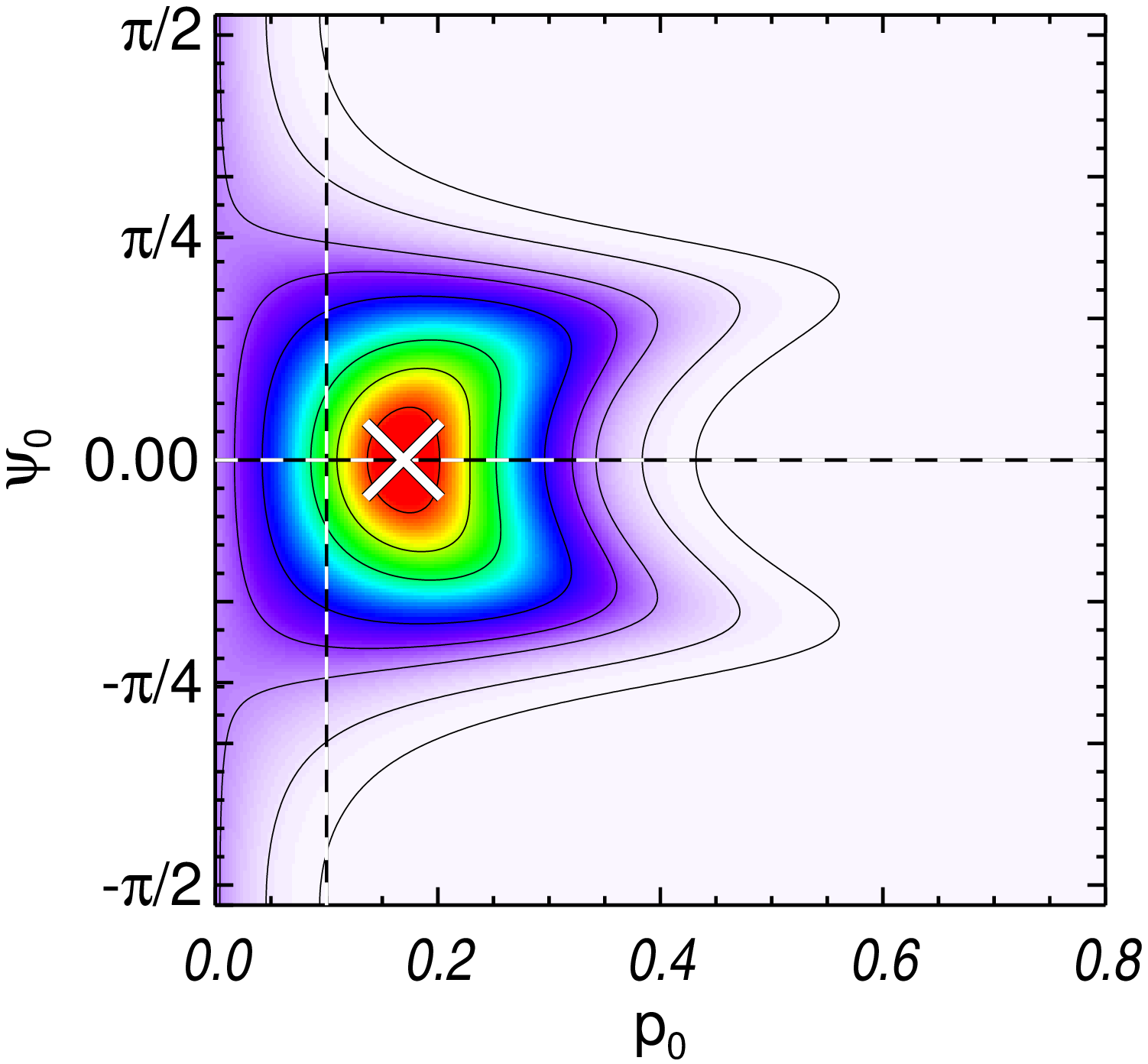} \end{minipage} &
\begin{minipage}[c]{.15\linewidth} \includegraphics[width=3cm, viewport=200 0 600 400]{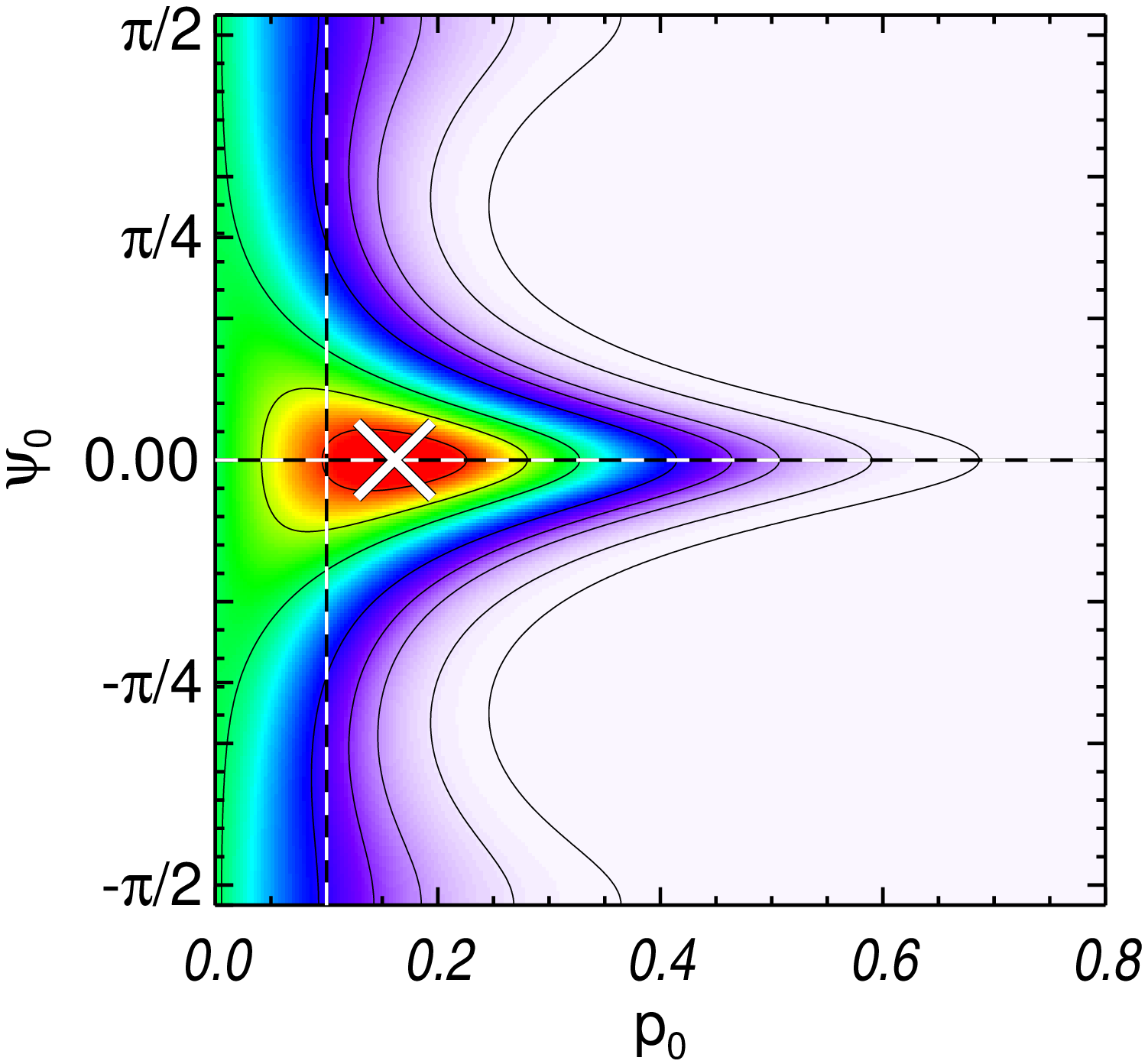} \end{minipage} &
\begin{minipage}[c]{.15\linewidth} \includegraphics[width=3cm, viewport=200 0 600 400]{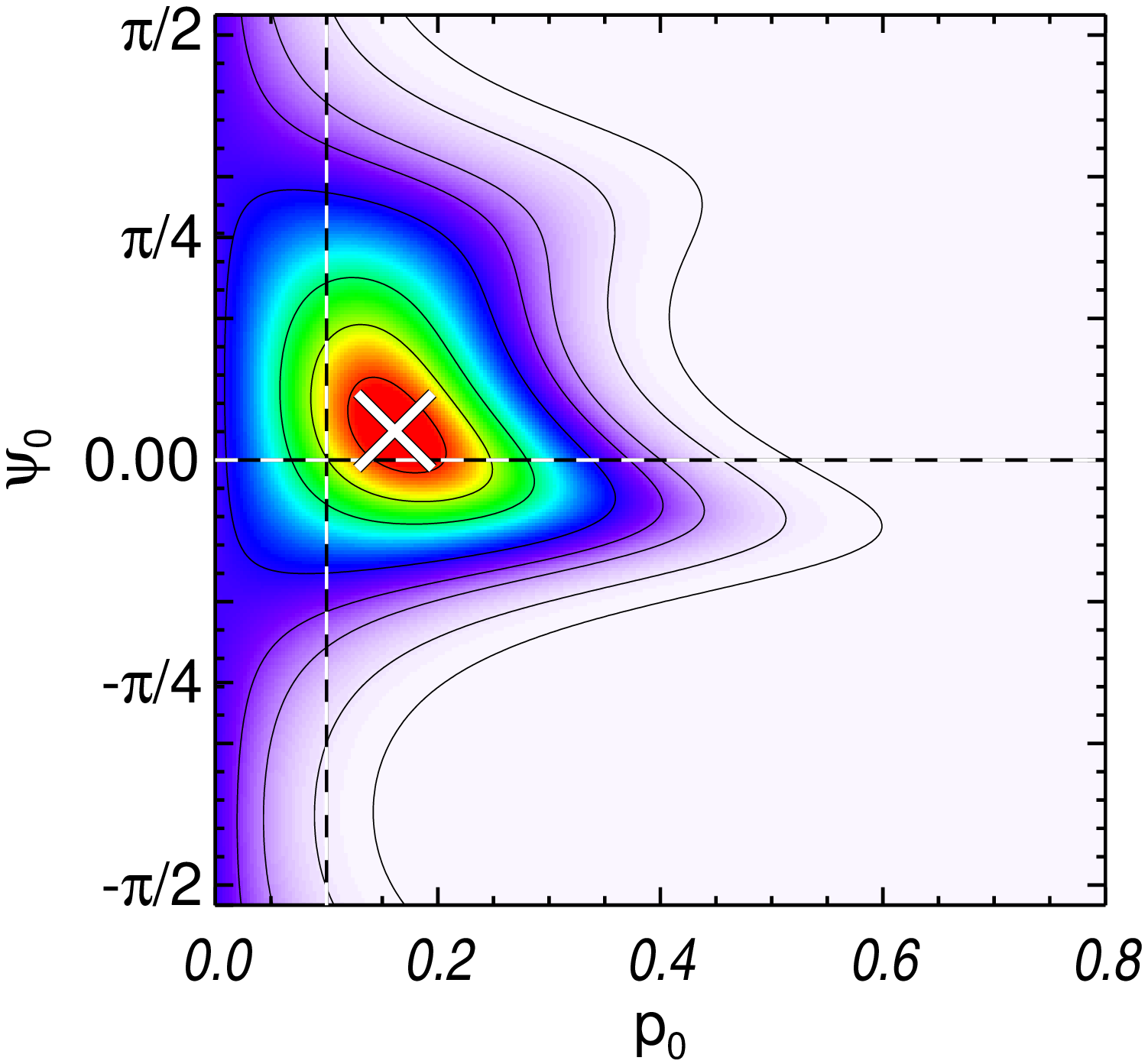} \end{minipage} &
\begin{minipage}[c]{.15\linewidth} \includegraphics[width=3cm, viewport=200 0 600 400]{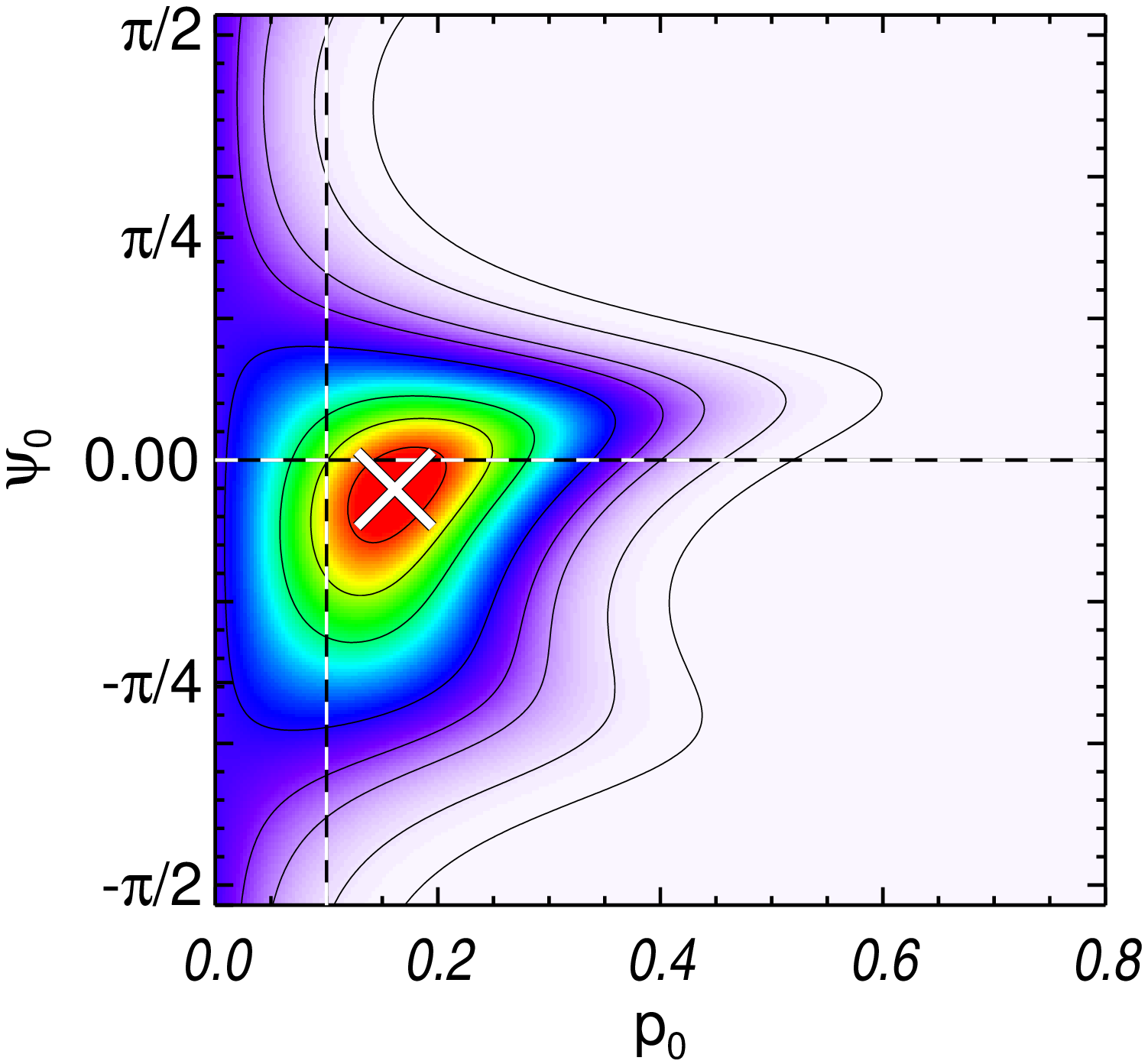} \end{minipage} \\

\begin{minipage}[c]{.13\linewidth}
 \begin{tabular}{l}
$p_0/\sigma_{p,G}=5$ 
\end{tabular} 
\end{minipage} &
\begin{minipage}[c]{.15\linewidth} \includegraphics[width=3cm, viewport=200 0 600 400]{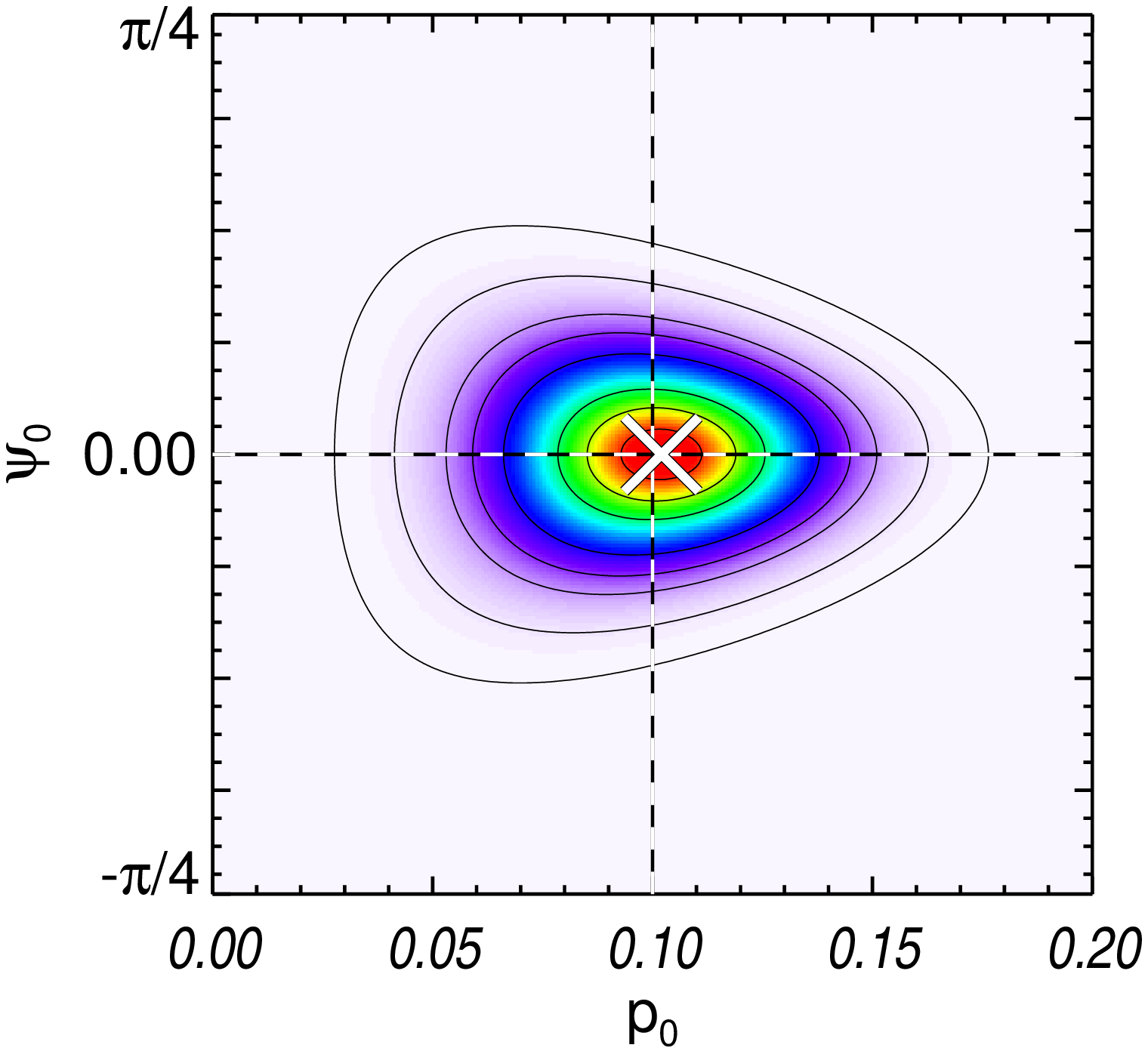} \end{minipage} &
\begin{minipage}[c]{.15\linewidth} \includegraphics[width=3cm, viewport=200 0 600 400]{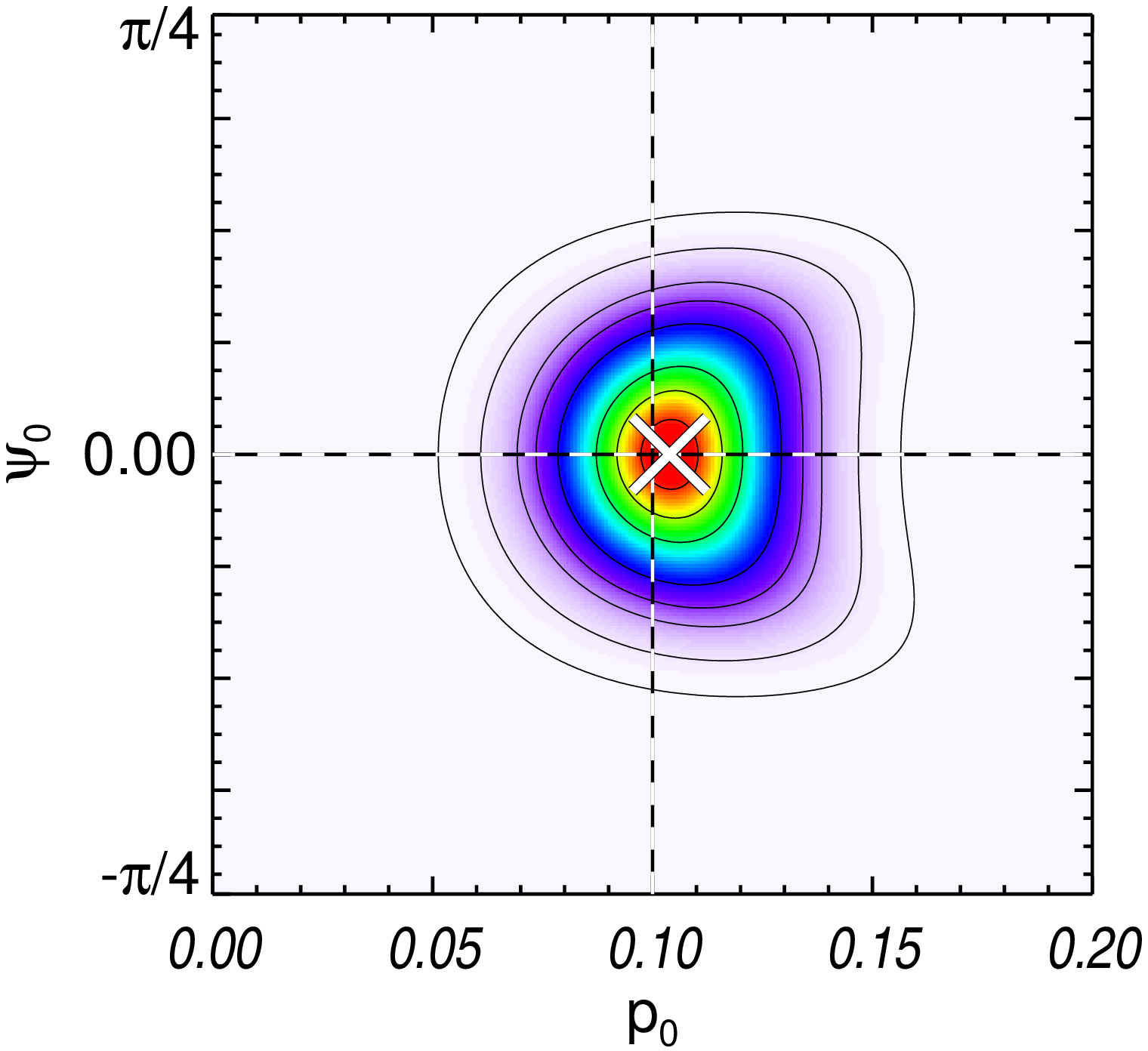} \end{minipage} &
\begin{minipage}[c]{.15\linewidth} \includegraphics[width=3cm, viewport=200 0 600 400]{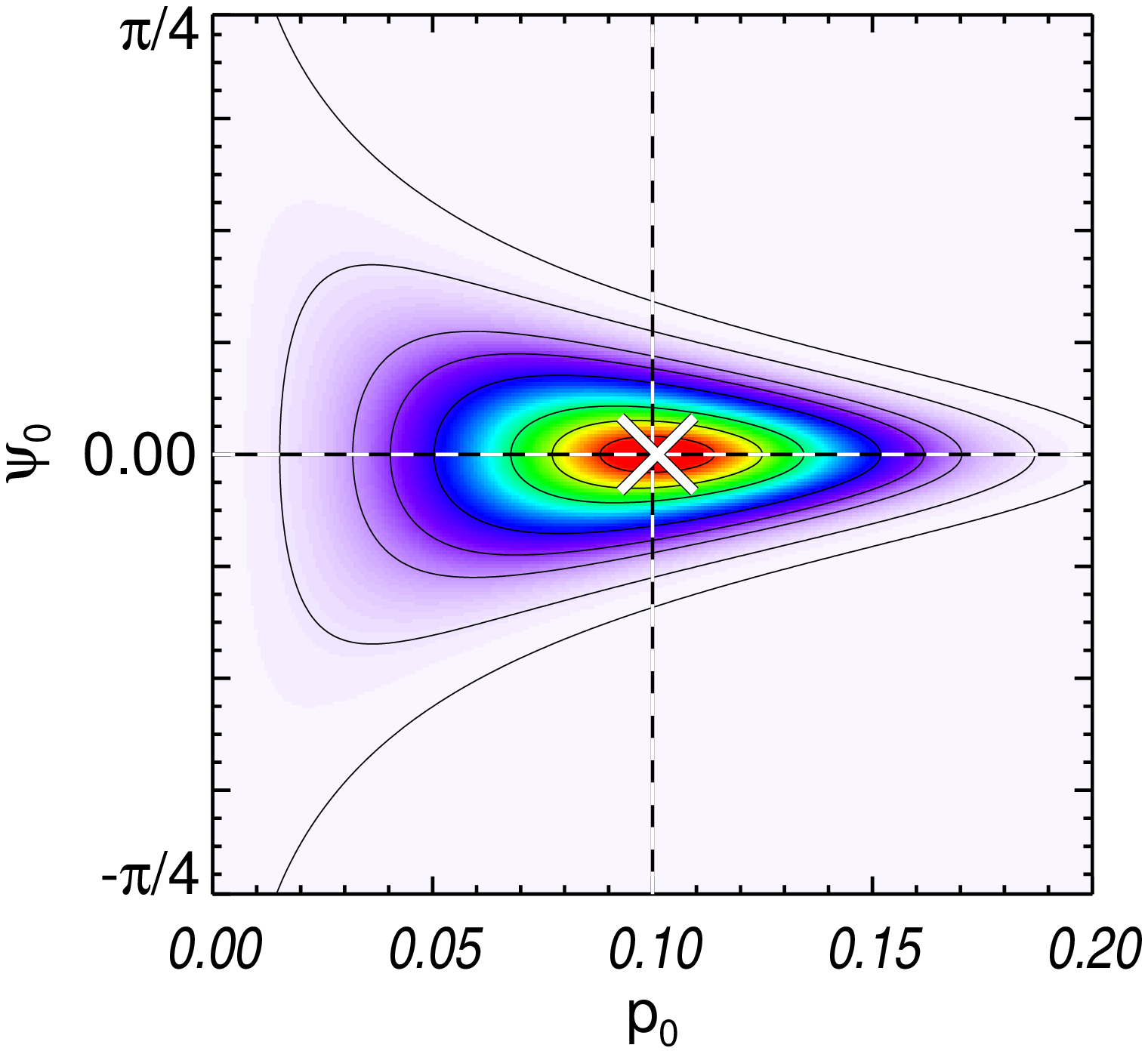} \end{minipage} &
\begin{minipage}[c]{.15\linewidth} \includegraphics[width=3cm, viewport=200 0 600 400]{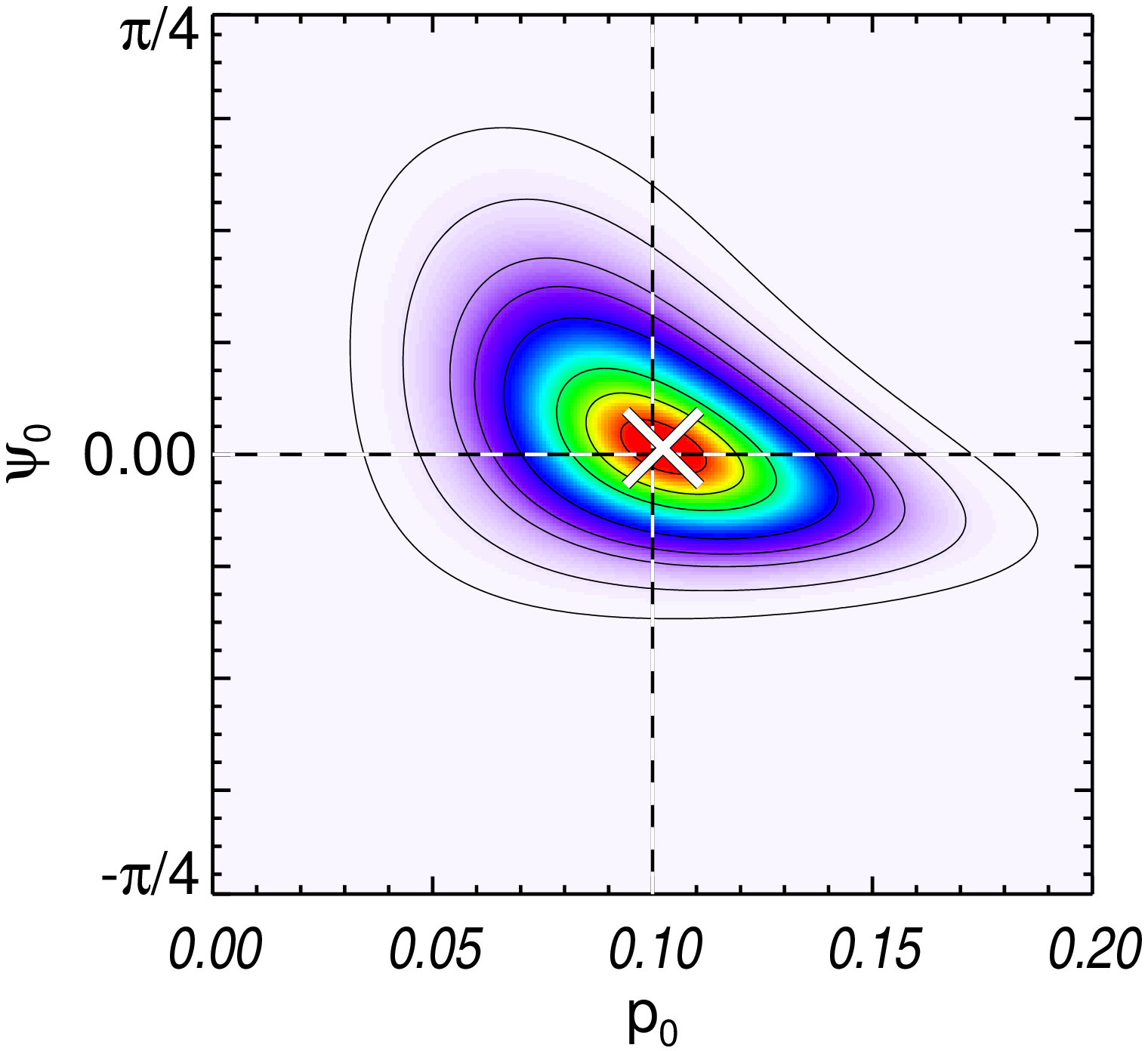} \end{minipage} &
\begin{minipage}[c]{.15\linewidth} \includegraphics[width=3cm, viewport=200 0 600 400]{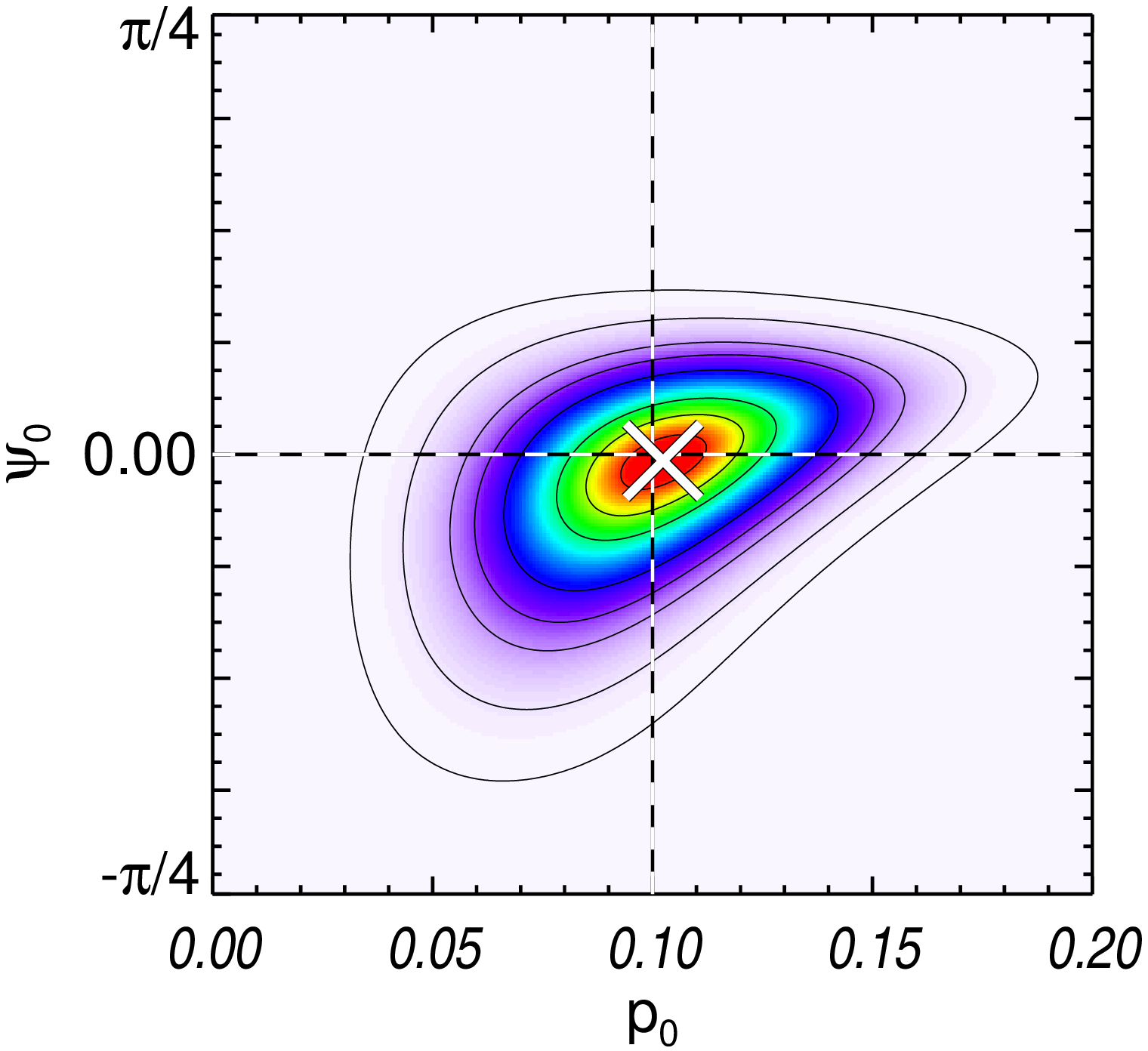} \end{minipage} \\

 \end{tabular} 
 \caption{Posterior probability density functions $B_{2D}(p_0,\psi_0\, | \, p, \psi, \tens{\Sigma}_p)$ computed for
the most probable measurements ($p$,$\psi$) of the $f_{2D}$ distribution (crosses), which 
were obtained for a given set of true polarization parameters $\psi_0=0^\circ$ and $p_0=0.10$ (dashed lines) and various configurations of the
covariance matrix, at four levels of SNR $p_0/\sigma_{p,G}=0.1, 0.5, 1$ and $5$  (top to bottom).}
\label{fig:pdf_posterior_impact_epsirho}
\end{figure*}

We illustrate the shape of the posterior pdf in Fig.~\ref{fig:pdf_posterior_impact_epsirho}, where $B_{2D}( p_0, \psi_0\, | \,p,\psi, \tens{\Sigma}_p)$ 
is shown at four levels of the SNR and five couples of ($\varepsilon$, $\rho$).
It is interesting to notice that the posterior pdf allows the polarization fraction to be null at low SNR, when these values were rejected by the pdf (see Appendix B of PMA I). 
Moreover the posterior pdf peaks at the location of the measurements used to compute it.
As largely emphasized in PMA I, we also recall that once the effective ellipticity of the covariance matrix departs  from the canonical simplification, 
the pdfs are sensitive to the initial true polarization angle $\psi_0$. 

\section{Mean Bayesian Posterior analytical expression}
\label{sec:meanposterior_analytical}

\begin{figure}
 \includegraphics[width=9cm]{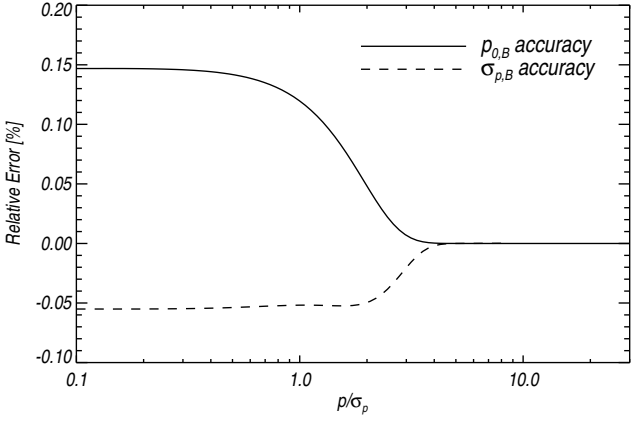}
 \caption{ Accuracy of the approximate analytical expression of the Bayesian estimates of the polarization fraction $\hat{p}_{\text{MB}}$ (solid line) and its associated
 uncertainty $\hat{\sigma}_{p,\text{MB}}$ (dashed line), as a function of the SNR of the measurement $p/\sigma_p$, where $\sigma_p$=$\sigma_{\rm Q}/I_0$=$\sigma_{\rm U}/I_0$.}
 \label{fig:bayesian_approx}
\end{figure}

\begin{figure}
\begin{tabular}{c}
 \includegraphics[width=9cm]{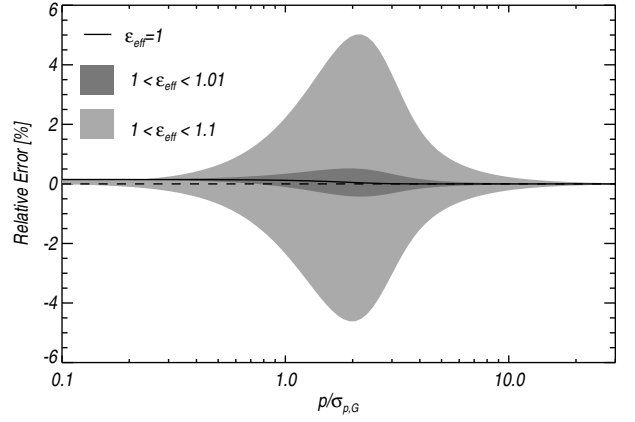}  \\
 \includegraphics[width=9cm]{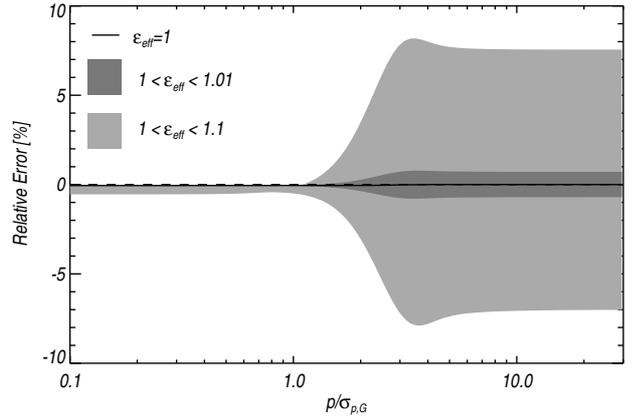}
 \end{tabular}
 \caption{ Accuracy of the generalized approximate analytical expression of the Bayesian estimates $\hat{p}_{\text{MB}}$ (top) and 
 $\hat{\sigma}_{p,\text{MB}}$ (bottom), taking into account the full covariance matrix components, in the {\it low} (light grey ) and {\it tiny} (dark grey) regimes.}
 \label{fig:bayesian_approx_fullcov}
\end{figure}

In the canonical case, the MB estimator of the polarization fraction $p$ 
takes a simple analytical expression. The Bayesian posterior on $p$ is given in this case by:
\begin{equation}
  B_p(p_0\, |\, p, \tens{\Sigma}_p)=\frac{R(p \, | \, p_0,\tens{\Sigma}_p) \cdot \kappa(p_0)}{\int_0^1 R(p\, | \,p'_0,\tens{\Sigma}_p) \, \kappa(p'_0) \,dp'_0} \, ,
\end{equation}
where $\kappa$ 
is the prior chosen equal to 1 over the definition range ([0,1]), and $R$ denotes the \citet{Rice1945}  function which is defined by
  \begin{equation}
 R(p\, | \, p_0, \tens{\Sigma}_p) = \frac{p}{\sigma_p^2}  \mathrm{exp} \left( -\frac{(p^2 + p_0^2)}{2\sigma_p^2} \right) \mathcal{I}_0\left( \frac{p p_0}{\sigma_p^2}\right) \, ,
 \label{eq:rice}
 \end{equation}
 where $\mathcal{I}_0(x)$ is the zeroth-order modified Bessel function of the first kind \citep{GradshteynRyzhik2007} 
 and $\sigma_p=\sigma_{\rm Q}/I_0=\sigma_{\rm U}/I_0$ is the characteristic noise level of the polarization fraction.

The MB estimator and the posterior variance take the following forms
\begin{equation}
\hat{p}_{\text{MB}} = \frac{  \int_0^1 p_0 \, e^{(-p_0^2/2\sigma_p^2)} \mathcal{I}_0\left(\frac{pp_0}{\sigma_p^2}\right)dp_0}
{ \int_0^1e^{(-{p_0}^2/2\sigma_p^2)} \mathcal{I}_0\left(\frac{pp_0}{\sigma_p^2}\right)dp_0  }
\end{equation}
and
\begin{equation}
\hat{\sigma}_{p,\text{MB}} = \frac{  \int_0^1 (p_0- \hat{p}_{\text{MB}})^2\, e^{(-p_0^2/2\sigma_p^2)} \mathcal{I}_0\left(\frac{pp_0}{\sigma_p^2}\right)dp_0}
{ \int_0^1e^{(-{p_0}^2/2\sigma_p^2)} \mathcal{I}_0\left(\frac{pp_0}{\sigma_p^2}\right)dp_0  }\, .
\end{equation}
If we assume in a first approximation that the integral of $p_0$ over $[0,1]$ can be taken over $[0,+\infty[$ (which is fine at high SNR), 
and we use the formula of~\cite{Prudnikov1986a}
\begin{equation}
\int_0^{\infty} x^{a-1} e^{-bx^2} \mathcal{I}_0(cx) dx = \frac{1}{2}b^{-a/2} \Gamma(a/2) _1F_1 \left( \frac{a}{2}, 1, \frac{c^2}{4b} \right) \, ,
\end{equation}
where $\Gamma$ is the Gamma function and $_1F_1$ is the confluent hypergeometric function of the first kind, we can derive that
\begin{eqnarray}
\int_0^{\infty}  e^{(-p_0^2/2\sigma_p^2)} \mathcal{I}_0\left(\frac{pp_0}{\sigma_p^2}\right)dp_0
& = &  \frac{1}{2} \left(\frac{1}{2\sigma_p^2}\right)^{-\frac{1}{2}} \Gamma\left(\frac{1}{2}\right)\,_1F_1 \left( \frac{1}{2}, 1, \frac{p^2}{2\sigma_p^2} \right) \nonumber \\
& = & \sqrt{\pi/2} \,  \sigma_p \, e^{p^2 / 4\sigma_p^2} \mathcal{I}_0\left( p^2 / 4\sigma_p^2 \right) \nonumber \\
\end{eqnarray}
and
\begin{eqnarray}
\int_0^{\infty} p_0 \, e^{(-p_0^2/2\sigma_p^2)} \mathcal{I}_0\left(\frac{pp_0}{\sigma_p^2}\right)dp_0
& = &  \sigma_p^2 \,_1F_1 \left( 1, 1, \frac{p^2}{2\sigma_p^2} \right) \nonumber \\
& = & \sigma_p^2 e^{p^2 / 2\sigma_p^2} \, ,
\end{eqnarray}
and finally
\begin{eqnarray}
\hspace{-0.5cm}
\int_0^{\infty} p_0^2 \, e^{(-p_0^2/2\sigma_p^2)} \mathcal{I}_0\left(\frac{pp_0}{\sigma_p^2}\right)dp_0
& = &  \frac{1}{2} \left(\frac{1}{2\sigma_p^2}\right)^{-\frac{3}{2}} \Gamma\left(\frac{3}{2}\right) \,_1F_1 \left( \frac{3}{2}, 1, \frac{p^2}{2\sigma_p^2} \right) \nonumber \\
& = & \sqrt{\pi/2} \, \sigma_p^3\, _1F_1 \left( \frac{3}{2}, 1, \frac{p^2}{2\sigma_p^2} \right) \, .
\end{eqnarray}
We finally obtain the simple expression of the MB estimator and the associated Bayesian variance:
\begin{equation}
\label{eq:p0_bayesian_formula}
\hat{p}_{\text{MB}} = \frac{ \sigma_p \sqrt{\frac{2}{\pi}} \, \mathrm{exp}\left( \tfrac{p^2}{4\sigma_p^2} \right) }
{ \mathcal{I}_0\left( \frac{p^2}{4\sigma_p^2} \right)}  
\end{equation}
and 
\begin{equation}
\label{eq:sigp0_bayesian_formula}
\hat{\sigma}_{p,\text{MB}} =   \hat{p}_{\text{MB}}  
\sqrt{ \frac{\pi}{2}  \mathrm{exp}\left( \tfrac{-3 p^2}{4\sigma_p^2} \right) \mathcal{I}_0\left( \frac{p^2}{4\sigma_p^2} \right) \,_1F_1 \left( \frac{3}{2}, 1, \frac{p^2}{2\sigma_p^2} \right) -1} \, .
\end{equation}

As shown in Fig.~\ref{fig:bayesian_approx}, this analytical approximation gives less than 0.15\% of relative error at low 
SNR compared to the exact $\hat{p}_{\text{MB}}$ estimate and less than 0.05\% for the associated uncertainty. This small departure quickly 
tends to 0 for a SNR$>$4. Thus these expressions may be used to speed up the computing time when the canonical simplification may be assumed.

We explore in Fig.~\ref{fig:bayesian_approx_fullcov} to extent to which the canonical simplification 
may be done in the presence of an effective ellipticity  of the covariance matrix. 
In this more general case, we suggest changing $\sigma_p$ into $\sigma_{p,G}$ in the Eqs.~\ref{eq:p0_bayesian_formula} and \ref{eq:sigp0_bayesian_formula}. 
The relative error between the approximate estimate and the exact bayesian estimate has been explored in two regimes of the covariance matrix, the
{\it low} (1$<$$\varepsilon_\mathrm{eff}$$<$1.1) and {\it tiny}  (1$<$$\varepsilon_\mathrm{eff}$$<$1.01) regimes. 
Three domains are observed in the top panel of Fig.~\ref{fig:bayesian_approx_fullcov} dealing with the accuracy of the $\hat{p}_{\text{MB}}$ estimate: 
i) at low SNR ($<$1) the bias on $p$ is so large that the presence of an effective ellipticity does not affect significantly the estimate in comparison; 
ii) for an intermediate range of the SNR (1$<$SNR$<$4), the effective ellipticity of the $\tens{\Sigma}_p$ significantly affects the Bayesian 
estimate so that the departure of the analytical approximation from the exact estimate becomes important;
iii) at high SNR($>$4) the noise is so low that the Bayesian estimate is not sensitive to the asymmetry of the covariance matrix anymore.
Consequently, the approximate analytical expression provides very good estimates of $\hat{p}_{\text{MB}}$ for SNR$<$1 and SNR$>$4, 
and 5\% to 0.5\% of relative error for intermediate 1$<$SNR$<$4 in the {\it low} and {\it tiny} regimes of the covariance matrix, respectively.
Notice that in the {\it extreme} regime of the covariance matrix the relative error increases up to 20\%.

Concerning the accuracy of the Bayesian approximate estimate $\hat{\sigma}_{p,\text{MB}}$  of the polarization fraction uncertainty (bottom panel), 
the agreement is better than 0.1\% for SNR$<$1, and about 8\% SNR$>$1 in the {\it low} regime, and 1\% in the {\it tiny} regime.
Because the uncertainty becomes small compared to the polarization fraction at high SNR, 
up to 8\% of error in $\hat{\sigma}_{p,\text{MB}}$ is still acceptable for this approximation.

\bibliographystyle{aa}
\bibliography{biblio_v1.8,planck_bib_v6.0}

\end{document}